\documentclass[twocolumn,superscriptaddress,showpacs,floatfix,aps,prb,longbibliography,10pt]{revtex4-1}

\pdfoutput=1
\usepackage{graphicx}
\usepackage{amsmath}
\usepackage{amssymb}
\usepackage{amsbsy}

\usepackage[pdftitle={},
            pdfsubject={},
            pdfauthor={Roser Valenti},
            ]{hyperref}

\usepackage{color}
\usepackage{soul}

\graphicspath{{FIG/}}

\begin{document}
\author{Stephen M. Winter}
\affiliation{Institut f\"ur Theoretische Physik, Goethe-Universit\"at Frankfurt, Max-von-Laue-Str. 1, 60438 Frankfurt am Main, Germany}

\author{Alexander A. Tsirlin}
\affiliation{Experimental Physics VI, Center for Electronic Correlations and Magnetism,
University of Augsburg, 86159 Augsburg, Germany}

\author{Maria Daghofer}
\affiliation{Institut f\"ur Funktionelle Materie und 
Quantentechnologien, Universit\"at Stuttgart,
Pfaffenwaldring 57, 70569 Stuttgart, Germany}

\author{Jeroen~van~den~Brink}
\affiliation{Institute for Theoretical Solid State Physics, IFW Dresden, Helmholtzstrasse 20, 01069 Dresden, Germany}
\affiliation{Institute for Theoretical Physics, TU Dresden, 01069 Dresden, Germany}

\author{Yogesh Singh}
\affiliation{Indian Institute of Science Education and Research Mohali, Sector 81, S. A. S. Nagar, Manauli PO 140306, India}

\author{Philipp Gegenwart}
\affiliation{Experimental Physics VI, Center for Electronic Correlations and Magnetism,
University of Augsburg, 86159 Augsburg, Germany}

\author{Roser Valent\'{\i}}
\email[]{valenti@itp.uni-frankfurt.de}
\affiliation{Institut f\"ur Theoretische Physik, Goethe-Universit\"at Frankfurt, Max-von-Laue-Str. 1, 60438 Frankfurt am Main, Germany}


\title{Models and Materials for Generalized Kitaev Magnetism}

\begin{abstract}

 The exactly solvable Kitaev model  on the honeycomb
lattice has  recently received   
enormous attention linked to the hope of achieving novel spin-liquid
states with fractionalized Majorana-like excitations.  
In this review, we
analyze the mechanism proposed by G. Jackeli and G. Khaliullin to
identify Kitaev materials based on  spin-orbital dependent bond
interactions and provide a comprehensive overview of its implications
in real materials. We set the focus on
 experimental results and current theoretical understanding of planar
honeycomb systems (Na$_2$IrO$_3$, $\alpha$-Li$_2$IrO$_3$, and
$\alpha$-RuCl$_3$), three-dimensional Kitaev materials ($\beta$- and
$\gamma$-Li$_2$IrO$_3$), and other potential candidates, completing the review
with the list of open questions awaiting new insights.

\end{abstract}

\maketitle

\section{Introduction}

One of the most sought after states of matter in magnetic materials is 
a quantum spin liquid with its highly uncommon properties, such as fractionalized excitations and non-trivial entanglement. The realization of quantum spin liquid states remains, however, elusive with very few known candidates (for reviews, see Refs.~\onlinecite{Norman2016QSL} and~\onlinecite{RevModPhys.89.025003}).
The hope for finding new candidates experienced in the last decade a considerable boost triggered by (i) the formulation by Alexei Kitaev in 2006 of an exactly solvable model on the hexagonal (honeycomb) lattice with a quantum spin liquid ground state and fractionalized Majorana-like excitations,~\cite{kitaev2006anyons} and (ii) the proposal by George Jackeli and Giniyat Khaliullin in 2009 of a mechanism for designing appropriate Kitaev exchange interaction terms in spin-orbit-coupled $4d$ and $5d$ transition-metal-based insulators.\cite{Jackeli2009} Since then, an enormous amount of theoretical and experimental work has been devoted to understanding the properties of such so-called Kitaev systems
and, at the same time, it has opened new fields of research.

In this review, we present an extensive theoretical and experimental overview of the models and materials related to the Jackeli-Khaliullin mechanism, and discuss our present understanding of their properties as well as future directions.

\section{Theoretical Considerations}
\label{sec:theory}
\subsection{The Kitaev Honeycomb Model}

\begin{figure}[t]
\includegraphics[width=\linewidth]{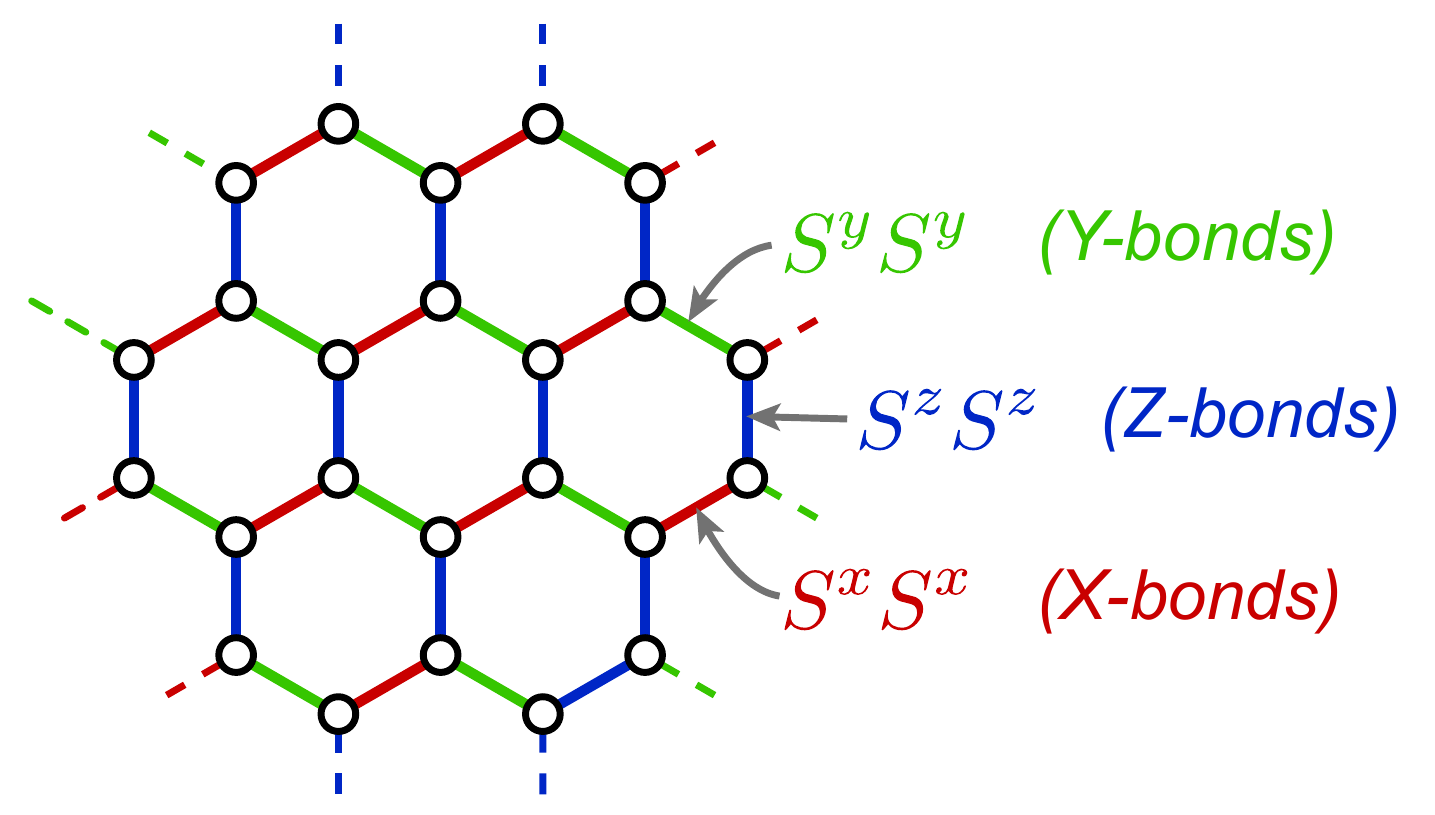}
\caption{\label{fig-lattices}(Color online) Definition of the interactions in Kitaev's honeycomb model. The so-called X-, Y-, and Z-bonds host orthogonal Ising interactions. }
\end{figure}

We begin with a brief review of Kitaev's much-studied honeycomb model, and its exact solution.\cite{kitaev2006anyons} A more in-depth review can be found, for example, in Refs.~\onlinecite{kitaev2006anyons,knolle2016dynamics,PhysRevB.86.224417}. The model belongs to a larger class of so-called quantum compass Hamiltonians,\cite{nussinov2015} in which spin-spin interactions along each bond are anisotropic, and depend on the orientation of the bond. For Kitaev's, there are three flavours of bonds emerging from each site on the honeycomb lattice; these bonds host orthogonal Ising interactions:
\begin{align}
\mathcal{H} = \sum_{\langle ij\rangle} S_i^\gamma S_j^\gamma
\end{align}
where $\gamma = \{x,y,z\}$. Such bonds are labelled X-, Y- and Z-bonds, respectively, as shown in Fig.~\ref{fig-lattices}. Exact solution of the model is accomplished through representation of the spin operators in terms of four types of Majorana fermions $\{b_i^x,b_i^y,b_i^z,c_i\}$, such that $S_i^\gamma = \frac{i}{2} b_i^\gamma c_i$. The Hamiltonian is then written:
\begin{align}
\mathcal{H} = \frac{1}{4}\sum_{\langle ij\rangle} b_i^\gamma b_j^\gamma c_i c_j
\end{align}
From this form, it can be seen that the $b^\gamma$ fermions are completely local entities, since bonds of any given type are disconnected from other bonds of the same type. For this reason, $u_{ij} = ib_i^\gamma b_j^\gamma = \pm 1$ is a constant of motion. In this sense, the $b^\gamma$ operators associated with each bond can be replaced by their (self-consistently determined) expectation values, providing the quadratic Hamiltonian:
\begin{align}
\mathcal{H} = \frac{-i}{4}\sum_{ij} \langle u_{ij} \rangle c_i c_j
\end{align}
This form can be exactly diagonalized for a given configuration of $\langle u_{ij} \rangle$. The states in this representation are therefore defined by the configuration of ``flux'' variables $u_{ij}$ and ``matter'' $c$ fermions. Since the Majorana basis is an over-complete representation, one must, however, be careful to identify gauge distinct configurations. 

The description of the ground state was given by Kitaev,\cite{kitaev2006anyons} with reference to earlier work by Lieb.\cite{PhysRevLett.73.2158} The ground state possesses long-range order in the emergent flux degrees of freedom described by the gauge-invariant plaquette operator $W_p = 2^6 S_1^x S_2^y S_3^z S_4^x S_5^y S_6^z = \prod_{i=1}^{6} S_i^\gamma S_{i+1}^\gamma = \prod_{i=1}^6 u_{i,i+1}$. On the honeycomb lattice, the lowest energy corresponds to the ``flux-free'' condition with $W_p = +1$ on every six-site hexagonal plaquette. Since $W_p$ does not commute with the local spin operators, this ``flux-ordered'' ground state cannot exhibit any long-range spin order, and 
instead is a $\mathbb{Z}_2$ spin-liquid with only short range nearest neighbour spin-spin correlations. Much of the interest in this phase arises from Kitaev's observation that the gapped phase appearing in finite magnetic field displays anyonic excitations that may be relevant to applications in topological quantum computing.\cite{kitaev2006anyons} 

From the theoretical side, the availability of an exact solution has facilitated a significant understanding of the model, with major advancements in descriptions of the dynamics, and topological properties.\cite{kitaev2006anyons,knolle2016dynamics,PhysRevB.92.115127,PhysRevLett.112.207203,perreault2016resonant,PhysRevLett.113.187201,perreault2015} These aspects have been reviewed elsewhere.\cite{trebst2017kitaev,kitaev2010topological,hermanns2017physics} From the experimental perspective, the relative simplicity of the Kitaev model has inspired the possibility for realization in real materials. Indeed, only a few years after Kitaev's work, a mechanism for designing the required Ising terms in Mott insulators with heavy transition metals that exhibit strong spin-orbit coupling was put forward by Jackeli and Khaliullin.\cite{Jackeli2009} This mechanism is discussed in the next section.

\subsection{The Jackeli-Khaliullin Mechanism}
\label{sec:jackeli}

\begin{figure}[t]
\includegraphics[width=\linewidth]{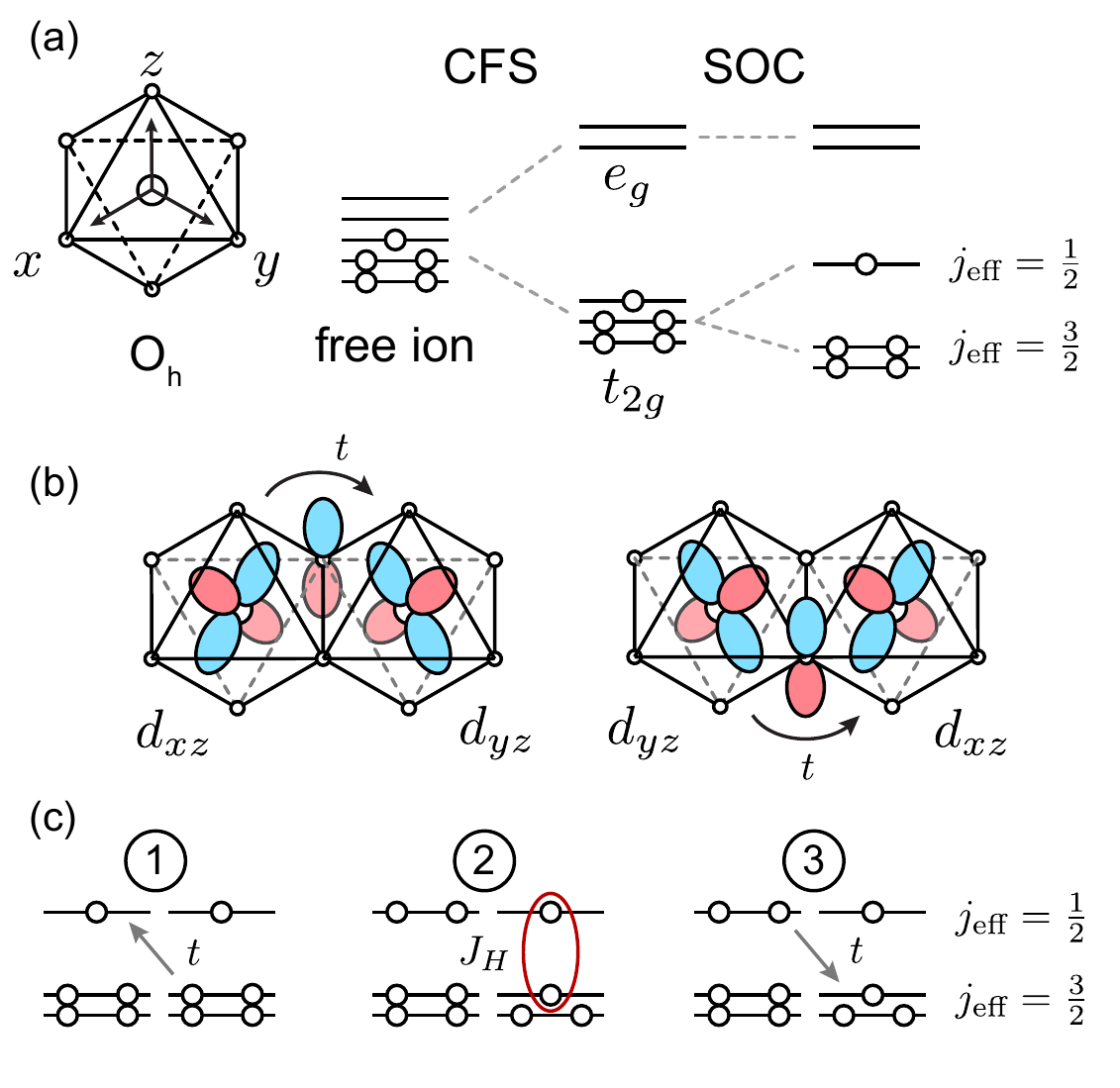}
\caption{\label{fig-JK1}(Color online) (a) Combined effect of crystal field splitting and spin-orbit coupling (SOC) on the local $d$-orbital states. (b) Summary of hopping paths considered in the idealized edge-sharing model of Jackeli and Khaliullin. (c) Schematic view of virtual processes that lead to the emergence of the Kitaev interactions for this case. }
\end{figure}
Khaliullin\cite{khaliullin2005orbital} and later Jackeli and Khaliullin\cite{Jackeli2009} studied the magnetic interactions between spin-orbital coupled $d^5$ ions in an octahedral environment. In this case, the crystal field splits the $d$-orbitals into an empty $e_g$ pair, and a triply degenerate $t_{2g}$ combination, containing one hole (Fig.~\ref{fig-JK1}(a)). The unquenched $t_{2g}$ orbital degree of freedom can lead to a variety of complex effects.\cite{khaliullin2005orbital} For heavy $4d$ and $5d$ transition metals, the direct coupling of the spin and orbital moments of the hole via $\mathcal{H} = \lambda\mathbf{L}_{t_{2g}}\cdot \mathbf{S}$ can split the $t_{2g}$ states into those with total effective angular momentum $j_\text{eff} = \frac{1}{2}$ and $\frac{3}{2}$ described by:
\begin{align}
|j_{1/2}\rangle = \left\{\begin{array}{cc}\frac{1}{\sqrt{3}}(-|xy,\uparrow\rangle - i|xz,\downarrow\rangle - |yz,\downarrow\rangle)&(m_j = +\frac{1}{2})\\
\frac{1}{\sqrt{3}}(|xy,\downarrow\rangle + i|xz,\uparrow\rangle - |yz,\uparrow\rangle)&(m_j = -\frac{1}{2})
 \end{array} \right. 
 \end{align}
 and
 \begin{align}
|j_{3/2}\rangle = \left\{\begin{array}{cc}\frac{1}{\sqrt{2}}(-i|xz,\uparrow\rangle - |yz,\uparrow\rangle)&(m_j = +\frac{3}{2})\\
\frac{1}{\sqrt{6}}(2|xy,\uparrow\rangle - i|xz,\downarrow\rangle - |yz,\downarrow\rangle)&(m_j = +\frac{1}{2})\\
\frac{1}{\sqrt{6}}(2|xy,\downarrow\rangle - i|xz,\uparrow\rangle +|yz,\uparrow\rangle)&(m_j = -\frac{1}{2})\\
\frac{1}{\sqrt{2}}(-i|xz,\downarrow\rangle + |yz,\downarrow\rangle)&(m_j = -\frac{3}{2})
 \end{array} \right. 
\end{align}
In the limit of large Hubbard $U$, one hole is localized on each $d^5$ metal atom, and the low-energy degrees of freedom are the local $j_\text{eff} = \frac{1}{2}$ local magnetic moments. Given their spin-orbital nature, the interactions between such local moments are generally highly anisotropic\cite{PhysRev.120.91} and can be cast into the form:
\begin{align}
\mathcal{H} = \sum_{ij} J_{ij} \ \mathbf{S}_i \cdot \mathbf{S}_j + \mathbf{D}_{ij} \cdot (\mathbf{S}_i \times \mathbf{S}_j ) + \mathbf{S}_i \cdot \mathbf{\Gamma}_{ij}\cdot \mathbf{S}_j
\end{align}
where $J_{ij}$ is the isotropic Heisenberg coupling, $\mathbf{D}_{ij}$ is the Dzyaloshinskii-Moriya (DM) vector, and $\mathbf{\Gamma}_{ij}$ is the symmetric pseudo-dipolar tensor. Realization of the pure Kitaev model requires that $J_{ij},\mathbf{D}_{ij} \rightarrow 0$ for every bond, while only one component of the $\mathbf{\Gamma}_{ij}$ tensor must remain nonzero (i.e. $\Gamma_{zz} \neq 0$ for the Z-bond). 

At first, such strict conditions may appear difficult to engineer in real materials, particularly because the leading contributions to the interactions (i.e. at order $t^2/U$) are known to satisfy a hidden symmetry\cite{PhysRevLett.69.836,PhysRevB.52.10239} $\mathbf{\Gamma}_{ij} \propto \mathbf{D}_{ij} \otimes \mathbf{D}_{ij}$. This hidden symmetry is only violated by higher order contributions, for example, at order $t^2J_H/U^2$, where $J_H$ is the strength of Hund's coupling. As a result, for those bonds where the DM interaction vanishes by symmetry, $\mathbf{\Gamma}_{ij}$ also tends to be small. Inversion-symmetric bonds are therefore typically dominated by isotropic Heisenberg terms $J_{ij} \sim t^2/U$ unless special circumstances are achieved. This result applies equally for the limits of both weak and strong spin-orbit coupling.

For $d^5$ filling, the inclusion of Hund's coupling within the $t_{2g}$ orbitals allows particular compass terms to appear in the absence of DM-interactions in both corner-sharing\cite{khaliullin2001} and edge-sharing\cite{Jackeli2009} geometries. Essentially, spin-orbit entanglement transfers the bond-directional nature
of orbitals into that of pseudospins.\cite{khaliullin2005orbital} Investigation of this effect led Khaliullin\cite{khaliullin2005orbital} and later Jackeli and Khaliullin\cite{Jackeli2009} to particularly important conclusions in the context of the Kitaev exchange. These authors showed, for idealized {\it edge-sharing} octahedra with inversion symmetry, that (i) {\it all} leading order contributions $\sim t^2/U$ to the interactions vanish, (ii) $J_{ij}$ and $\mathbf{D}_{ij}$ are identically zero up to the next higher order $\sim t^2J_H/U^2$, and (iii) the only nonzero component of $\mathbf{\Gamma}_{ij}$ arising from these higher order $\sim t^2J_H/U^2$ effects is precisely the desired Kitaev term. This amazing insight spawned the entire field of research reviewed in this work.

In particular, Jackeli and Khaliullin considered the case where hopping between edge-sharing metal sites occurs only via hybridization with the intervening ligand $p$-orbitals. In this case, the hopping paths shown in Fig.~\ref{fig-JK1}(b) interfere, so that hopping of holes between \mbox{$j_\text{eff} = \frac{1}{2}$} states vanishes. In fact, the only relevant hopping takes a hole from a $j_\text{eff} = \frac{1}{2}$ state to an $m_j = \pm \frac{3}{2}$ component of the $j_\text{eff} = \frac{3}{2}$ quartet on an adjacent site (Fig.~\ref{fig-JK1}(c)). In such a virtual configuration, with two holes on a given site, Hund's coupling ($J_H$) acts between the $j_\text{eff} = \frac{1}{2}$ and excited $\frac{3}{2}$ moments, ultimately generating ferromagnetic interactions in the ground state $\propto t^2J_H/U^2$. Importantly, since only the extremal $m_j = \pm\frac{3}{2}$ components contribute, these couplings become Ising-like $S_i^\gamma S_j^\gamma$, with principle axis ($\gamma$) perpendicular to the plane of the bond. This renders precisely the desired Kitaev interaction. For edge-sharing octahedra, the three bonds emerging from each metal site naturally have orthogonal Ising axes.

While experimental studies, reviewed below, demonstrate the validity of Jackeli and Khaliullin's observations, it remains essential to  understand the modifications to the Jackeli-Khaliullin picture in real materials. Deviations from the ideal scenario result in a variety of complex phenomena.

\subsection{Extensions for Real Materials}
\label{ssec_real_mat}

Microscopically, plausible extensions of the Jackeli-Khaliullin mechanism to real materials are based mostly
on two observations: (i) a more accurate consideration of the coupling on each bond must include the effects of 
local distortions of the crystal field, direct $d$-$d$ hopping, and mixing with higher lying states outside the $t_{2g}$
manifold, and (ii) the $4d$ and $5d$ orbitals are spatially rather extended, which may generate substantial
longer-range exchange beyond nearest neighbours. In this section, we review the current understanding of each of these effects.

In the most general case, anisotropic magnetic interaction between sites $i$ and $j$ is described by the Hamiltonian:
\begin{align}
\mathcal{H}_{ij} = \mathbf{S}_i \cdot \mathbf{J}_{ij} \cdot \mathbf{S}_j
\end{align}
where $\mathbf{J}_{ij}$ is a $3 \times 3$ exchange tensor. There are different schemes to parametrize this tensor, which are appropriate for different local symmetries. 
Assuming local $C_{2h}$ symmetry of the $ij$-bond, the convention is to write the interactions:
\begin{align}
\mathcal{H}_{ij} =& \  J_{ij} \ \mathbf{S}_i \cdot \mathbf{S}_j + K_{ij} \ S_i^\gamma S_j^\gamma + \Gamma_{ij} \left(S_i^\alpha S_j^\beta + S_i^\beta S_j^\alpha \right) \nonumber \\
& \ + \Gamma_{ij}^\prime \left(S_i^{\gamma} S_j^\alpha + S_i^\gamma S_j^\beta + S_i^\alpha S_j^\gamma + S_i^\beta S_j^\gamma \right) \label{eqn-2}
\end{align}
where $\{\alpha,\beta,\gamma\} = \{y,z,x\},\{z,x,y\}$ and $\{x,y,z\}$, for the
X-, Y-, and Z-bonds, respectively. For lower symmetry local environments,
further terms may also be required to fully parameterize the interactions. For
example, a finite Dzyaloshinskii-Moriya interaction $\mathbf{D}_{ij} \cdot
\left( \mathbf{S}_i \times \mathbf{S}_j \right)$ is symmetry permitted for
second-neighbour interactions in all Kitaev candidate lattices, as well as
certain first-neighbour bonds in the 3D materials, discussed in Sec.~\ref{sec:beta-gamma}. 

Before reviewing the origin of these additional interactions, we remark that the phase diagram of Eq.~\eqref{eqn-2} has been studied in detail in various parameter regimes. The first works considered the simplest extension to Kitaev's model on the honeycomb lattice, namely the addition of a nearest neighbour $J_1$ term to yield the Heisenberg-Kitaev (HK) model, which has now been studied at the classical and quantum levels, both at zero,\cite{Chaloupka2010,PhysRevLett.110.097204,PhysRevB.90.195102,PhysRevB.95.024426,yamaji2016} and finite temperature,\cite{Reuther11,PhysRevLett.109.187201,PhysRevB.88.024410} as well as finite magnetic field.\cite{PhysRevB.83.245104,PhysRevLett.117.277202,PhysRevB.95.144427} The effects of finite off-diagonal nearest-neighbour interactions $\Gamma_1$ and $\Gamma_1^\prime$ were later considered,\cite{rau2014trigonal,PhysRevLett.110.097204,PhysRevB.92.024413,janssen2017} along with longer range second neighbour Kitaev $K_2$ terms,\cite{PhysRevX.5.041035} and Heisenberg $J_2,J_3$ interactions.\cite{PhysRevB.84.180407,PhysRevB.90.155126} These works have revealed, in addition to the Kitaev spin-liquid states appearing for large nearest neighbour Kitaev $|K_1|$ interactions, a complex variety of interesting magnetically ordered states, which are selected by the various competing anisotropic interactions. A relatively comprehensive view of these phases, in relation to the real materials, has now emerged from detailed analysis of the parameter regimes thought to be relevant to various materials.\cite{Nishimoto2014,katukuri2015strong,yadav2016kitaev,katukuri2016,katukuri2014kitaev,PhysRevLett.113.107201,PhysRevB.93.214431} The interested reader is referred to these works. Finally, significant interest in Kitaev-like models on other lattices has been prompted by the study of materials detailed in sections \ref{sec:beta-gamma} and \ref{sec:othermat}. For example, a variety of theoretical works focusing on the 3D honeycomb derivatives\cite{kimchi2014,lee2014a,lee2015,nasu2014,kimchi2015,lee2016} have now appeared, along with studies on the 2D triangular lattice,\cite{khaliullin2005orbital,li2015,rousochatzakis2016,PhysRevB.91.155135,PhysRevB.92.184416} and others.\cite{PhysRevB.89.014414}

\subsubsection{Local Distortions}
\label{sssec_distort}

 In real materials, distortion of the local crystal field environment away from perfect octahedral geometry reduces the point group symmetry at each metal atom from the ideal $O_h$ to $C_2$ or $C_3$, for example. Such lattice distortions lift the degeneracy of the $t_{2g}$ orbitals and partially quench
 the orbital angular momentum. This effect alters the nature of the $4d$ and $5d$ holes from spin-orbit entangled $j_\text{eff}=\frac{1}{2}$ states to states favouring a different mixture of spin and orbital character. Accordingly, the effective magnetic couplings also interpolate between different regimes, depending on the strength of spin-orbit coupling in relation to the magnitude of the induced $t_{2g}$ splitting. For example, for distortions that completely lift the $t_{2g}$ degeneracy, the local moments are continuously deformed into conventional pure $s=\frac{1}{2}$ states, which exhibit nearly isotropic Heisenberg interactions, as the orbital angular momentum is progressively quenched. Otherwise, coupling of the spin to a partially quenched orbital momentum may produce alternate anisotropic exchange interactions beyond the ideal Kitaev terms. 

\begin{figure}[t]
\includegraphics[width=\linewidth]{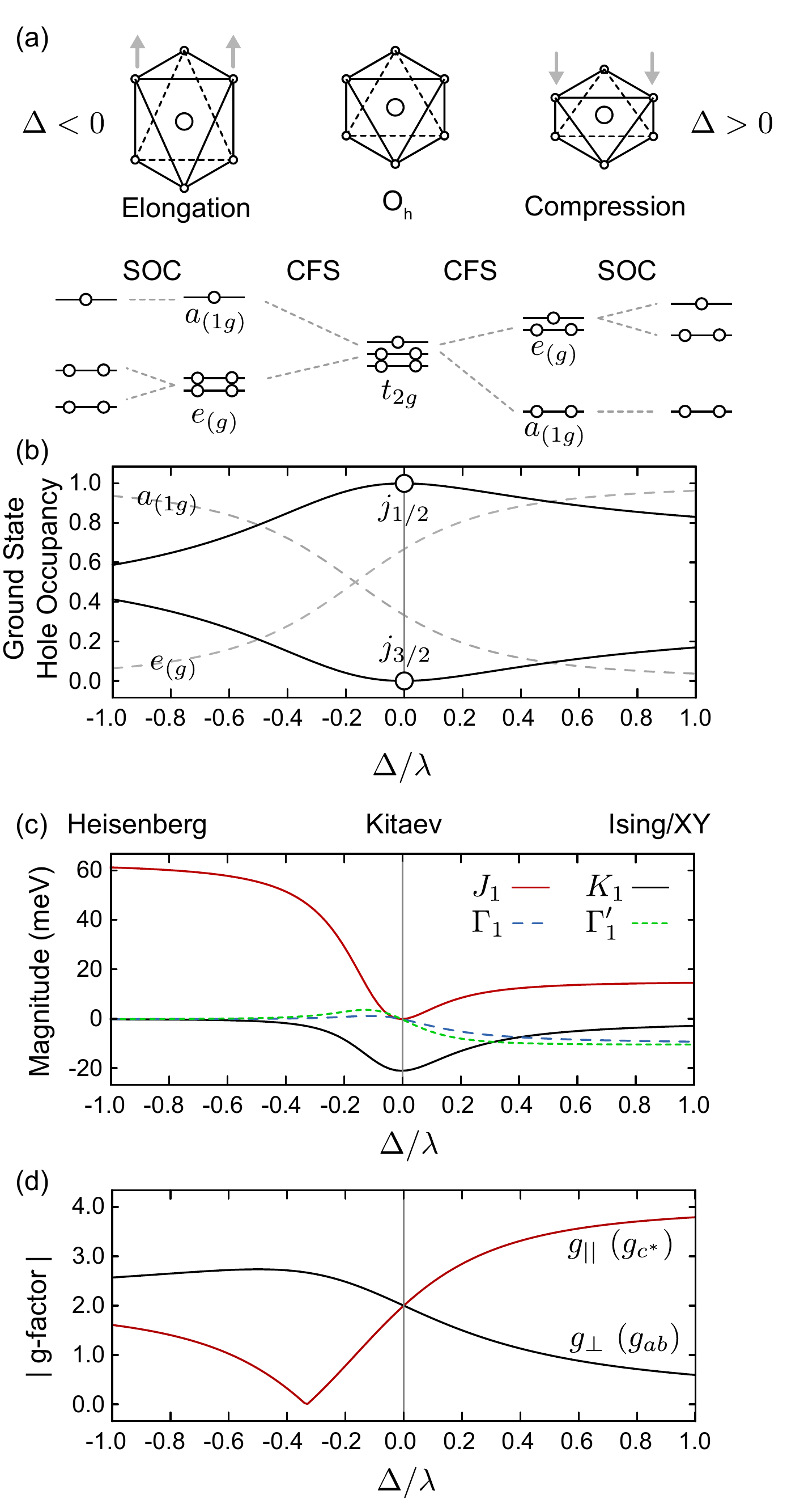}
\caption{\label{fig-CFS1b}(Color online) (a) Effects of trigonal distortion on the $d$-orbital states. (b) Evolution of the composition of the $t_{2g}$ hole with crystal field splitting. The contribution from $j_\text{eff} = \frac{1}{2}$ states remains large over a wide range. (c) Modification of the nearest neighbour interactions for pure ligand-assisted ($t_2$) hopping. (d) Induced anisotropy of the $g$-factor; $||$ refers to the normal of the honeycomb plane.}
\end{figure}

The effects of local distortions of the crystal field can be illustrated by reviewing the simplest relevant case where $C_3$ symmetry is retained, such as considered in Ref.~\onlinecite{PhysRevB.94.064435,rau2014trigonal,PhysRevB.90.155126}. Such distortions include trigonal compression or elongation of the octahedra, as shown in Fig.~\ref{fig-CFS1b}(a). In this case, the $t_{2g}$ manifold is split into singly degenerate $a_{(1g)}$ and doubly degenerate $e_{(g)}$ orbitals (for $\lambda = 0$).
For $\lambda \neq 0$, Fig.~\ref{fig-CFS1b}(b) shows
the ground state hole occupancy as a function of $\Delta/\lambda$
expressed in both,  the $j_\text{eff}$ and the $t_{2g}$  basis.
 For a distortion with a [111] principal axis, in terms of the cubic $\{x,y,z\}$ axes,\cite{PhysRevB.90.155126} the $a_{(1g)}$ and $e_{(g)}$ orbitals are:
\begin{align}
|a_{(1g)}\rangle = \frac{1}{\sqrt{3}} \left(|xy\rangle + |xz\rangle + |yz\rangle \right) \ \ , \ \ E_{a1} = -2\Delta \\
|e_{(g)}\rangle = \left\{\begin{array}{c} \frac{1}{\sqrt{6}} \left( 2|xy\rangle -|xz\rangle -|yz \rangle  \right) \\ \frac{1}{\sqrt{2}}\left(|xz\rangle - |yz\rangle \right) \end{array}  \right\} \ \ , \ \ E_{e} = +\Delta
\end{align}

For $\Delta > 0$, the $4d$ or $5d$ hole mostly occupies the $e$ orbitals, resulting in unquenched orbital angular momentum that couples to the spin, splitting the $e$ orbitals into two spin-orbital doublets. The limit of large distortion $\Delta \gg \lambda$ was studied in Refs.~\onlinecite{khaliullin2005orbital,1367-2630-14-7-073015} for the case of pure ligand-assisted hopping. In this case, the nearest neighbour Kitaev coupling vanishes ($K_1 \rightarrow 0$), to be replaced by large off-diagonal interactions $\Gamma_1 = \Gamma_1^\prime$, as shown in Fig.~\ref{fig-CFS1b}(c). 

After a coordinate rotation, the Hamiltonian of Eq.~\eqref{eqn-2} becomes, in this limit:
\begin{align}
\mathcal{H} = \sum_{\langle ij \rangle} J \ \mathbf{S}_i \cdot \mathbf{S}_j  \ + B \ S_i^{\hat{n}} \ S_j^{\hat{n}}
\end{align}
where $\hat{n} \ || \ [111]$ for every bond. This is nothing more than the Heisenberg-Ising model with Ising axis perpendicular to the honeycomb plane. This regime is characterized by a strongly anisotropic $g$-factor,\cite{PhysRevB.94.064435} with $g_{||} \gg g_{\perp}$, where $||$ refers to the $[111]$ direction (Fig.~\ref{fig-CFS1b}(d)). 

For $\Delta < 0$, the $4d$ or $5d$ hole instead mostly occupies the nondegenerate $a_{1}$ orbital, completely quenching the orbital angular momentum for large $|\Delta|$. For the limit $-\Delta \gg \lambda$, all anisotropic interactions are therefore suppressed, resulting in pure spin doublets coupled by Heisenberg interactions (Fig.~\ref{fig-CFS1b}(a,c)). This regime is associated with $g_{\perp} > g_{||}$\cite{PhysRevB.94.064435} (Fig.~\ref{fig-CFS1b}(d)).

It is worth noting that even a small a trigonal crystal field splitting $\Delta/\lambda
\sim 0.2$ may result in a significant modification of the local magnetic
interactions. For this reason, quantification of $\Delta$ through estimates of
the anisotropic $g$-tensor and through RIXS measurements\cite{PhysRevB.91.155104} of
the $d$-$d$ transition energies provides vital information about the
composition of the low-energy magnetic degrees of freedom. Controlling the
ratio $\Delta/\lambda$ represents a significant synthetic goal in designing
Kitaev-Jackeli-Khaliullin materials.

\subsubsection{General hopping scenario} 
\label{sssec_ddhop}

\begin{figure}[t]
\includegraphics[width=0.95\linewidth]{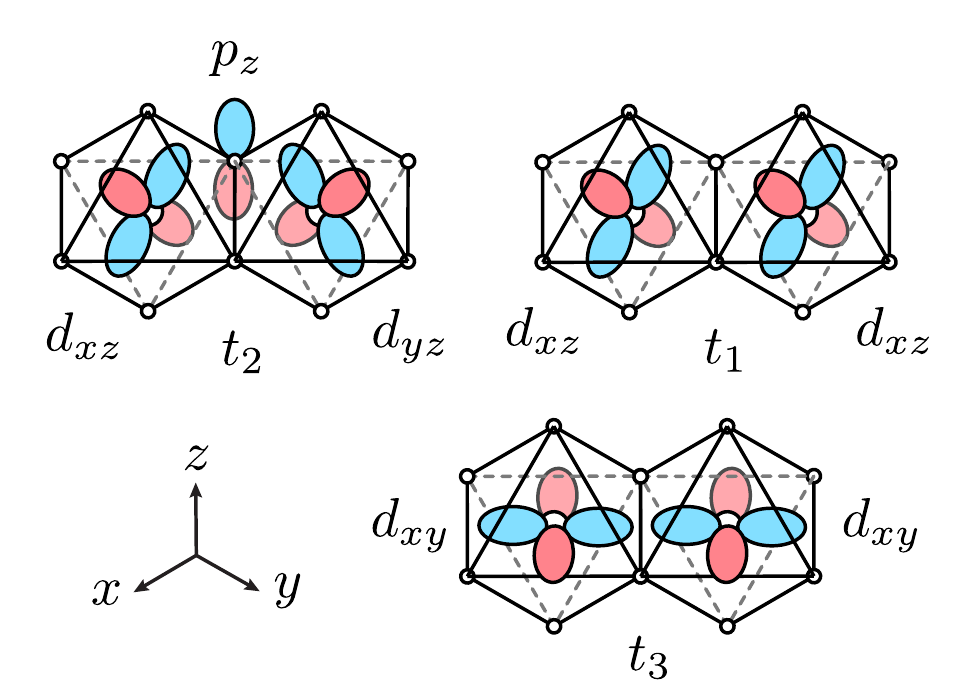}
\caption{\label{fig-hop1}(Color online) Contributions to nearest neighbour hopping interactions in edge-sharing octahedra (Z-bond). While $t_2$ is dominated by ligand-assisted hopping, $t_1$ and $t_3$ arise mainly from direct metal-metal hopping.}
\end{figure}

As discussed in Refs.~\onlinecite{PhysRevLett.112.077204,rau2014trigonal,PhysRevB.93.214431,PhysRevB.90.155126}, additional magnetic interactions arising from non-ligand assisted direct $d-d$ hopping may also induce significant deviations from the pure Kitaev interactions in real materials. This is particularly true because the heavy $4d$ and $5d$ elements possess rather diffuse orbitals, which may have a significant direct overlap. For the Z-bond, assuming $C_{2h}$ symmetry, the $d$-$d$ hopping matrix may generally be written (in the notation of Ref.~\onlinecite{PhysRevLett.112.077204}):
\begin{align}
\begin{array}{c|ccc} 
& d_{i,yz} & d_{i,xz} & d_{i,xy} \\ \hline
d_{j,yz}&t_1 &t_2&t_4\\
d_{j,xz}&t_2 &t_1&t_4\\
d_{j,xy}&t_4&t_4&t_3
 \end{array}
\end{align}
where $t_2$ is dominated by ligand-assisted hopping, while $t_1$ and $t_3$ arise primarily from direct metal-metal interactions (Fig.~\ref{fig-hop1}). The typically smaller $t_4$ vanishes for perfect $O_h$ local geometry, and is therefore associated with local distortions of the metal octahedra discussed above.\cite{rau2014trigonal} In terms of these hopping integrals, the magnetic interactions, up to second order,\cite{PhysRevB.93.214431,PhysRevLett.112.077204} are given by:
\begin{align}
J_{ij} =& \  \frac{4\mathbb{A}}{9} \left( 2t_1 + t_3\right)^2 - \frac{8\mathbb{B}}{9} \left\{ 9t_4^2 + 2(t_1 - t_3)^2\right\} \\
K_{ij} =& \  \frac{8\mathbb{B}}{3} \left\{ (t_1 - t_3)^2 + 3t_4^2 - 3t_2^2 \right\} \\
\Gamma_{ij}=& \ \frac{8\mathbb{B}}{3} \left\{ 2t_2(t_1 - t_3) + 3t_4^2 \right\}  \\
\Gamma_{ij}^\prime = & \ \frac{8\mathbb{B}}{3} \left\{ t_4 (3t_2 + t_3 - t_1) \right\} 
\end{align}
for $\mathbb{A} \sim 1/U\gg \mathbb{B} \sim J_H/(3U^2)$, in terms of the local Coulomb repulsion $U$ and Hund's coupling $J_H$. As discussed above, the presence of an inversion center between sites $i$ and $j$ forbids low-order contributions $\propto \mathbb{A}$ to the anisotropic $K,\Gamma$ and $\Gamma^\prime$ terms. The anisotropic exchange arises completely from the effects of Hund's coupling, as in the Jackeli-Khaliullin mechanism.

\begin{figure}[t]
\includegraphics[width=0.8\linewidth]{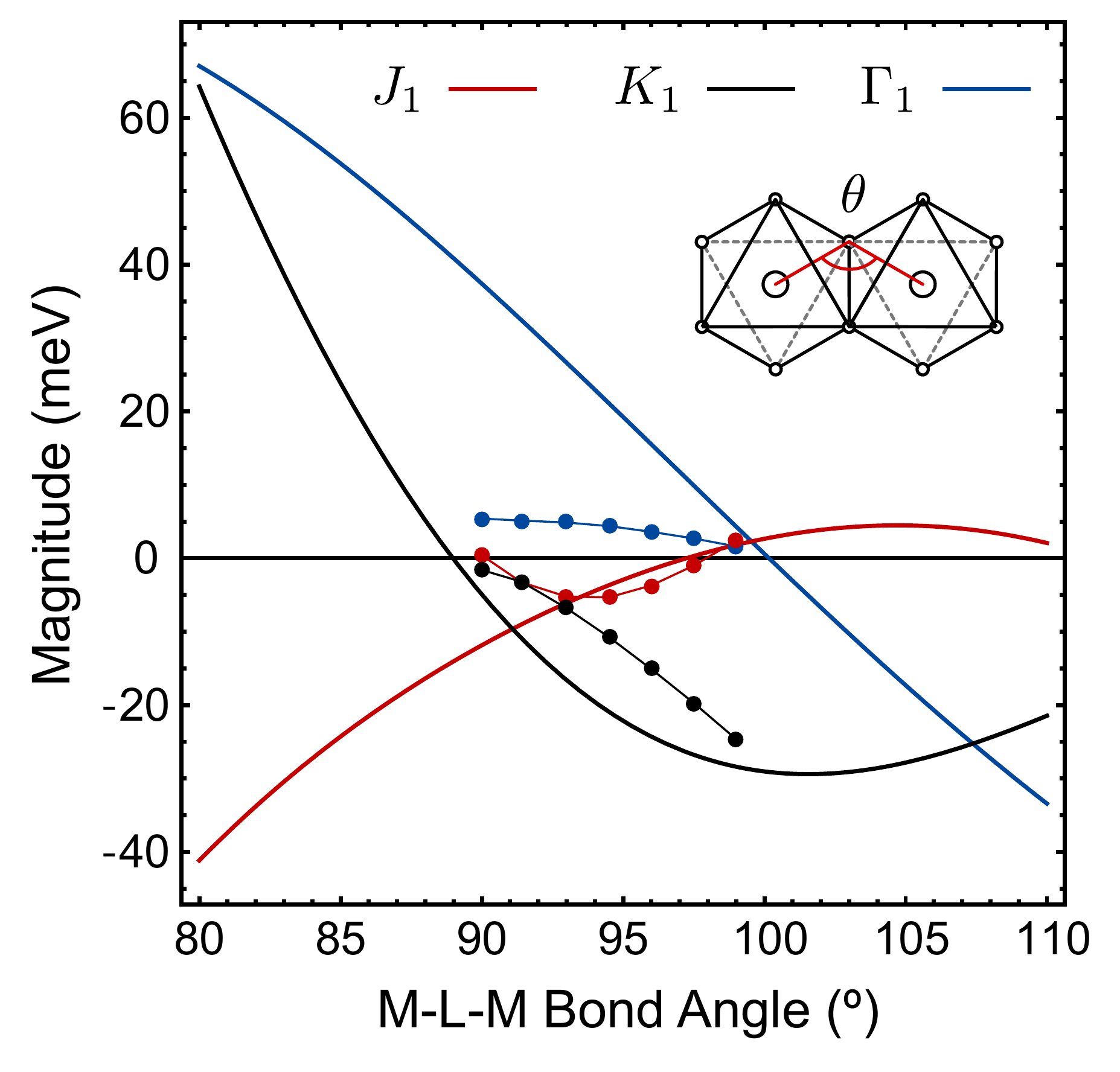}
\caption{\label{fig-hop}(Color online) Dependence of the nearest neighbour magnetic interactions on metal-ligand-metal (M-L-M) bond angle $\theta$. The solid lines represent data from perturbation theory in Ref.~\onlinecite{PhysRevB.93.214431}, and do not include the effects of crystal field splitting or $t_{2g}-e_g$ mixing. Points represent data from quantum chemistry calculations in Ref.~\onlinecite{Nishimoto2014} for an idealized structure of $\alpha$-Li$_2$IrO$_3$. Both methods show similar trends: $K_1<0$ is dominant only near $\theta \sim 100^\circ$.}
\end{figure}

The effects of direct metal-metal hopping on the interactions are controlled
primarily by the metal-metal bond distance, or alternately the
metal-ligand-metal (M-L-M) bond angle, which modulates the strength of $t_1$
and $t_3$ hopping.\cite{PhysRevB.93.214431} For the large M-L-M bond angles $>
90^\circ$ typically found in real materials, $t_1$ and $t_3$ are partly
suppressed, leading to dominant ferromagnetic Kitaev interactions $K_1 < 0$ as
proposed in the original Jackeli-Khaliullin mechanism. In contrast, small M-L-M
bond angles (large $t_1$ and $t_3$) may provide instead an antiferromagnetic
Kitaev term $K_1 > 0$, and large $\Gamma_1 > 0$ and $J_1 > 0$ (Fig.~\ref{fig-hop}). It can be
expected that the real materials lie somewhere between these two extremes,
suggesting the relevant interactions for real materials include a ferromagnetic
nearest neighbour Kitaev term, supplemented by finite $J_1$ and $\Gamma_1$.
This expectation has been confirmed by various {\it ab-initio} studies on a
variety of Kitaev materials.\cite{PhysRevB.93.214431,katukuri2014kitaev,Nishimoto2014,yadav2016kitaev,PhysRevB.91.241110}
As discussed in Refs.~\onlinecite{PhysRevLett.112.077204,lee2015}, this region
of nearest-neighbour interactions supports, on various lattices, both collinear
zigzag antiferromagnetic order, and incommensurate noncollinear orders, which
are consistent with the observed ground states in the known Kitaev candidate
materials (discussed in detail below).  The application of external pressure is
generally expected to compress the metal-metal bonds, suppressing $K_1$, and
shifting the materials away from the Kitaev spin-liquid.\cite{kim2016} 

\subsubsection{Higher Order Nearest Neighbour Terms} 
\label{sssec_highordernn}

There also exist additional contributions to the above nearest neighbour interactions that arise from $t_{2g}$-$e_g$ mixing and metal-ligand hybridization.\cite{khaliullin2005orbital,PhysRevLett.110.097204} Combined, these higher order effects produce interactions of the form:
\begin{align}
\mathcal{H}_{ij} = I_{ij} \left( 2 S_i^\gamma S_j^\gamma - \mathbf{S}_i \cdot \mathbf{S}_j\right)
\end{align}
where:
\begin{align}
I_{ij} \sim \frac{4t_2^2}{9}\left( \frac{t_{pd\sigma}^2}{t_2\Delta_p} \frac{\tilde{J}_H}{\Delta_{eg}^2}- \frac{U_p-J_p}{\Delta_{p}^2}\right)
\end{align}
which therefore modify the Kitaev and Heisenberg couplings. Here, $\Delta_{eg}$ and $\Delta_{p}$ are the charge-transfer energies from the $t_{2g}$ to $e_g$ and ligand $p$-orbitals, respectively; $t_{pd\sigma}$ is the ligand-metal hopping integral in Slater-Koster notation, $U_p$ and $J_p$ are the ligand Coulomb parameters, and $\tilde{J}_H$ is the effective Hund's coupling between $t_{2g}$ and $e_g$ orbitals. Estimation of the microscopic parameters suggests that the two contributions to $I_{ij}$ are generally comparable and have opposite sign, therefore reducing the effects of such higher order terms. Based on Ref.~\onlinecite{Foyevtsova2013}, it is suggested that $I_{ij} > 0$, slightly shifting the real materials away from the ferromagnetic Kitaev point.

\subsubsection{ Longer Range Interactions} 
\label{sssec_longrange}

\begin{figure}[t]
\includegraphics[width=0.95\linewidth]{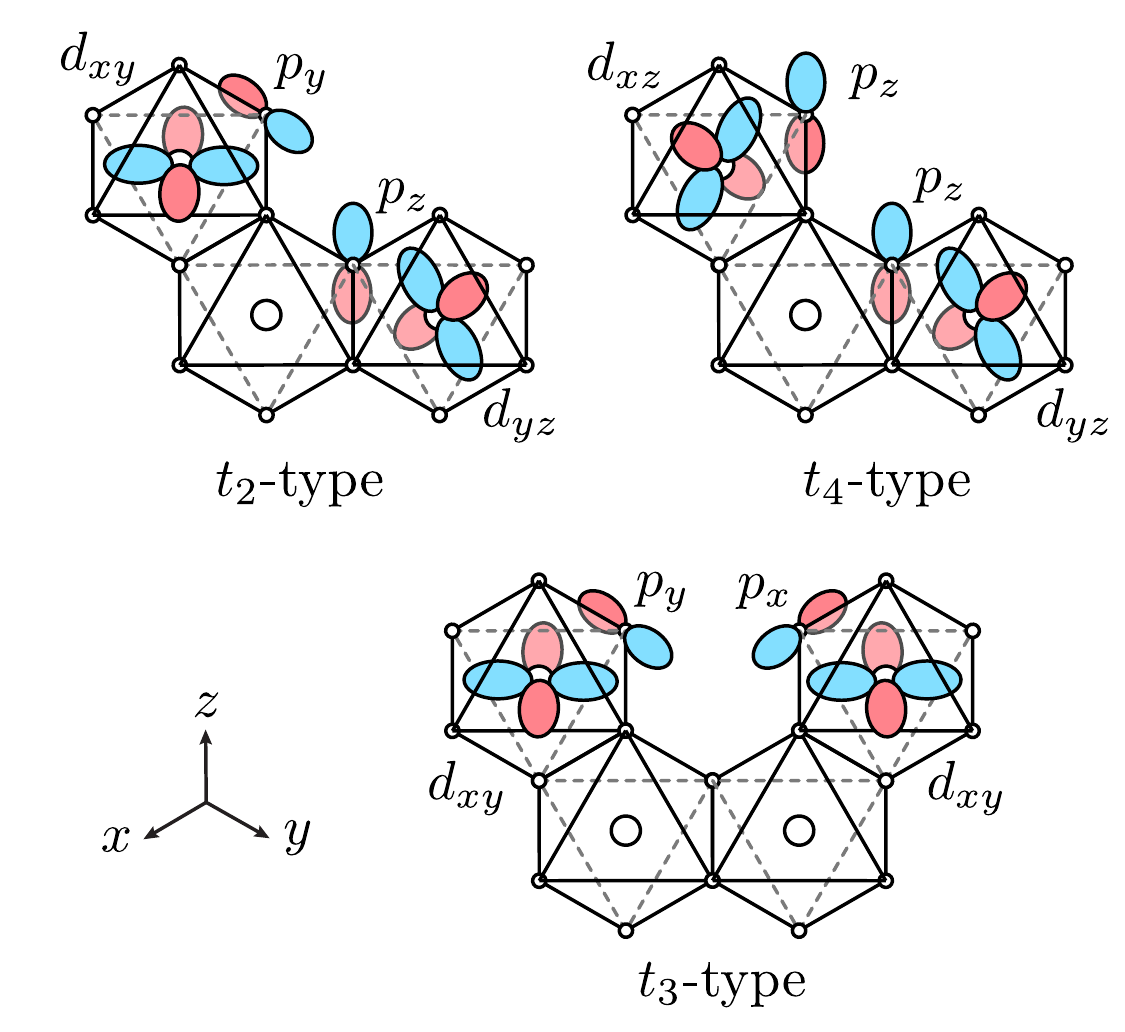}
\caption{\label{fig-hop2}(Color online) Main contributions to second and third neighbour hopping interactions from metal-ligand-ligand-metal (M-L-L-M) hopping paths. For second neighbour bonds, the dominant hopping integrals are of the $t_2$ and $t_4$ type, resulting in primarily anisotropic magnetic interactions. In contrast, third neighbour paths are of $t_3$-type, resulting in Heisenberg interactions.}
\end{figure}

A key feature of the Jackeli-Khaliullin mechanism is that the dominant Kitaev $K_1 \propto \mathbb{B}$ interactions emerge only due to strong suppression of the typically large $J_1 \propto \mathbb{A}$ couplings via carefully tuned bonding geometry. However, even if such a geometry is realised, there is no mechanism to suppress further neighbour interactions, which may remain sizable compared to the nearest-neighbour Kitaev term.\cite{PhysRevB.93.214431} For this there are two reasons: i) the $4d$ and $5d$ holes may be only weakly localized due to large $t/U$ ratios, and ii) significant long-range hopping terms arise in the real materials from various M-L-L-M hopping pathways occasioned by short ligand-ligand distances within the van der Waals radii. 

For second neighbour bonds, the largest M-L-L-M hopping integrals are of the $t_2$ and $t_4$ type (Fig.~\ref{fig-hop2}). This, combined with the typical absence of an inversion centre, allows large anisotropic terms to appear at low-order $K_2,\Gamma_2,\mathbf{D}_2\propto \mathbb{A}$. Of these, the presence of a finite Dzyaloshinkii-Moriya interaction $\mathbf{D}_2 \cdot (\mathbf{S}_i \times \mathbf{S}_j)$ has been suggested to play a role in stabilizing the incommensurate spiral orders observed in $\alpha,\beta,\gamma$-Li$_2$IrO$_3$.\cite{PhysRevB.93.214431} Otherwise, only the effects of second neighbour $J_2$ and $K_2$ terms have been studied in detail (see, e.g. Refs.~\onlinecite{PhysRevB.90.155126,PhysRevX.5.041035}).

For third neighbour bonds across a honeycomb plaquette, the largest M-L-L-M hopping integrals are of the $t_3$ type. This fact, combined with the typical presence of an inversion center, allows only low-order contributions to the Heisenberg coupling, resulting in large $J_3$ interactions. This latter interaction tends to stabilize the zigzag order observed in $\alpha$-RuCl$_3$ and Na$_2$IrO$_3$, as discussed below in Section \ref{sssec_ir_rh_thermag} and \ref{sssec_rucl_thermag}.

\section{Honeycomb Lattice Materials and Derivatives}
\label{sec_exp_rev}

\subsection{First candidates: \\ Na$_2$IrO$_3$, $\alpha$-Li$_2$IrO$_3$, and Li$_2$RhO$_3$}
\label{ssec_honeycomb_ir_rh}
The edge-sharing octahedra of $d^5$ ions required by the Jackeli-Khaliullin
mechanism are commonly found in A$_2$MO$_3$-type compounds. In this case,
octahedrally coordinated tetravalent M$^{4+}$ ions form honeycomb planes
interleaved by monovalent A$^+$ ions. Historically, Na$_2$IrO$_3$ was the first
Kitaev material extensively studied at low temperatures in
2010,\cite{singh2010} nearly six decades after its original synthesis in
1950's.\cite{scheer1955,mcdaniel1974} Two isostructural and isoelectronic
compounds, $\alpha$-Li$_2$IrO$_3$~\cite{kobayashi2003} and
Li$_2$RhO$_3$,\cite{scheer1955,todorova2011} were identified shortly
afterwards.\cite{singh2012,luo2013} These honeycomb materials serve as focus
of this section.

\subsubsection{Synthesis and Structure}
\label{sssec_ir_rh_syn}
Crystal growth of iridates and rhodates is notoriously difficult. Floating-zone techniques are inapplicable, because feasible oxygen pressures are not high enough to stabilize Ir$^{4+}$ and Rh$^{4+}$ during growth.\cite{omalley2008} Chloride fluxes routinely used for perovskite-type iridates~\cite{cao1998} could not be adapted for honeycomb iridates with alkaline metals.\cite{freund2016} On the other hand, vapor transport proved to be efficient, but is often employed in an open system, in stark contrast to the conventional realization of the method. 

For example, while polycrystalline samples of Na$_2$IrO$_3$ are synthesized by annealing Na$_2$CO$_3$ and IrO$_2$, single crystals are obtained by a technique as simple as further annealing the resulting polycrystals in air.\cite{singh2010} Minor excess of IrO$_2$ facilitates the growth.\cite{Mannithesis} The detailed mechanism of this process remains to be understood, but it seems plausible that sodium and iridium oxides evaporate and react to produce Na$_2$IrO$_3$ single crystals with the linear dimensions of several mm on the surface of a polycrystalline sample.\cite{singh2010} 

\begin{figure}[t]
\includegraphics[width=\linewidth]{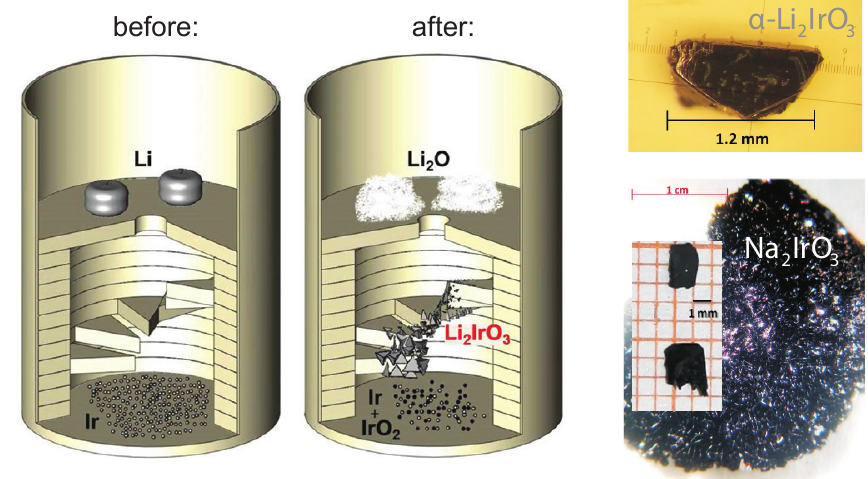}
\caption{\label{fig-ligrowth}(Color online) Crystal growth procedure for $\alpha$-Li$_2$IrO$_3$. Li and Ir educts are separated in space, whereas small single crystals grow on spikes placed in the middle of the crucible. The resulting $\alpha$-Li$_2$IrO$_3$ crystal is shown in the upper right panel. For comparison, in the bottom right panel we show Na$_2$IrO$_3$ crystals grown on the polycrystalline bed by simple annealing in air. The figure is adapted from Refs.~\onlinecite{freund2016} and~\onlinecite{Mannithesis}.}
\end{figure}

For growing $\alpha$-Li$_2$IrO$_3$ crystals, additional arrangements are required (Fig.~\ref{fig-ligrowth}). Li metal and Ir metal are placed in different parts of the growth crucible. Upon annealing in air, they form, respectively, gaseous lithium hydroxide and iridium oxide that meet to form crystals of $\alpha$-Li$_2$IrO$_3$ on spikes deliberately placed in the middle.\cite{freund2016} Synthesis of $\alpha$-Li$_2$IrO$_3$ is always a trade-off between increasing temperature to alleviate structural defects and decreasing it to avoid formation of the $\beta$-polymorph that becomes stable above 1000\,$^{\circ}$C (see below). Twinning poses a further difficulty, because $\alpha$-Li$_2$IrO$_3$ is unfortunate to suffer from several twinning mechanisms.\cite{freund2016} High-quality mono-domain crystals of $\alpha$-Li$_2$IrO$_3$ have typical sizes well below 1\,mm; larger crystals are doomed to be twinned. Whereas single crystals could be prepared by vapor transport only, the best polycrystalline samples are, somewhat counter-intuitively, obtained from chloride flux.\cite{Mannithesis} The flux reduces the annealing temperature by facilitating diffusion without leading to the actual crystal growth. Structural (dis)order of the $\alpha$-Li$_2$IrO$_3$ samples should be carefully controlled, because stacking faults effectively wash magnetic transitions out
~\cite{Mannithesis} and lead to the apparent paramagnetic behavior that was confusingly reported in early studies of this material.\cite{kobayashi2003} 

Synthesis of Li$_2$RhO$_3$ is even more complicated, to the extent that no single crystals were obtained so far. Although lithium rhodate does not form high-temperature polymorphs, its thermal stability is severely limited by the fact that Rh$^{4+}$ transforms into Rh$^{3+}$ upon heating.\cite{Mannithesis} 

It should be noted that the honeycomb iridates and rhodates are air-sensitive. On a time scale of several hours, they react with air moisture and CO$_2$ producing alkali-metal carbonates while changing the oxidation state of iridium.\cite{krizan2014} Despite the retention of the honeycomb structure and only minor alterations of lattice parameters, both peak shapes in x-ray diffraction and low-temperature magnetic behavior change drastically.\cite{krizan2014} Appreciable (although non-crucial) variations in structural parameters and low-temperature properties reported by different groups may be rooted in such sample deterioration. Storing samples in dry or completely inert atmosphere is thus essential.

\begin{figure}
\includegraphics[width=\linewidth]{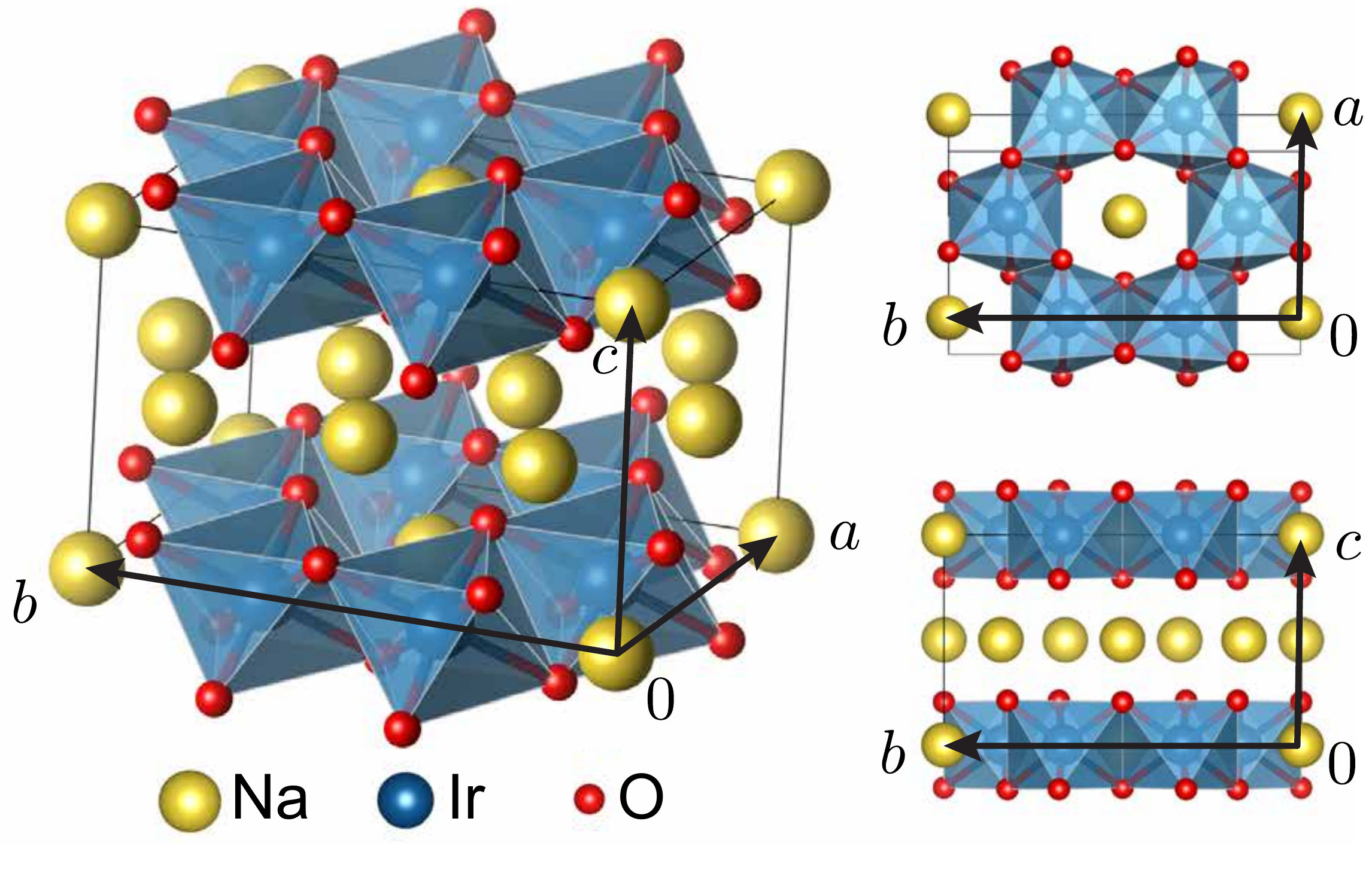}
\caption{Different views of the $C2/m$ unit cell of Na$_2$IrO$_3$; $\alpha$-Li$_2$IrO$_3$ and Li$_2$RhO$_3$ are isostructural. The structure can be described as an ordered variant of the rock salt structure containing cation layers that alternate between pure $A$ layers and mixed metal $A$Ir$_2$O$_6$ layers. Within the $A$Ir$_2$O$_6$ layers, edge sharing IrO$_6$ octahedra form an almost perfect honeycomb lattice, while the $A$ atoms occupy voids between the IrO$_6$ octahedra. }
\label{fig-nair-struc}
\end{figure}

Crystallographic work established monoclinic structures (space group $C2/m$) for both Na$_2$IrO$_3$ and $\alpha$-Li$_2$IrO$_3$, with a single crystallographic position of Ir and three nonequivalent Na/Li sites (Fig.~\ref{fig-nair-struc}). 
Several other $A_2T$O$_3$ ($A = $~Li, Na, and $T = $~Mn, Ru, Ir, Pd) type materials are also known to adopt a similar structure.\cite{Breger2005,Kobayashi1995,kobayashi2003,Panina2007} 
Like all layered structures, honeycomb iridates are prone to stacking disorder, which led to initial confusion in some early papers that described these crystals as having the $C2/c$ space group with a different stacking sequence~\cite{kobayashi2003,singh2010} or featuring the antisite Na(Li)/Ir(Rh) disorder within the $C2/m$ space group.\cite{omalley2008,todorova2011,ye2012} Such assignments are likely due to artifacts arising from the description of stacking disorder within a given crystallographic symmetry, which this disorder violates. The most accurate crystallographic information for Na$_2$IrO$_3$~\cite{choi2012} and $\alpha$-Li$_2$IrO$_3$~\cite{freund2016} was obtained by x-ray diffraction on single crystals with low concentration of stacking faults.\footnote{On the other hand, Ref.~\onlinecite{Gretarsson2013} reports local disorder in Na$_2$IrO$_3$ based on total-scattering experiments, an observation, which is difficult to reconcile with the single-crystal data of Ref.~\onlinecite{choi2012}.} While an equally accurate structure determination for Li$_2$RhO$_3$ is pending availability of single crystals, a similar $C2/m$ structure can be envisaged based on the x-ray powder data~\cite{todorova2011,luo2013} and \textit{ab-initio} results.\cite{MazinRh2013}

\subsubsection{Electronic properties}
\label{sssec_ir_rh_elec}

\begin{table*}[t]
\caption{ \label{table_2DElec} Summary of electronic parameters for honeycomb
materials Na$_2$IrO$_3$, $\alpha$-Li$_2$IrO$_3$, Li$_2$RhO$_3$, and
$\alpha$-RuCl$_3$. The latter material is discussed in section \ref{ssec_rucl}.
The source(s) of each estimate is indicated; RIXS = ``Resonant Inelastic X-ray
Scattering'', PE = ``Photoemmission'', $\Delta_{\rm ch}$ refers to the charge gap,
while $\Delta$ refers to the trigonal crystal field splitting, as defined in section
\ref{sssec_distort}.} \begin{ruledtabular}
\begin{tabular}{c|ccc|cc|cc}
Property&
Na$_2$IrO$_3$&
$\alpha$-Li$_2$IrO$_3$&&
Li$_2$RhO$_3$&&
$\alpha$-RuCl$_3$\\[1.0ex]
\hline
$\Delta_{\rm ch}$\footnote{Estimates of $\Delta_{\rm ch}$ based only on $\rho(T)$ may be unreliable.}&
$\sim$ 0.35\,eV& 
$\sim$ 0.15\,eV&&
$\sim$ 0.08\,eV&&
$1.1-1.9$\,eV\footnote{Analysis of $\rho(T)$ for $\alpha$-RuCl$_3$ yields $\Delta_{ch}\sim$ 0.15\,eV, which is likely far underestimated; see discussion in the text.} \\
&(PE\cite{Comin2012},  $\sigma(\omega)$\cite{Comin2012,Sohn2013}, $\rho(T)$\cite{Mannithesis})&
($\rho(T)$\cite{singh2012})&&
($\rho(T)$\cite{MazinRh2013})&&
(PE\cite{PhysRevB.94.161106,PhysRevLett.117.126403,sinn2016electronic}, $\sigma(\omega)$\cite{binotto1971optical,PhysRevB.94.195156})
\\[1.0ex] \hline
$\lambda$&
\multicolumn{3}{c|}{$0.4-0.5$\,eV}&
$0.1-0.15$\,eV&&
$0.1-0.15$\,eV  \\
&\multicolumn{3}{c|}{(RIXS\cite{Gretarsson2013,BHKim2014}, $\sigma(\omega)$\cite{BHKim2014})}&
({\it ab-initio}\cite{katukuri2015strong})&&
($\sigma(\omega)$\cite{PhysRevB.91.241110})\\
$\Delta$&
\multicolumn{3}{c|}{$20-50$\,meV}&
$\sim$ 60\,meV&&
$\sim$ 20\,meV\\
&\multicolumn{3}{c|}{(RIXS\cite{Gretarsson2013,BHKim2014}, {\it ab-initio}\cite{PhysRevB.93.214431,PhysRevLett.113.107201})}&
({\it ab-initio}\cite{katukuri2015strong})&&
({\it ab-initio}\cite{yadav2016kitaev,PhysRevB.93.214431})
\\
$10Dq$&
\multicolumn{3}{c|}{$\sim$ 3.3\,eV}&
$-$&&
$\sim$ $2.0-2.2$\,eV \\
&\multicolumn{3}{c|}{(RIXS\cite{BHKim2014,Ir213_rixs_gretarsson_2013})}
&&&
(PE\cite{sinn2016electronic}, XAS\cite{PhysRevB.90.041112}, $\sigma(\omega)$\cite{PhysRevB.90.041112,PhysRevB.94.195156}, {\it ab-initio}\cite{PhysRevB.91.241110,PhysRevB.90.041112})
\\ \hline
$J_H$&
\multicolumn{3}{c|}{ $0.25-0.30$\,eV}&
$-$&&
$\sim$ 0.4\,eV
 \\
&\multicolumn{3}{c|}{($\sigma(\omega)$\cite{BHKim2014}, {\it ab-initio}\cite{PhysRevLett.113.107201})}
&&&
($\sigma(\omega)$\cite{PhysRevB.94.195156})
\\
$U$&
\multicolumn{3}{c|}{$1.3-1.7$\,eV}&
 $-$&&
 $\sim$ 2.4\,eV\\
&\multicolumn{3}{c|}{($\sigma(\omega)$\cite{BHKim2014}, {\it ab-initio}\cite{PhysRevLett.113.107201})}
&&&
($\sigma(\omega)$\cite{PhysRevB.94.195156})
\end{tabular}
\end{ruledtabular}
\end{table*}

The iridate and rhodate compounds discussed in this section are robust magnetic insulators.\cite{singh2010,singh2012,luo2013,MazinRh2013} The bulk electrical resistivities of Na$_2$IrO$_3$ and $\alpha$-Li$_2$IrO$_3$ display insulating behavior with large room-temperature values of order $20-35~\Omega$\,cm, a pronounced increase upon
 cooling,\cite{singh2010,singh2012} and strong directional anisotropy.\cite{Mannithesis} Arrhenius behavior is observed in a limited temperature range near room temperature,\cite{singh2012, Mannithesis,MazinRh2013} allowing a rough estimation of the charge gaps, summarized in Table~\ref{table_2DElec}. All three systems display a three-dimensional variable range hopping temperature dependence of the electrical resistivity between 100 and 300\,K. 

The insulating nature of Na$_2$IrO$_3$ has been further probed by angle-resolved photoemission (ARPES) studies.\cite{Comin2012,alidoust2016,Lupke2015} These revealed that the filled $t_{2g}$ bands are essentially dispersionless, and show little variation in photoemission intensity with momentum, suggesting relatively localized electronic states. The character of the surface states remains somewhat controversial. Historically, early electronic structure studies of Na$_2$IrO$_3$ considered the possibility of quantum spin Hall effect and predicted metallic states on the surface.\cite{Shitade2009} A metallic linear-like surface band feature crossing the Fermi level at the $\Gamma$-point has been deduced in one ARPES study.\cite{alidoust2016} On the other hand, a scanning tunneling microscopy study on in-situ cleaved single crystals found two different reconstructed surfaces with Na deficiency and charge gaps exceeding the bulk value.\cite{Lupke2015} Surface etching facilitates crossover between different conductivity regimes along with metal-insulator transitions as a function of temperature.\cite{mehlawat2016,mehlawat2017b} That being said, attempts to estimate the bulk charge gap from photoemission yielded a value of 340\,meV, consistent with the DC resistivity measurements. 
 
The origin of the bulk charge gap in these materials has been a matter of significant discussion.\cite{MazinRh2013,kim2014antiferromagnetic,Mazin2012} On the one hand, $d^5$ rhodates are often found to be correlated metals (such as the Ruddlesden-Popper series\cite{PhysRevB.64.224424,PhysRevB.66.134431,perry2006sr2rho4,PhysRevLett.101.226402,PhysRevB.76.100402}) due to the relative weakness of Coulomb repulsion in the diffuse $4d$ orbitals. On the other hand, strong spin-orbit coupling in the $d^5$ iridates may assist in establishing an insulating state \cite{PhysRevLett.101.076402,cao2017challenge}. In either case, the appearance of a robust Mott-insulating state in the honeycomb Rh and Ir materials is not completely obvious, and several pictures have been advanced to explain this behaviour. Interestingly, such conditions indeed exist in both limits of weak and strong spin-orbit coupling.
 
\begin{figure}
\includegraphics[width=0.95\linewidth]{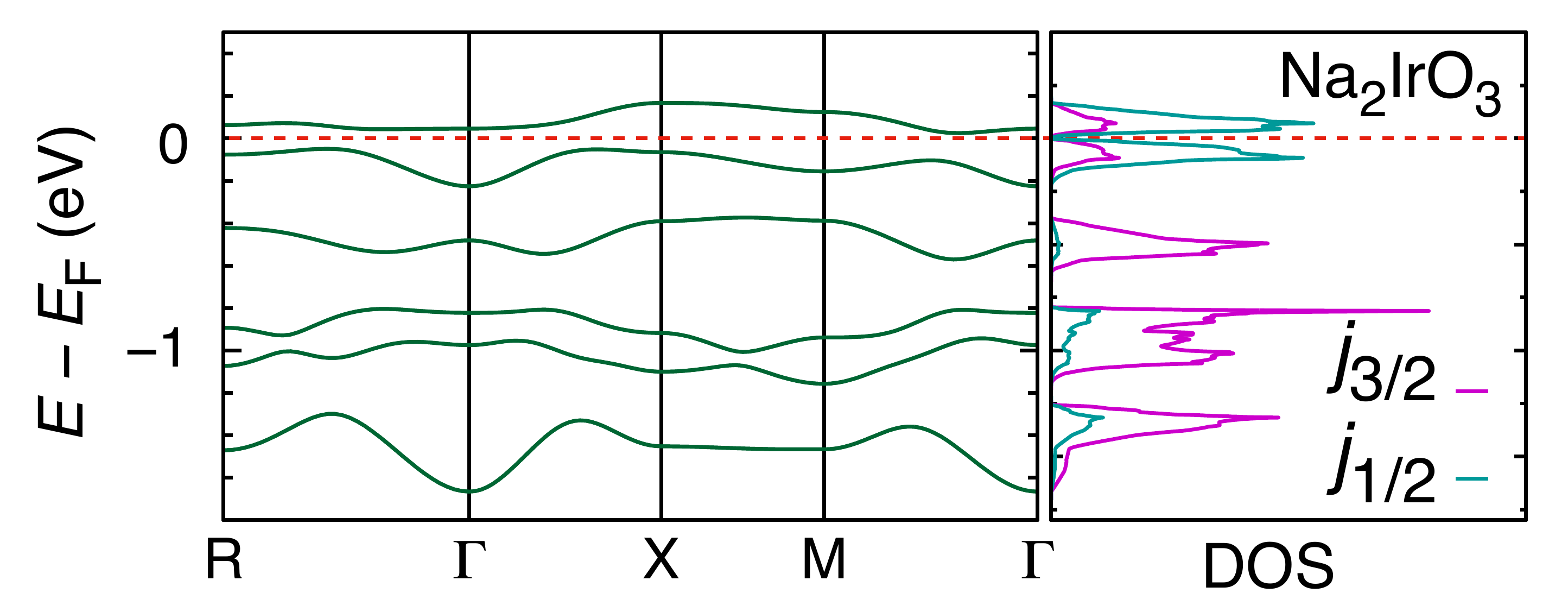}
\caption{Relativistic band structure and density of states (DOS) for Na$_2$IrO$_3$ computed at the GGA+SO level. The narrow bands near the Fermi level are predominantly $j_{\rm eff} = \frac{1}{2}$ in character.}
\label{fig-nair-bands}
\end{figure}

For Na$_2$IrO$_3$ and $\alpha$-Li$_2$IrO$_3$, strong spin-orbit coupling is now thought to play the essential role in establishing the charge gap. 
For purely oxygen-mediated ($t_2$) hopping, the hopping between $j_\text{eff} = \frac{1}{2}$ orbitals vanishes, resulting in exceedingly flat bands at the Fermi level. This condition is nearly realized in the honeycomb materials, as shown in Fig.~\ref{fig-nair-bands} for Na$_2$IrO$_3$. In fact, this is precisely the mechanism that minimizes the nearest neighbour Heisenberg couplings in the large-$\lambda,U$ limit described by Jackeli and Khaliullin. In such ``spin-orbit'' assisted Mott insulators, the $j_\text{eff}=\frac{1}{2}$ states are easily localized, even for weak Coulomb repulsion. The bands near the Fermi energy only become dispersive through mixing of the $j_\text{eff} = \frac{1}{2}$ and $\frac{3}{2}$ states.

\begin{figure}
\includegraphics[width=0.95\linewidth]{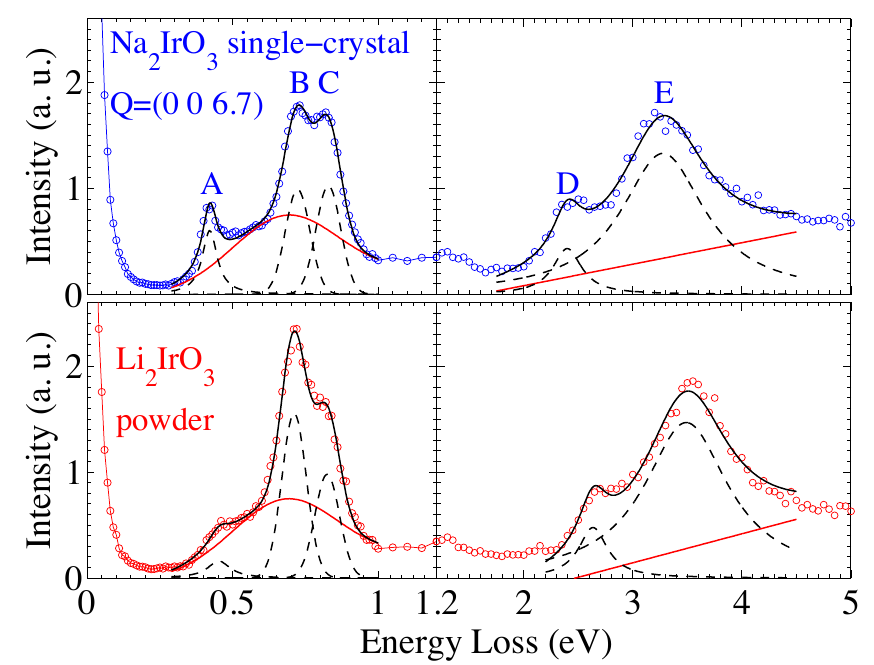}
\caption{Resonant inelastic x-ray scattering (RIXS) spectra of a Na$_2$IrO$_3$ single crystal (upper panel) and $\alpha$-Li$_2$IrO$_3$ powder (lower panel). The black and red lines represent fitted peaks and the background, respectively. The small separation of peaks B and C results from crystal field splitting of $\Delta/\lambda \sim 0.1$, indicating the validity of the $j_\text{eff}=\frac{1}{2}$ picture. Reproduced from Ref.~\onlinecite{Ir213_rixs_gretarsson_2013} by permission from the American Physical Society: \textcopyright \ 2013. }
\label{EP3}
\end{figure}

Evidence for this $j_\text{eff}$ picture in Na$_2$IrO$_3$ and Li$_2$IrO$_3$ has been obtained through detailed measurements of the crystal-field splitting of Ir $5d$ states using RIXS.\cite{Ir213_rixs_gretarsson_2013} Five characteristic peaks are found arising primarily from local $d-d$ excitations (Fig.~\ref{EP3}). Of these, peaks labelled B and C result from transitions within the $t_{2g}$ manifold from the filled $j_{\rm eff}=\frac{3}{2}$ to higher lying empty $j_{\rm eff}=\frac{1}{2}$ states.\cite{BHKim2014} Their splitting arises primarily from the trigonal distortion of the IrO$_6$ octahedra discussed in section \ref{sssec_distort}. From the position of such peaks, and the small splitting, one can estimate the trigonal crystal-field splitting $\Delta/\lambda \sim 0.1$.\cite{BHKim2014} Since $\lambda \gg \Delta$, the A$_2$IrO$_3$ systems are expected to be well described by the $j_{\rm eff}=\frac{1}{2}$ Mott insulator scenario.\cite{Ir213_rixs_gretarsson_2013} Naively, this is supported by the fact that the IrO$_6$ octahedra are not far from being regular, although in iridates distant neighbors may affect crystal-field levels significantly.\cite{Bogdanov2015}

 \begin{figure}
\includegraphics[width=0.9\linewidth]{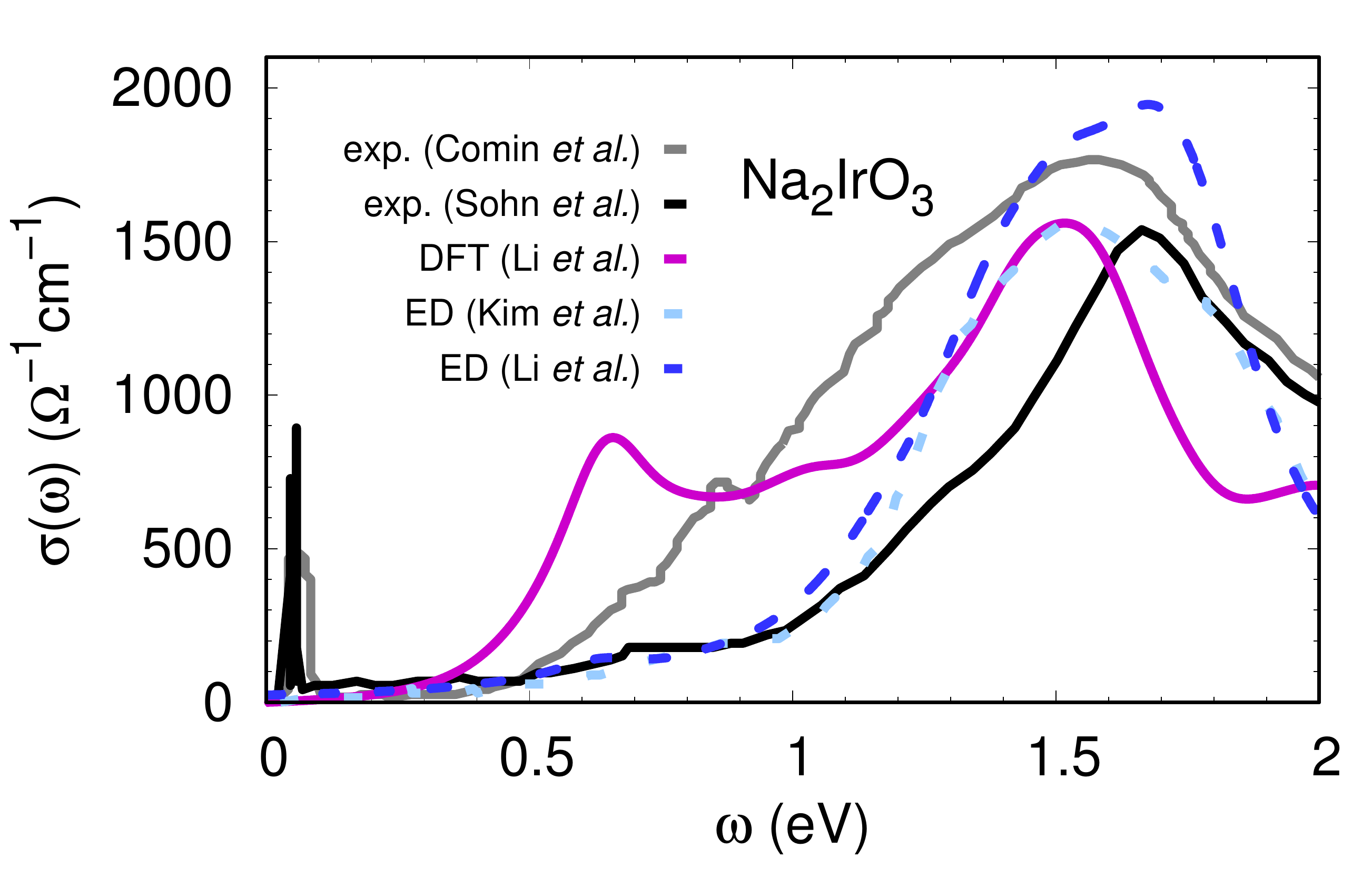}
\caption{Experimental and calculated optical conductivity of Na$_2$IrO$_3$ from Ref.: \onlinecite{Comin2012} (experiment 1), \onlinecite{Sohn2013} (experiment 2),  
\onlinecite{Yingli2015} (theory: DFT), \onlinecite{BHKim2014} (theory: ED), and \onlinecite{PhysRevB.95.045129} (theory: ED).}
\label{EP4}
\end{figure}

The optical conductivity of Na$_2$IrO$_3$ (Fig.~\ref{EP4}) displays a broad peak near 1.5\,eV and smaller features in the range between 0.5 and 1\,eV.\cite{Comin2012,Sohn2013} The onset of spectral intensity is compatible with a bulk gap of order 0.35\,eV.\cite{Comin2012} These results are well captured within the local $j_\text{eff}$ picture.\cite{BHKim2014,PhysRevB.95.045129,PhysRevLett.117.187201} The lowest energy excitations, appearing near $\omega \sim 3\lambda/2\sim 0.6-0.8$ eV, consist of local promotion of an electron from the filled $j_\text{eff} = \frac{3}{2}$ states to an empty $j_\text{eff}=\frac{1}{2}$ state at the same atomic site. These spin-orbital excitons are optically forbidden for single photon measurements when the transition-metal ion is located at an inversion center. However, they may be accessed through coupling to inversion symmetry breaking intersite excitations or phonons, leading to weak intensity at the bottom of the charge gap. The lowest energy intersite excitations consist of the transfer of electrons between $j_\text{eff} = \frac{1}{2}$ orbitals on adjacent sites, and are centered around $\omega \sim U-4J_H/3 \sim 1.1 - 1.3$\,eV. The spectral weight associated with these excitations tends to be spread across a wide energy range, and is suppressed by the small transfer integrals between such states. Thus, the dominant optical intensity appears centered around $\omega \sim U + 3\lambda/2 - 2J_H \sim 1.5-1.7$\,eV, corresponding to intersite $d^5-d^5$ $\rightarrow$ $d^4-d^6$ transitions. This observation can be taken as proof of dominant oxygen-assisted hopping. Analysis of the optical response, together with {\it ab-initio} calculations, have thus been instrumental in establishing the magnitude of the microscopic parameters, summarized in Table~\ref{table_2DElec}.


The validity of the $j_\text{eff} = \frac{1}{2}$ picture for Li$_2$RhO$_3$ is considerably more questionable than for the iridates. The smaller strength of spin-orbit coupling in the $4d$ element may lead to significant mixing of the $j_\text{eff} = \frac{1}{2}$ and $\frac{3}{2}$ states through
trigonal crystal field $\Delta/\lambda\sim 0.5$ and intersite hopping terms. Indeed, based on a preliminary crystal structure, the authors of Ref.~\onlinecite{katukuri2015strong} noted that the low-energy states are significantly perturbed from the ideal $j_\text{eff} = \frac{1}{2}$ composition in quantum chemistry calculations. 

\begin{figure}[t]
\includegraphics[width=0.85\linewidth]{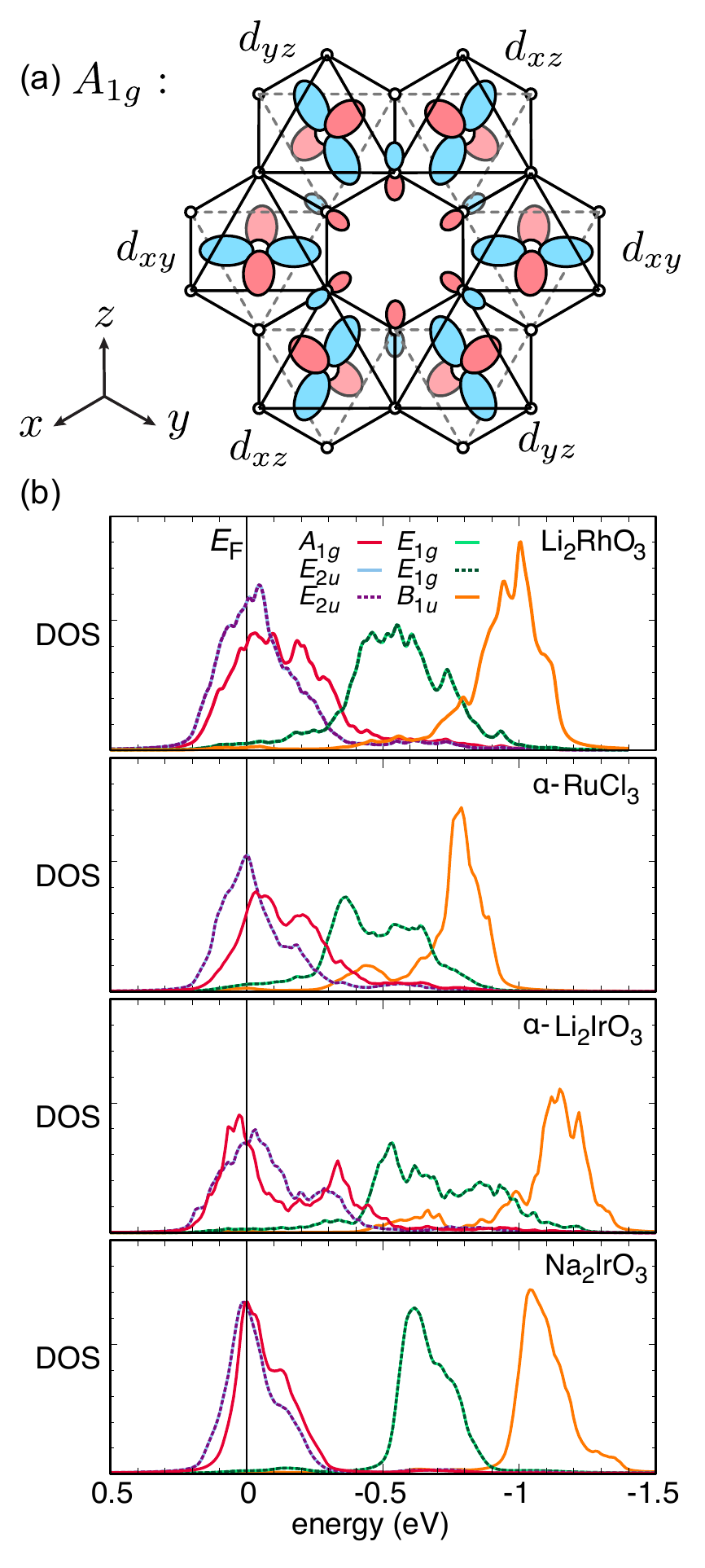}
\caption{\label{fig-qmo1}(Color online) (a) Relevant metal and ligand orbitals for constructing the basis of quasimolecular orbitals (QMOs) showing the hopping path within a single hexagon. The orbitals are pictured with phases corresponding to the totally symmetric $a_{1g}$ QMO combination. (b) Nonrelativistic DOS computed at the GGA lavel for honeycomb materials Li$_2$RhO$_3$, $\alpha$-RuCl$_3$, $\alpha$-Li$_2$IrO$_3$ and Na$_2$IrO$_3$ showing contributions from the six QMOs of different symmetry.  }
\end{figure}

In this context, in Ref.~\onlinecite{Mazin2012,Foyevtsova2013} it was pointed out that the non-relativistic ($\lambda \rightarrow 0$) electronic structure of the honeycomb iridates and rhodates also features weakly dispersing bands due to entirely different mechanisms than in the $j_\text{eff}$ picture. Instead, the dominant oxygen-mediated hopping confines the electrons to local hopping paths of the type $d_{xy}$-O$p_x$-$d_{xz}$-O$p_z$-$d_{yz}$-O$p_y$-$d_{xy}$, shown in Fig.~\ref{fig-qmo1}. Following such a hopping path, each $t_{2g}$ hole can only traverse a local hexagon formed by six metal sites in the $\lambda \rightarrow 0$ limit. In this way, all states become localized to such hexagons even at the single-particle level! In analogy with molecular benzene, the nonrelativistic $t_{2g}$ bands are split into six nearly flat bands described in the basis of quasi-molecular orbitals (QMOs) built from linear combinations of the six $t_{2g}$ orbitals shown in Fig.~\ref{fig-qmo1}. Such a QMO-based insulating state can be distinguished from the $j_{\rm eff}=\frac12$ state using experimental observables, including optical conductivity and RIXS data, with the honeycomb iridates lying on the $j_{\rm eff}=\frac12$ side of the phase diagram.\cite{PhysRevLett.117.187201} 

Interestingly, the QMOs form a natural basis for many layered honeycomb systems with $4d$ ions, as in Li$_2$RhO$_3$~\cite{MazinRh2013} and SrRu$_2$O$_6$.\cite{PhysRevB.92.134408,PhysRevB.94.205148} These QMOs states are, however, very sensitive to changes in the crystal structure.\cite{Foyevtsova2013} Further investigation of these issues related to Li$_2$RhO$_3$ currently await detailed RIXS and optical conductivity measurements, which have so-far been hampered by unavailability of high quality single crystals.
  
\subsubsection{Magnetic Properties}
\label{sssec_ir_rh_thermag}
\begin{figure}[t]
\includegraphics[width=0.95\linewidth]{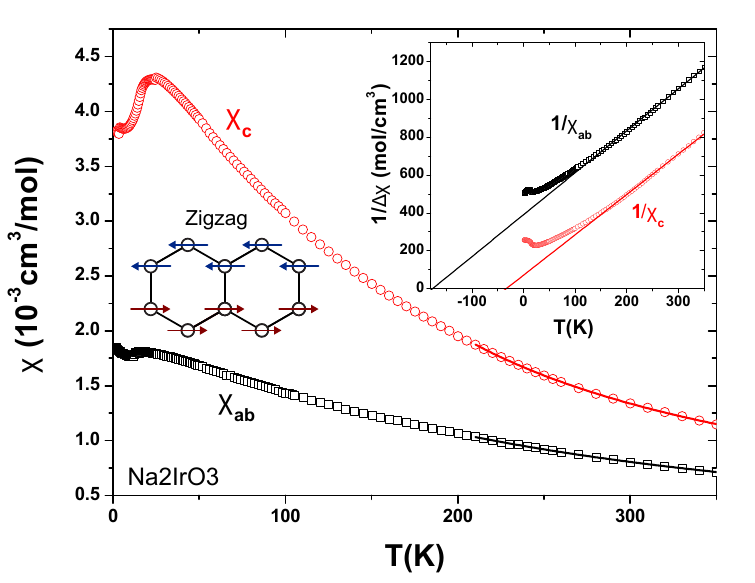}
\caption{\label{fig-nairomag1}(Color online) Magnetic susceptibility of Na$_2$IrO$_3$. The kink at low temperatures signifies the onset of collinear zigzag magnetic order at $T_N \sim 13 - 18$ K. Inset: Curie-Weiss fitting of the inverse susceptibility showing anisotropy in the measured Curie-Weiss temperatures. In contrast, $\alpha$-Li$_2$IrO$_3$ adopts an incommensurate spiral order pictured in Fig.~\ref{fig-spiralmag}. Figure adapted from Ref.~\onlinecite{Mannithesis}.}
\end{figure}

At high temperatures, the magnetic susceptibilities\cite{singh2012,Mannithesis,freund2016} of Na$_2$IrO$_3$ and $\alpha$-Li$_2$IrO$_3$ follow the Curie-Weiss law with effective moments close to 1.73\,$\mu_B$, consistent with the $j_{\rm eff}=\frac12$ scenario suggested by RIXS and optical measurements. Whereas the effective moments are weakly dependent on the field direction (owing to a small anisotropy in the $g$-tensor), the magnetic susceptibility is strongly anisotropic following strong directional dependence of the Curie-Weiss temperature $\Theta$ (Fig.~\ref{fig-nairomag1}). Opposite flavors of the anisotropy (Table \ref{table_2DMag}), reflect salient microscopic differences between the two iridates.

The N\'eel temperatures ($T_N$) are reported to be 15\,K in $\alpha$-Li$_2$IrO$_3$~\cite{singh2012,williams2016} and ranging from 13 to 18\,K in Na$_2$IrO$_3$,\cite{singh2010,liu2011,ye2012} presumably due to differences in sample quality. The suppression of the ordering temperatures far below the Weiss temperatures in both systems is an indicator of strong frustration via the standard criterion of the $\Theta/T_N$ ratio,\cite{ramirez1994} which turns out to be between 5 and 10 for the iridates.\footnote{Absolute values of the Curie-Weiss temperatures should be taken with caution, because they depend on the temperature range of the fitting.} Further signatures of the frustration include large release of the magnetic entropy above $T_N$~\cite{mehlawat2017} and significant reduction in the ordered moments, 0.22(1)\,$\mu_B$ in Na$_2$IrO$_3$~\cite{ye2012} and 0.40(5)\,$\mu_B$ in $\alpha$-Li$_2$IrO$_3$,\cite{williams2016} both well below 1\,$\mu_B$ expected for $j_{\rm eff}=\frac12$, although covalency effects should also play a role here.



Below $T_N$, Na$_2$IrO$_3$ develops zigzag order~\cite{liu2011,choi2012,ye2012} with the
propagation vector $\mathbf k=(0,1,\frac12)$ and spins lying at the
intersection of the crystallographic $ac$-plane, and the cubic
$xy$-plane.\cite{chun2015} The onset of long-range magnetic order below $T_N
\approx 15$~K is also confirmed via zero-field muon-spin rotation
experiments.\cite{choi2012} This zigzag state may arise from several
microscopic scenarios, including Heisenberg interactions beyond nearest
neighbors,\cite{fouet2001} leading to significant discussion regarding the
underlying magnetic interactions in Na$_2$IrO$_3$. Experimentally, diffuse
resonant x-ray scattering has provided direct evidence for the relevance of the
Kitaev terms in the spin Hamiltonian by pinpointing predominant correlations
between $S_x$, $S_y$, and $S_z$ components on different bonds of the
honeycomb.\cite{chun2015} 

\begin{table}[t]
\caption{ \label{table_2DMag} Summary of magnetic parameters for honeycomb Na$_2$IrO$_3$, $\alpha$-Li$_2$IrO$_3$, Li$_2$RhO$_3$, and $\alpha$-RuCl$_3$. The latter material is discussed in section \ref{ssec_rucl}. See text for relevant references.}
\begin{ruledtabular}
\begin{tabular}{c|cc|c|c}
Property&Na$_2$IrO$_3$&$\alpha$-Li$_2$IrO$_3$&Li$_2$RhO$_3$&$\alpha$-RuCl$_3$\\[1.0ex]
\hline
$\mu_\text{eff} \ (\mu_B)$&1.79&1.83  & 2.03&2.0 to 2.7\\
$\Theta_{iso}$ (K)&$\sim$ $-120$&$-33$ to $-100$&$\sim$ $-50$&$\sim$ +40\\
$\Theta_{ab}$ (K)&-176& $\Theta_{ab} > \Theta_c$& $-$ & +38 to +68 \\
$\Theta_{c}$ (K)&$-40$&$-$& $-$ & $-100$ to $-150$
\\[1.0ex] \hline
$T_N$ (K)&$13-18$& $\sim$ 15&(6)&7 to 14\\
Order&Zigzag&Spiral&Glassy&Zigzag\\
$\mathbf k$-vector&$(0,1,\frac{1}{2})$&$(0.32,0,0)$&$-$&$(0,1,\frac{1}{2})$
\end{tabular}
\end{ruledtabular}
\end{table}

From the theoretical perspective, there have been several {\it ab-initio} calculations seeking to establish parameters of the $j_\text{eff} = \frac12$ spin Hamiltonian, employing differing methods from fully {\it ab-initio} quantum chemistry techniques\cite{katukuri2014kitaev} to perturbation theory\cite{PhysRevLett.113.107201} and exact diagonalization\cite{PhysRevB.93.214431} (based on hopping integrals derived from DFT and experimental Coulomb parameters). These results are summarized in Table~\ref{table_NaIr1}, and reviewed in Ref.~\onlinecite{PhysRevB.93.214431}. Initially, the observation of zigzag magnetic order and an antiferromagnetic Weiss constant led to the suggestion that the Kitaev term may become antiferromagnetic.\cite{PhysRevLett.110.097204} Indeed, a ferromagnetic Kitaev term is not compatible with zigzag order within the pure nearest neighbour Heisenberg-Kitaev model that was featured in many early theoretical works.\cite{Chaloupka2010,Reuther11,PhysRevB.83.245104} However, the {\it ab-initio} results tell a different story. 

In accordance with the original work of Jackeli and Khaliullin, the dominant oxygen-assisted hopping leads to a large ferromagnetic nearest neighbour Kitaev interaction ($K_1 < 0$). This is supplemented by several smaller interactions, which enforce the zigzag order, moment direction, and $\Theta < 0$. The most significant of such interactions is expected to be a third neighbour Heisenberg ($J_3 >0$) term coupling sites across the face of each hexagon.\cite{katukuri2014kitaev,PhysRevB.93.214431} This interaction is estimated to be as much as 30\% of the Kitaev exchange, as suggested by early analysis of the magnetic susceptibility,\cite{PhysRevB.84.180407} or even stronger according to inelastic neutron scattering results.\cite{choi2012} The direction of the ordered moment is then selected\cite{PhysRevB.94.064435} by the off-diagonal $\Gamma_1$ and $\Gamma_1^\prime$ terms, on the order of 10\% of $K_1$. The ordering wavevector, parallel to the $b$-axis within the plane, is favoured by small bond-dependency of the Kitaev term, i.e. $|K_1^Z| > |K_1^{X,Y}|$. In this sense, the key aspects of the magnetic response of Na$_2$IrO$_3$ appear to be well understood: the Jackeli-Khaliullin mechanism applies, leading to dominant Kitaev interactions at the nearest neighbour level. However, zigzag magnetic order is ultimately established at low temperatures by additional interactions. 

\begin{table}[t]
\caption{ \label{table_NaIr1} Bond-averaged values of the largest magnetic interactions (in units of meV) within the plane for Na$_2$IrO$_3$ computed using various methods. ``Pert. Theo.'' refers to second order perturbation theory (Sec.~\ref{sssec_ddhop}), ``QC'' = quantum chemistry methods, ``ED'' = exact diagonalization.}
\begin{ruledtabular}
\begin{tabular}{ccccccc}
Method&$J_1$&$K_1$&$\Gamma_1$&$\Gamma_1^\prime$&$K_2$&$J_3$\\ 
\hline
Pert. Theo.\cite{PhysRevLett.113.107201}&+3.2&$-29.4$&+1.1&$-3.5$&$-0.4$&+1.7\\[1.0ex]
QC (2-site)\cite{katukuri2014kitaev}&+2.7&$-16.9$&  +1.0&$-$&$-$&$-$\\[1.0ex]
ED (6-site)\cite{PhysRevB.93.214431}& +0.5& $-16.8$& +1.4&$-2.1$&$-1.4$&+6.7\\[1.0ex]
\end{tabular}
\end{ruledtabular}
\end{table}

\begin{table*}[t]
\caption{ \label{table_LiIr1} Values of the largest magnetic interactions (in units of meV) within the plane for $\alpha$-Li$_2$IrO$_3$ obtained from various methods. ``QC'' = quantum chemistry methods, ``ED'' = exact diagonalization.}
\begin{ruledtabular}
\begin{tabular}{ccccccccccc}
Method&$J_1^Z$&$J_1^X$&$K_1^Z$&$K_1^X$&$\Gamma_1^Z$&$\Gamma_1^X$&$K_2$&$\Gamma_2$&$|\mathbf{D}_2|$&$J_3$\\ 
\hline
QC (2-site)\cite{katukuri2014kitaev}& $-19.2$&+0.8&$-6.0$&$-11.6$&+1.1&$-4.2$&$-$&$-$&$-$&$-$\\[1.0ex]
ED (6-site)\cite{PhysRevB.93.214431}& $-3.1$&$-2.5$&$-6.3$&$-9.8$&+9.4&+8.7&$-3.7$&+3.4&+2.7&+6.0\\[1.0ex]
\end{tabular}
\end{ruledtabular}
\end{table*}

\begin{figure*}
\includegraphics[width=\linewidth]{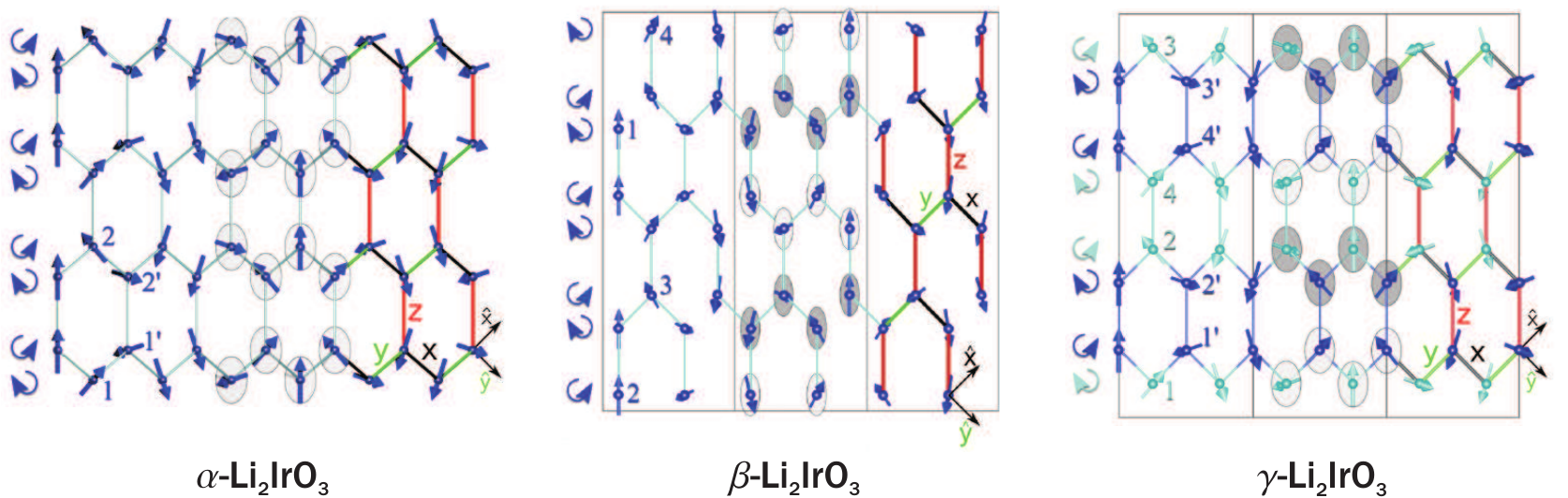}
\caption{Magnetic structures of $\alpha$-, $\beta$-, and $\gamma$-Li$_2$IrO$_3$ showing common counter-rotating spiral order\cite{williams2016,biffin2014b,biffin2014g}. Figures adapted, with permission, from Ref.~\onlinecite{williams2016} and \onlinecite{biffin2014b}.}
\label{fig-spiralmag}
\end{figure*}

In the case of $\alpha$-Li$_2$IrO$_3$, indications for anisotropic bond-dependent interactions are ingrained in the spin arrangement itself. The N\'eel temperature of about 15\,K marks a transition to an incommensurate state,~\cite{williams2016} with the propagation vector $\mathbf k=(0.32(1),0,0)$. RXS studies have established that the magnetic structure is described by the basis vector combination $(-iA_x,F_y,-iA_z)$ that in real space corresponds to counter-rotating spirals for the Ir1 and Ir2 atoms in the unit cell (shown in Fig.~\ref{fig-spiralmag}).\cite{williams2016} This counter-rotation requires a large Kitaev term in the spin Hamiltonian, but leaves a multiple choice for other interactions.\cite{williams2016} 

There have been at least two proposals consistent with the observed order. The authors of Ref.~\onlinecite{kimchi2015} noted that the spiral state might emerge from significantly bond-dependent interactions allowed within the crystallographic $C2/m$ symmetry. They introduced a three parameter ($J,K,I_c$) Hamiltonian, where $I_c$ controls the degree of bond-dependence; this is equivalent to the choice $(J_1, K_1)= (J,K)$ for the nearest neighbour X- and Y-bonds, while $(J_1, K_1, \Gamma_1) =  (J + \frac{1}{2}I_c, K - \frac{1}{2}I_c, -\frac{1}{2}I_c)$ for the Z-bond. For dominant ferromagnetic Kitaev $K<0$ and bond-dependent $I_c <0$ terms, the ground state was found to be an incommensurate state consistent with the experiment. This view was challenged by the authors of Ref.~\onlinecite{lee2016}, who argued that incommensurate states also arise in the Kitaev materials if the bond-dependence is removed, but the off-diagonal $\Gamma_1>0$ and large $K_1<0$ couplings are retained on all bonds. Indeed, the bond-isotropic $(J_1,K_1,\Gamma_1)$ honeycomb model features the observed incommensurate state.\cite{PhysRevLett.112.077204} However, it is likely that these two limits are smoothly connected to one another, rendering the distinction somewhat arbitrary. 

From the perspective of {\it ab-initio} studies, the resolution of the interactions in $\alpha$-Li$_2$IrO$_3$ has been severely complicated by the absence of high quality structural information, until recently. Results are summarized in Table \ref{table_LiIr1}. Early quantum chemistry studies\cite{Nishimoto2014} were based on crystal structures obtained by analysis of powder samples, and suggested significant bond-anisotropy at the nearest neighbour level.
  More recent studies\cite{PhysRevB.93.214431} considered also longer-ranged interactions and the
effects of relaxing the powder structure within the DFT
framework.\cite{manni2014a} Ref.~\onlinecite{PhysRevB.93.214431} suggested a relatively non-local
spin Hamiltonian with significant terms at first, second, and third neighbour.
In particular, large second neighbour $K_2$ and $\Gamma_2$ were identified,
along with a second neighbour Dzyaloshinkii-Moriya $\mathbf{D}_2\cdot
(\mathbf{S}_i \times \mathbf{S}_j)$ interaction (which is allowed by symmetry).
The authors argued that this latter interaction likely also plays a role in
establishing the incommensurate state. Presently, it is firmly established that
the largest interactions in $\alpha$-Li$_2$IrO$_3$ must include a ferromagnetic
Kitaev term, in agreement with the Jackeli-Khaliullin mechanism. However, the
role of additional interactions remains less clear than for Na$_2$IrO$_3$.

It is worth noting that the \textit{ab-initio} studies also reveal the origin of anisotropic Curie-Weiss temperatures in Na$_2$IrO$_3$ and $\alpha$-Li$_2$IrO$_3$. The difference between $\Theta_{ab}$ and $\Theta_c$ is rooted in the off-diagonal terms $\Gamma_1$ and $\Gamma_1'$, as well as in the bond-dependency of the Kitaev term, $K_1^Z\neq K_1^{X,Y}$. The difference between $\Theta_{ab}$ and $\Theta_c$ is thus a rough measure of the deviation from the Heisenberg-Kitaev regime, where Curie-Weiss temperature would be isotropic.

Finally, let us briefly mention that Li$_2$RhO$_3$ is somewhat different from the honeycomb iridates considered so far. At high temperatures, the magnetic susceptibility follows a Curie-Weiss law with an enhanced effective moment $\mu_\text{eff} = 2.03\,\mu_B$ associated with intermediate spin-orbit coupling~\cite{luo2013} (see section \ref{sssec_rucl_thermag} below). While Li$_2$RhO$_3$ displays a sizeable Weiss temperature $\Theta \sim -50$ K, it lacks any magnetic ordering, and instead shows spin freezing around 6\,K.\cite{luo2013} The glassy state is gapless with $T^2$ behavior of both zero-field specific heat and nuclear magnetic resonance (NMR) spin-lattice relaxation rate.\cite{khuntia2015} Spin freezing may obscure the intrinsic physics in Li$_2$RhO$_3$, possibly due to the structural disorder.\cite{katukuri2015strong} However, further investigation pends availability of single crystals of this material.

\subsubsection{Doping experiments}

\label{sssec_ir_rh_dope}
The distinct differences between Na$_2$IrO$_3$ and $\alpha$-Li$_2$IrO$_3$ triggered multiple doping attempts. Despite an early report of the continuous Na/Li substitution,\cite{cao2013b} detailed investigation revealed a large miscibility gap.\cite{manni2014a} On the Na-rich side, only 25\,\% of Li can be doped, which is the amount of Li that fits into the Na position in the center of the hexagon.\cite{manni2014a} In contrast, no detectable doping on the Li-rich side could be achieved. 

Li doping into Na$_2$IrO$_3$ leads to a systematic suppression of $T_N$, whereas the powder-averaged Curie-Weiss temperature increases, approaching that of $\alpha$-Li$_2$IrO$_3$.\cite{manni2014a} With the maximum doping level of about 25\,\%, one reaches $T_N=5.5$\,K without any qualitative changes in thermodynamic properties.\cite{manni2014a} On the other hand, even the 15\,\% Li-doped sample shows magnetic excitations that are largely different from those of the zigzag phase of pure Na$_2$IrO$_3$,\cite{rolfs2015} which may indicate a change in the magnetic order even upon marginal Li doping.

Doping on the Ir site yields a much broader range of somewhat less interesting solid solutions that generally show glassy behavior at low temperatures. Non-magnetic dilution via Ti$^{4+}$ doping~\cite{manni2014b,PhysRevB.90.205112} leads to the percolation threshold at 50\,\% in $\alpha$-Li$_2$IrO$_3$ compared to only 30\,\% in Na$_2$IrO$_3$. The isoelectronic doping of $\alpha$-Li$_2$IrO$_3$ with rhodium gives rise to a similar dilution effect, because non-magnetic Rh$^{3+}$ is formed, triggering the oxidation of iridium toward Ir$^{5+}$, which is also non-magnetic.\cite{kumari2016}

Ru$^{4+}$ doping is also possible and introduces holes into the system, but all doped samples remain robust insulators.\cite{mehlawat2015} Similar to the Ti-doped case, glassy behavior is observed at low temperatures.\cite{mehlawat2015} Electron doping was realized by Mg substitution into Na$_2$IrO$_3$, resulting in the glassy behavior again.\cite{wallace2015b} This ubiquitous spin freezing triggered by even low levels of the disorder can be seen positively as an indication for the strongly frustrated nature of both Na$_2$IrO$_3$ and $\alpha$-Li$_2$IrO$_3$. It probably goes hand in hand with random charge localization that keeps the materials insulating upon both hole and electron doping. 

Another doping strategy is based on the cation (de)intercalation. Chemical deintercalation facilitates removal of one Na atom out of Na$_2$IrO$_3$ and produces NaIrO$_3$ that shows mundane temperature-independent magnetism due to the formation of non-magnetic Ir$^{5+}$.\cite{wallace2015} The more interesting intermediate doping levels seem to be only feasible in electrochemical deintercalation.\cite{mccalla2015,perez2016} Although the battery community pioneered investigation of the honeycomb iridates~\cite{kobayashi1997,kobayashi2003} long before the Kitaev model became the topic of anyone's interest, no low-temperature measurements on partially deintercalated samples were performed as of yet, possibly due to the small amount of deintercalated materials and their unavoidable contamination during the electrochemical treatment.

\subsection{$\alpha$-RuCl$_3$: a proximate spin-liquid material?}
\label{ssec_rucl}

Despite the intensive study of the iridates reviewed in the previous section, a complete picture of the magnetic excitations has remained elusive due to severe complications associated with inelastic neutron studies on the strongly neutron absorbing Ir samples.\cite{choi2012}  Raman studies have been possible on the iridates,\cite{glamazda2016,0295-5075-114-4-47004} but probe only $\mathbf{k}=0$, while RIXS measurements\cite{Gretarsson2013} still suffer from limited resolution. For this reason, there has been significant motivation to search for {\it non}-Ir based Kitaev-Jackeli-Khaliullin materials. Following initial investigations in 2014,\cite{PhysRevB.90.041112} $\alpha$-RuCl$_3$ has now emerged as one of the most promising and well-studied systems, due to the availability of high quality samples, and detailed dynamical studies. These are reviewed in this section.

\subsubsection{Synthesis and Structure}
\label{sssec_rucl_syn}
\begin{figure}
\includegraphics[width=\linewidth]{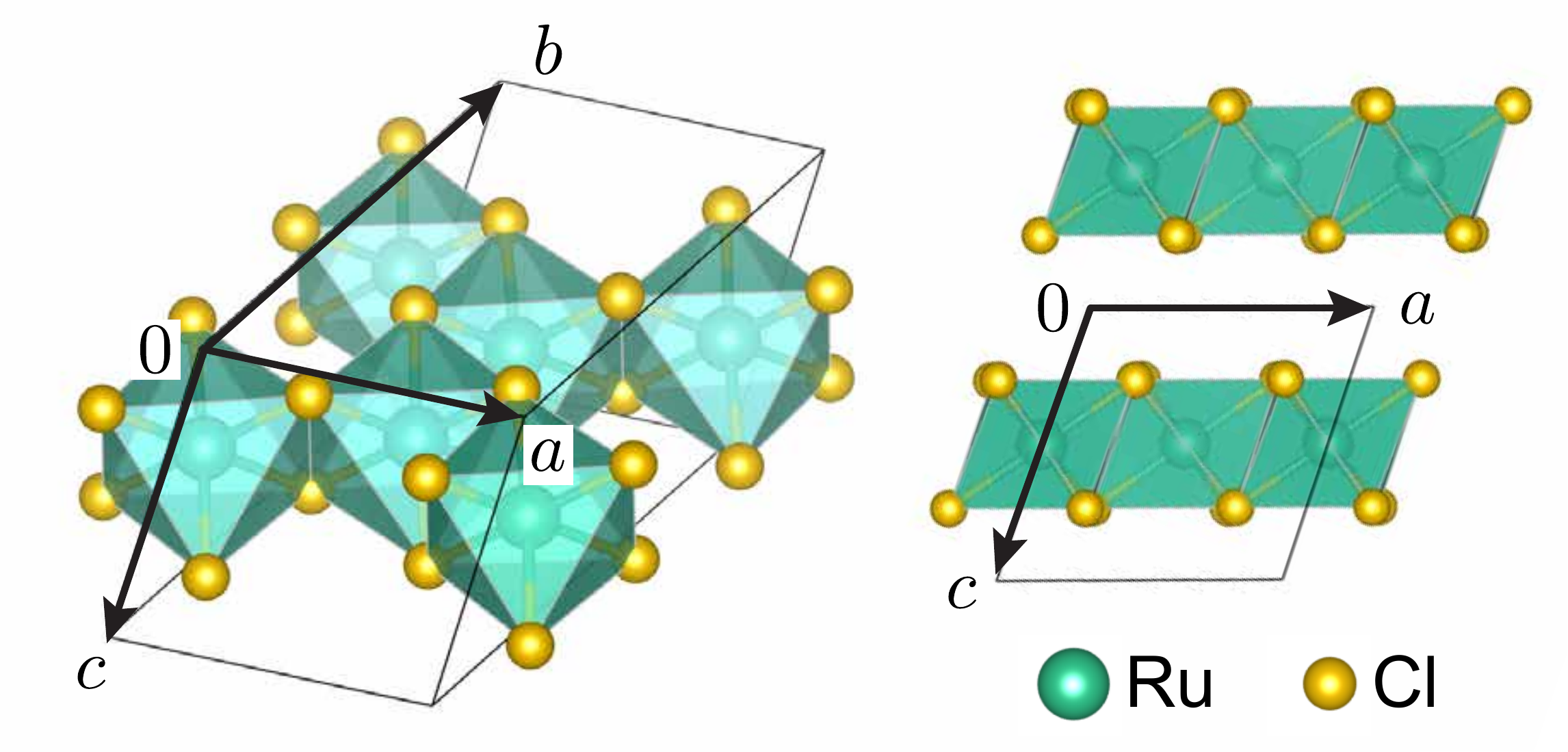}
\caption{Different views of the $C2/m$ unit cell of $\alpha$-RuCl$_3$. The material suffers significantly from stacking faults due to the weakly van der Waals bound layers, somewhat complicating assignment of the space group.\cite{PhysRevB.92.235119,PhysRevB.93.134423} }
\label{fig-rucl3-struc}
\end{figure}

Ruthenium trichloride was likely first prepared in 1845 from the direct reaction of Ru metal with Cl$_2$ gas at elevated temperatures,\cite{clausrucl,remy1924beitrage,fletcher1967x} which yields a mixture of allotropes.\cite{hyde1965alpha} The $\beta$-phase is obtained as a brown powder, and crystallizes in a $\beta$-TiCl$_3$-type structure, featuring one-dimensional chains of face-sharing RuCl$_6$ octahedra. The $\alpha$-phase, of recent interest in the context of Kitaev physics, crystallizes in a honeycomb network of edge-sharing octahedra (Fig.~\ref{fig-rucl3-struc}). Annealing the mixture above 450 $^\circ$C under Cl$_2$ converts the $\beta$-phase irreversibly to the $\alpha$-phase, which appears as shiny black crystals. Historically, RuCl$_3$ has been widely employed in organic chemistry primarily as an oxidation catalyst, or a precursor for organoruthenium compounds.\cite{griffith1975rut,griffith2010ruthenium} However, commercially available ``RuCl$_3\cdot$xH$_2$O'' is typically obtained by dissolving RuO$_4$ in concentrated hydrochloric acid, and contains a complex mixture of oxochloro and hydroxychloro species of varying oxidation states.\cite{hyde1965alpha,cotton2012chemistry} Pure samples of $\alpha$-RuCl$_3$ suitable for physical studies are therefore generated by purification of commercial samples. This may proceed, for example, via vacuum sublimation under Cl$_2$ with a temperature gradient between 650\,$^\circ$C and 450\,$^\circ$C, to ensure crystallization in the $\alpha$-phase.\cite{PhysRevB.93.134423,PhysRevB.92.235119} Further details regarding synthesis can be found, for example, in Refs.~\onlinecite{hill1950ruthenium,fletcher1963anhydrous}.

The structure of $\alpha$-RuCl$_3$ has been a matter of some debate. Similar layered materials are known to adopt a variety of structures, including BiI$_3$-type ($R\bar{3}$), CrCl$_3$-type ($P3_112$), and AlCl$_3$-type ($C2/m$).\cite{hulliger2012structural,douglas2007structure} Distinguishing between such structures is made difficult by the presence of stacking faults between the weakly bound hexagonal layers. Early structural studies indicated a highly symmetric $P3_112$ space group.\cite{stroganovstruktur,fletcher1967x} Later studies questioned this assignment,\cite{brodersen1968struktur} and more recent works have established that the low-temperature structure is of $C2/m$ symmetry for the highest quality samples.\cite{PhysRevB.92.235119, PhysRevB.93.134423} However, it should be noted that {\it ab-initio} studies find only very small energy differences between the various candidate structures,\cite{PhysRevB.93.155143} consistent with the observation that some crystals also exhibit a phase transition in the region $100-150$\,K.\cite{park2016emergence,banerjee2016neutron,PhysRevB.93.134423,PhysRevB.91.094422,reschke2017} Moreover, several recent studies\cite{park2016emergence,do2017incarnation} have suggested instead an $R\bar{3}$ structure for the low-temperature phase, in analogy with CrCl$_3$. 

The older $P3_112$ and newer $C2/m$ and $R\bar{3}$ structures of $\alpha$-RuCl$_3$ differ substantially, which has led to some confusion regarding the magnetic interactions, as discussed below in section \ref{sssec_rucl_thermag}. In particular, the $P3_112$ structure features essentially undistorted RuCl$_6$ octahedra, with Ru-Cl-Ru bond angles $\sim 89^\circ$. This observation led to the original association of $\alpha$-RuCl$_3$ with Kitaev physics, as the authors of Ref.~\onlinecite{PhysRevB.90.041112} suggested that weak trigonal crystal field splitting might preserve a robust $j_\text{eff}$ character despite weaker spin-orbit coupling strength $\lambda \sim 0.15$ eV compared to the iridates. In contrast, the recent $C2/m$ and $R\bar{3}$ structures (themselves very similar) imply a significantly larger trigonal compression, with Ru-Cl-Ru bond angles $\sim 94^\circ$ - similar to the iridates. In this context, one can expect deviations from the ideal $j_\text{eff}$ picture, as discussed below.

Finally, we mention that a number of studies have probed structural modifications to $\alpha$-RuCl$_3$. The 2D layers can be exfoliated, which leads to structural distortions,\cite{ziatdinov2016atomic} and alters the magnetic response.\cite{weber2016magnetic} Similar to the iridates, substitutional doping has also been explored, for example, affecting the replacement of Ru with nonmagnetic Ir$^{3+}$ ($5d^6$), which suppresses the magnetic order above a percolation threshold of $\sim\!\!25\%$ substitution.\cite{lampen2016destabilization}

\subsubsection{Electronic Properties}
\label{sssec_rucl_elec}
Early resistivity measurements identified pure $\alpha$-RuCl$_3$ as a Mott insulator, with in-plane and out-of-plane resistivity on the order of $10^3$\,$\Omega$\,cm and $10^6$\,$\Omega$\,cm, respectively. The resistivity follows Arrhenius behaviour, with a small activation energy estimated to be $\sim 100$ meV.\cite{binotto1971optical} A much larger charge gap is implied by a number of other experiments, including photoconductivity,\cite{binotto1971optical} photoemission,\cite{PhysRevB.94.161106,PhysRevLett.117.126403,sinn2016electronic} and inverse photoemission,\cite{sinn2016electronic} which arrive at estimates of $1.2-1.9$\,eV. Insight can also be obtained from optical measurements.\cite{PhysRevB.94.195156,binotto1971optical,reschke2017} 
Given the relatively weak spin-orbit coupling, the authors of Ref.~\onlinecite{PhysRevB.94.195156} analyzed the splitting of such excitations in the non-relativistic limit, obtaining estimates of the electronic parameters shown in Table \ref{table_2DElec}. In contrast with the iridates, spin-orbit 
coupling  plays in $\alpha$-RuCl$_3$ a less dominant role.\cite{PhysRevB.92.235119}

\begin{figure}
\includegraphics[width=0.75\linewidth]{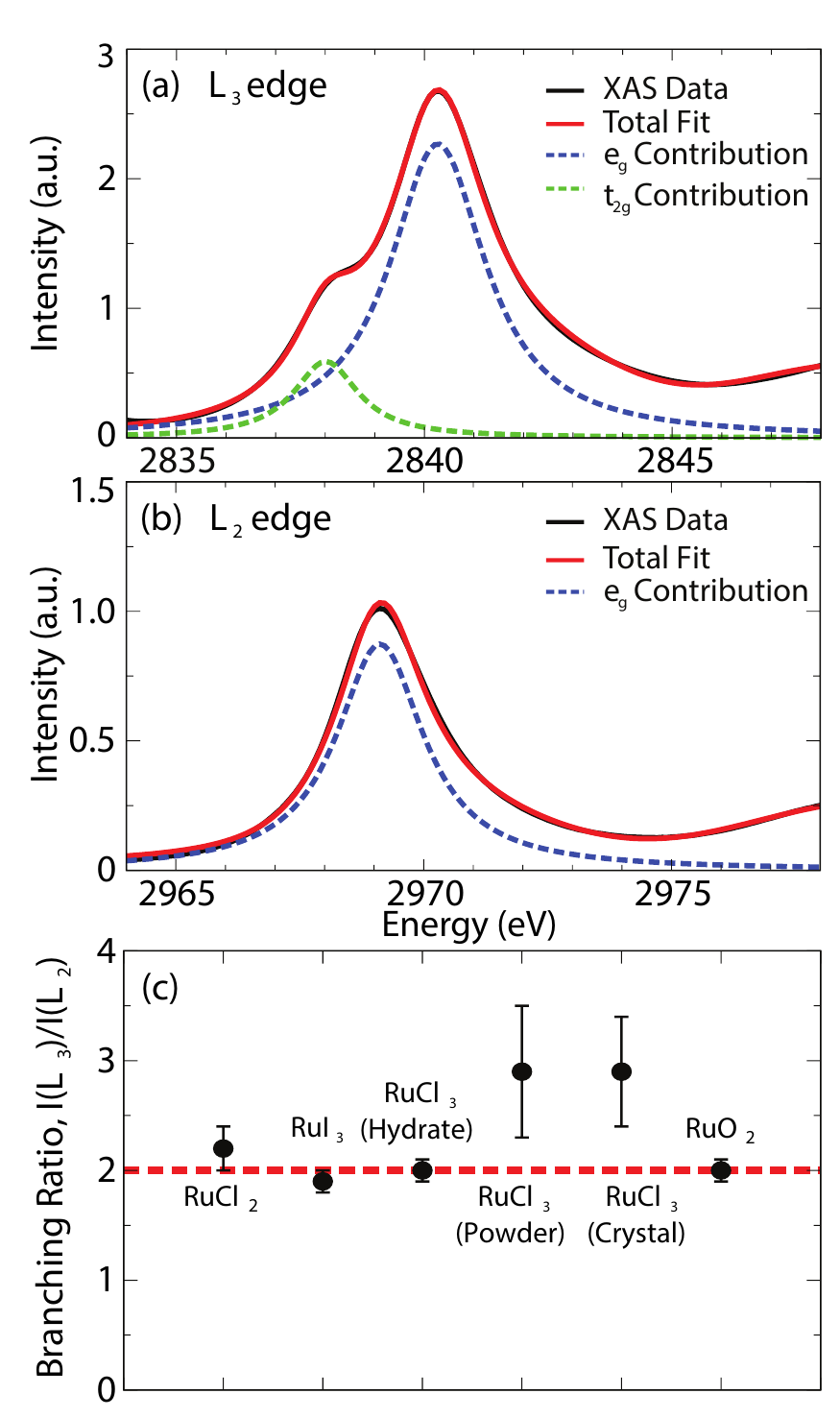}
\caption{\label{fig-rucl3-xas} X-ray absorption spectroscopy (XAS) data for $\alpha$-RuCl$_3$ at the (a) L$_2$ and (b) L$_3$ edges. The absence of $t_{2g}$ intensity at the L$_2$ edge (and therefore large branching ratio (c)) suggests significant $j_\text{eff}=\frac{1}{2}$ character of the $t_{2g}$ hole. Reproduced from Ref.~\onlinecite{PhysRevB.90.041112} by permission from the American Physical Society: \textcopyright \ 2014.}
\end{figure}

The first experimental indications of the $j_\text{eff}$ picture in
$\alpha$-RuCl$_3$ were based on x-ray absorption spectroscopy (XAS)
measurements,\cite{PhysRevB.90.041112,agrestini2017electronically} which are
consistent with electron energy loss spectroscopy (EELS)
data.\cite{PhysRevLett.117.126403} Such experiments probe excitations from
core-level Ru $2p$ to the valence $4d$ states. In the pure $j_\text{eff}$
picture, transitions to the empty $j_\text{eff} = \frac{1}{2}$ state from the
core $2p_{1/2}$ states (L$_2$ edge) are symmetry forbidden, while those from
the core $2p_{3/2}$ states (L$_3$ edge) are symmetry
allowed.\cite{de1994differences} The experimental absence of $t_{2g}$ intensity
at the L$_2$ edge (Fig.~\ref{fig-rucl3-xas}) can therefore be taken as a sign of significant $j_\text{eff}
= \frac{1}{2}$ character in the $t_{2g}$ hole. However, it should be noted that
the composition of the $t_{2g}$ hole is somewhat less sensitive to  trigonal
crystal field effects than the magnetic interactions, as discussed in section
\ref{sssec_distort}. Indeed, the Kitaev coupling can be strongly suppressed for
 trigonal crystal field terms as small as $|\Delta/ \lambda| \sim 0.2$, while the
$t_{2g}$ hole retains $\sim 90\%$ of the $j_\text{eff} = \frac{1}{2}$ character
in that case (see Fig.\ref{fig-CFS1b}(b-c)). In this sense, the spectroscopic measurements are promising, but
do not rule out deviations from the ideal Jackeli-Khaliullin
scenario. Direct measurements of the trigonal crystal-field splitting are therefore
highly desirable.

Additional evidence for the $j_\text{eff}$ picture can be seen in low-energy optical response.\cite{sandilands2016} In the range of $0.2-0.8$\,eV, the optical conductivity shows a series of excitations consistent with local spin-orbital excitons, as noted in Ref.~\onlinecite{PhysRevB.91.241110}. These peaks appear at multiples of $3\lambda/2$, allowing an estimation of $\lambda \sim 0.10-0.15$\,eV, consistent with the atomic value for Ru.

\subsubsection{Magnetic Properties}
\label{sssec_rucl_thermag}

\begin{figure}
\includegraphics[width=0.75\linewidth]{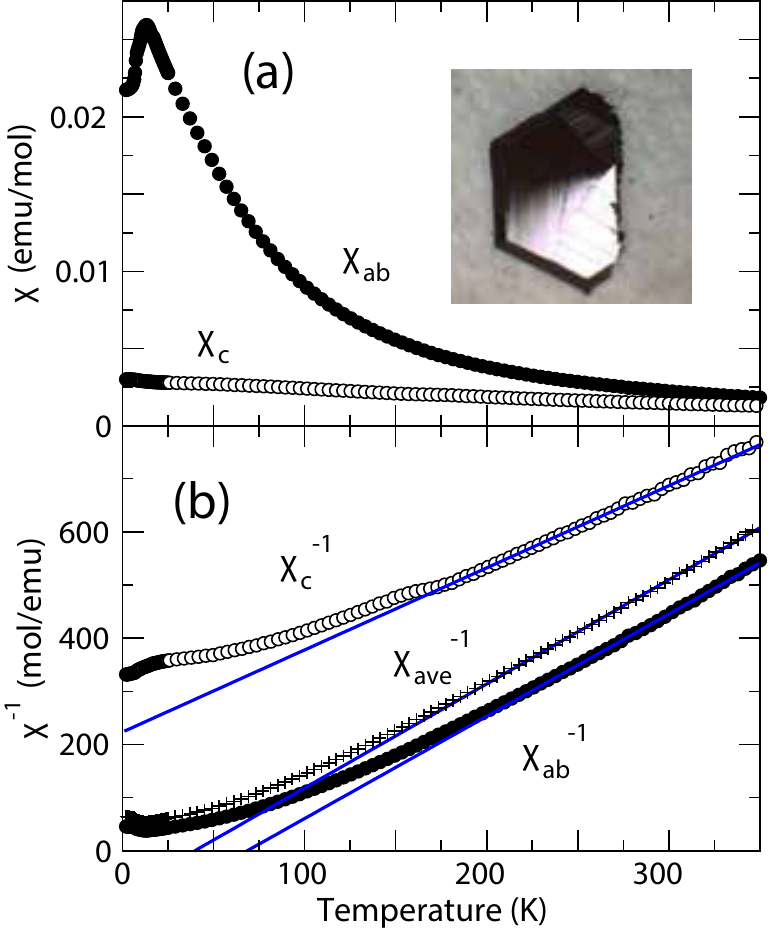}
\caption{\label{fig-rucl3-mag1} Temperature dependence of (a) magnetic susceptibilities and (b) inverse susceptibilities of $\alpha$-RuCl$_3$ with a field parallel to the $c$-axis and within the $ab$-plane. Blue lines indicate Curie-Weiss fits. Reproduced from Ref.~\onlinecite{PhysRevB.91.144420} by permission from the American Physical Society: \textcopyright \ 2015.}
\end{figure}

The magnetic susceptibility of $\alpha$-RuCl$_3$ has been reported by several groups.\cite{epstein1954magnetic,fletcher1967x,fletcher1963anhydrous,PhysRevB.91.094422,PhysRevB.91.180401,PhysRevB.91.144420,banerjee2016neutron} At high temperatures, it follows a Curie-Weiss law, with anisotropic effective moments of $2.0-2.4$\,$\mu_B$ for fields in the honeycomb $ab$-plane, and $2.3-2.7$\,$\mu_B$ for fields out of the plane (Fig.~\ref{fig-rucl3-mag1}). The enhancement of both values with respect to the spin-only or $j_\text{eff} = \frac{1}{2}$ value (1.73\,$\mu_B$) is a clear signature of intermediate spin-orbit coupling strength. This effect, sometimes attributed to Kotani,\cite{kotani1949magnetic} is well known in studies of $d^5$ metal complexes, and arises from thermal population of local $j_\text{eff} = \frac{3}{2}$ levels, i.e.~the spin-orbital excitons.\cite{figgis1964magnetic} Given that room temperature is roughly 20\% of $\lambda$, such population may be non-negligible. 
The anisotropy in $\mu_\text{eff}$ likely reflects an anisotropic $g$-value afforded by crystal field terms.\cite{yadav2016kitaev} Experimental\cite{PhysRevB.91.094422} and {\it ab-initio}\cite{yadav2016kitaev} estimates of the $g$-values have suggested $g_{ab} \sim 2.0-2.8$, while $g_c \sim 1.0-1.3$, which would be consistent with $|\Delta/\lambda|\sim 0.2$ (Fig.~\ref{fig-CFS1b}(c-d)). On the other hand, it was also suggested that the $g$-tensor anisotropy may be smaller, because large $\Gamma$ terms also produce strongly anisotropic magnetization, even with fully isotropic $g$-tensor.\cite{janssen2017}
The magnitude of $g$-anisotropy has called into question the precise relevance of the $j_\text{eff}$ picture. Indeed, significant deviations from ideal Kitaev interactions are strongly suggested by anisotropic Weiss constants; $\Theta_{ab} = +38$ to +68 K is ferromagnetic, while $\Theta_c = -100$ to $-150$ K is antiferromagnetic. The different signs of the Weiss constants are typically taken as evidence of significant $\Gamma_1$ interactions.\cite{PhysRevB.91.144420}

At low temperatures, kinks in the susceptibility signify the onset of zigzag magnetic order at $T_N = 7-14$\,K, depending on the character of the sample. The 14 K transition is commonly observed in powder samples and low-quality single crystals, and is associated with relatively broad features in the specific heat.\cite{PhysRevB.92.235119,PhysRevB.93.134423,PhysRevB.91.094422} Detailed analysis in Ref.~\onlinecite{banerjee2016proximate,PhysRevB.93.134423} identified this transition with regions of the sample exhibiting many stacking faults.
$\mu^+$SR measurements on powder confirmed a transition at 14\,K and find a second transition
at 11\,K.\cite{PhysRevB.94.020407}
 In contrast, high-quality single crystals exhibit a single transition at 7\,K,\cite{banerjee2016neutron,park2016emergence} with a sharply peaked specific heat. The appearance of zigzag order, in both cases, has been established by neutron diffraction studies.\cite{PhysRevB.91.144420,PhysRevB.92.235119,banerjee2016proximate} As with Na$_2$IrO$_3$, the ordering wavevector is parallel to the monoclinic $b$-axis, while the ordered moment lies in the $ac$-plane, with a magnitude of $0.4-0.7$\,$\mu_B$ -- likely greater than observed in the iridates.\cite{PhysRevB.92.235119,PhysRevB.93.134423} 
The reduced ordered moment (compared to 1\,$\mu_B$) has been noted as a sign of Kitaev physics, but is essentially in line with the expected values for unfrustrated interactions on the honeycomb lattice;\cite{0953-8984-1-10-007} such reductions are typical of magnets with low-dimensionality and reduced coordination number, which enhance quantum fluctuations.


More direct links to Kitaev physics have been suggested on the basis of inelastic probes, both Raman and neutron scattering. The Raman measurements reveal an unusual continuum of magnetic excitations,\cite{PhysRevLett.114.147201} which develops intensity below $100$ K (well above $T_N$), and extends over a wide energy range up to $20-25$\,meV. A similar continuum has been observed in pure and Li-doped Na$_2$IrO$_3$.\cite{0295-5075-114-4-47004} The appearance of the continuum is reminiscent of earlier predictions for the pure Kitaev model in the spin-liquid phase,\cite{PhysRevLett.113.187201} and the spectral shape remains essentially unchanged over a large temperature range, even below $T_N$. These observations are in contrast with the expected behaviour: while broad Raman features in two-dimensional systems are often observed in the paramagnetic phase above $T_N$,\cite{PhysRev.180.591,PhysRevB.85.144434,PhysRevB.91.144411} well-defined spin-wave excitations in the ordered phase often produce sharp two-magnon peaks in the Raman response for $T<T_N$. These peaks arise from the effects of magnon-magnon interactions,\cite{elliott1969effects} and/or van Hove singularities in the magnon density of states.\cite{cottam1986light} The absence of such sharp features below $T_N$ in $\alpha$-RuCl$_3$ (within the studied frequency range) has been suggested as evidence for unconventional magnetic excitations unlike ordinary magnons.\cite{PhysRevLett.114.147201,nasu2016fermionic} This exciting observation has prompted significant interest in the material.

\begin{figure}
\includegraphics[width=0.9\linewidth]{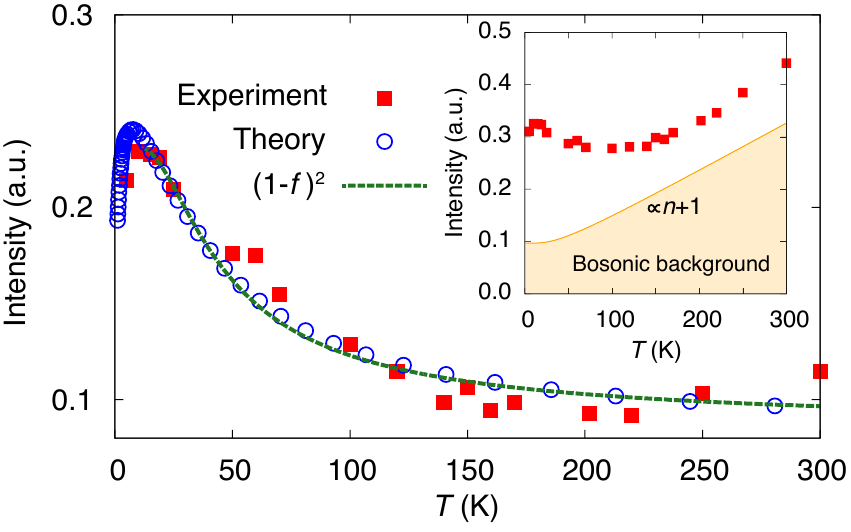}
\caption{\label{fig-rucl3-mag2} Comparison of the experimental Raman continuum intensity with theoretical results for the pure Kitaev model. The authors of Ref.~\onlinecite{nasu2016fermionic} suggested that direct evidence for fermionic excitations in $\alpha$-RuCl$_3$ can be taken from the $[1-f(\omega_0)]^2$ temperature dependence. Reprinted from Ref.~\onlinecite{nasu2016fermionic} by permission from Macmillan Publishers Ltd: \copyright \ 2016.}
\end{figure}

Intriguingly, the authors of Ref.~\onlinecite{nasu2016fermionic} suggested that direct evidence for unconventional {\it fermionic} excitations could be obtained by studying the temperature dependence of the continuum intensity in the paramagnetic phase. For the pure Kitaev model, Raman processes create pairs of Majorana fermions.\cite{perreault2016resonant} In the absence of other considerations, the intensity is therefore expected to decrease with increasing temperature as $\mathcal{I} \sim [1-f(\omega_0)]^2$, where $f(\omega_0)$ is the Fermi function evaluated at some characteristic frequency $\omega_0 \sim \mathcal{O}(K_1)$. Indeed, the authors of Ref.~\onlinecite{nasu2016fermionic} showed that the experimental intensity could be fit with a fermionic dependence (Fig.~\ref{fig-rucl3-mag2}), suggesting the possibility of nontrivial fermionic excitations in $\alpha$-RuCl$_3$! This observation remains to be fully established.\cite{trebst2017kitaev} Apart from experimental considerations, the key criticism is that the magnetic Raman intensity tends to have a relatively featureless temperature dependence above $T_N$. Here, it is sensitive primarily to short-range spin correlations that exist independent of the details of the magnetic interactions. Indeed, the evolution of the continuum intensity in $\alpha$-RuCl$_3$ is nearly indistinguishable (within current experimental resolution) from paramagnetic scattering observed in a range of materials; see, for example, Refs.~\onlinecite{PhysRevB.85.144434,PhysRevB.91.144411,0953-8984-24-43-435604,nakamura2014magnetic}. For this reason, further studies may be required to fully establish the character of the excitations.

\begin{figure}
\includegraphics[width=0.8\linewidth]{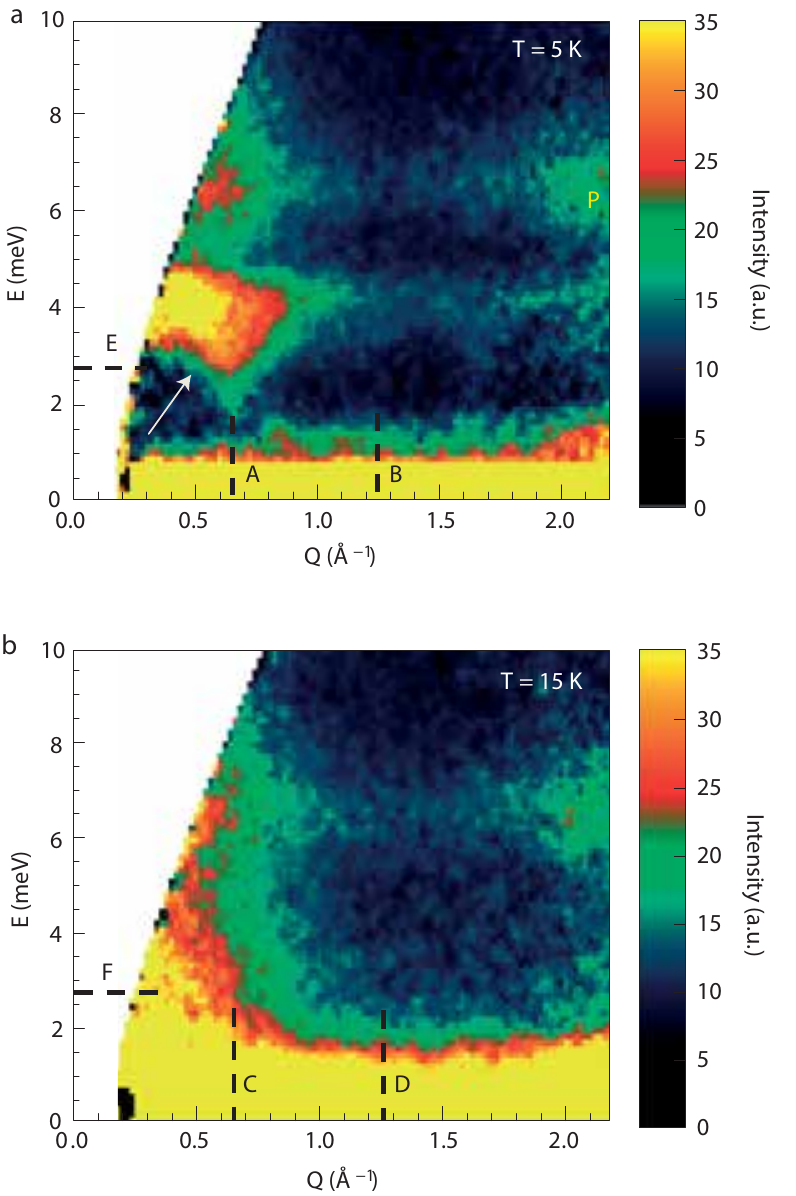}
\caption{\label{fig-rucl3-mag3} Comparison of the powder inelastic neutron scattering intensity above (bottom) and below (top) $T_N$. The magnetic order results in the appearance of well-defined dispersive modes at low energies. Adapted from Ref.~\onlinecite{banerjee2016proximate} by permission from Macmillan Publishers Ltd: \copyright \ 2016.}
\end{figure}

Further evidence for unconventional magnetism in $\alpha$-RuCl$_3$ comes from inelastic neutron scattering, which has provided a detailed view of the excitations in powder,\cite{banerjee2016proximate} and single-crystal samples.\cite{banerjee2016neutron,PhysRevLett.118.107203,do2017incarnation} The 2D character of the excitations has been confirmed by weak dispersion perpendicular to the honeycomb planes.\cite{banerjee2016neutron} Importantly, this allows the single-crystal experiments to probe the entire 2D Brillouin zone, by detecting scattered neutrons in higher 3D Brillouin zones with finite out-of-plane momentum. For this reason, a relatively complete view of the excitations has been possible. Above $T_N$, the paramagnetic continuum seen in Raman is also observed in the neutron response (Fig.~\ref{fig-rucl3-mag3}), extending up to $\sim 15-20$\,meV, with maximum intensity at the center of the 2D Brillouin zone.\cite{banerjee2016neutron,do2017incarnation} 
The continuum is broad in momentum space, but forms a characteristic six-fold star shape associated with well-developed correlations beyond nearest neighbours.\cite{banerjee2016neutron} These results contrast somewhat with the expectations for the pure Kitaev model, for which spin-spin correlations extend only to nearest neighbours at all temperatures.\cite{PhysRevB.92.115127,PhysRevLett.112.207203} Nonetheless, the observation that the continuum survives over a surprisingly broad temperature range $\lesssim 100$\,K (an order of magnitude larger than $T_N$) has led several groups to associate it with fractionalized excitations.\cite{do2017incarnation,banerjee2016neutron, banerjee2016proximate}

Below $T_N$, the onset of zigzag order is indicated by a major reconstruction of the low-energy intensity below 5\,meV, while the broad continuum persists essentially unchanged at high energies.\cite{banerjee2016proximate, banerjee2016neutron} In particular, the excitations above 6\,meV retain the broad six-fold star shape of the paramagnetic response.\cite{banerjee2016proximate,banerjee2016neutron} These excitations are indeed strongly inconsistent with the sharp magnons expected in conventional magnets. In contrast, the low-energy modes show clearer dispersion in momentum space (Fig.~\ref{fig-rucl3-mag3}), with sharp energy minima near the M-points of the honeycomb Brillouin zone.\cite{banerjee2016proximate,banerjee2016neutron,PhysRevLett.118.107203} Recent THz measurements have also identified a sharp magnetic excitation at the $\Gamma$-point.\cite{little2017antiferromagnetic} These are naturally identified with the lowest band of magnons associated with zigzag order.\cite{banerjee2016proximate,PhysRevLett.118.107203} The magnitude of the low-energy dispersion provides a clue regarding the size of the non-Kitaev interactions, since the scattering intensity of the pure Kitaev model is only weakly momentum dependent.\cite{PhysRevB.92.115127,PhysRevLett.112.207203} In particular, the authors of Ref.~\onlinecite{banerjee2016proximate} suggested the dispersing low-energy modes could be understood in terms of significant non-Kitaev terms (particularly, Heisenberg interactions). This finding brings into question the relevance of the Kitaev model for $\alpha$-RuCl$_3$. In this sense, identifying the specific magnetic interactions in $\alpha$-RuCl$_3$, and their relationship to the high-energy continuum, has become a key challenge for the field.

\begin{table*}[t]
\caption{ \label{table_RuCl} Bond-averaged values of the largest magnetic interactions (in units of meV) within the plane for $\alpha$-RuCl$_3$ obtained from various methods. For Ref.~\onlinecite{PhysRevB.93.155143}, the two numbers represent the range of values found in various relaxed structures. ``Pert. Theo.'' refers to second order perturbation theory, ``QC'' = quantum chemistry methods, ``ED'' = exact diagonalization, ``DFT'' = density functional theory total energy, ``Exp. An.'' = experimental analysis. See also Fig.~\ref{fig-rucl3-phasediag}.}
\begin{ruledtabular}
\begin{tabular}{cccccc}
Method&Structure&$J_1$&$K_1$&$\Gamma_1$&$J_3$\\ 
\hline
Exp. An.\cite{banerjee2016proximate}&$-$&$-4.6$&+7.0&$-$&$-$\\
\hline
Pert. Theo.\cite{PhysRevB.93.155143}&$P3_112$&$-3.5$&+4.6&+6.4&$-$\\
QC (2-site)\cite{yadav2016kitaev}&$P3_112$&$-1.2$&-0.5&+1.0&$-$\\
ED (6-site)\cite{PhysRevB.93.214431}&$P3_112$&$-5.5$&+7.6&+8.4&+2.3\\
\hline
Pert. Theo.\cite{PhysRevB.93.155143}&Relaxed&$-2.8/-0.7$&$-9.1/-3.0$&+3.7/+7.3&$-$\\
ED (6-site)\cite{PhysRevB.93.214431}&$C2/m$&$-1.7$&$-6.7$&+6.6&+2.7\\
QC (2-site)\cite{yadav2016kitaev}&$C2/m$&+0.7&$-5.1$&+1.2&$-$\\
DFT\cite{hou2016unveiling}&$C2/m$&$-1.8$&$-10.6$&+3.8&+1.3\\
\hline
Exp. An.\cite{winter2017breakdown}&$-$&$-0.5$&$-5.0$&+2.5&+0.5\\
\end{tabular}
\end{ruledtabular}
\end{table*}

\begin{figure}
\includegraphics[width=\linewidth]{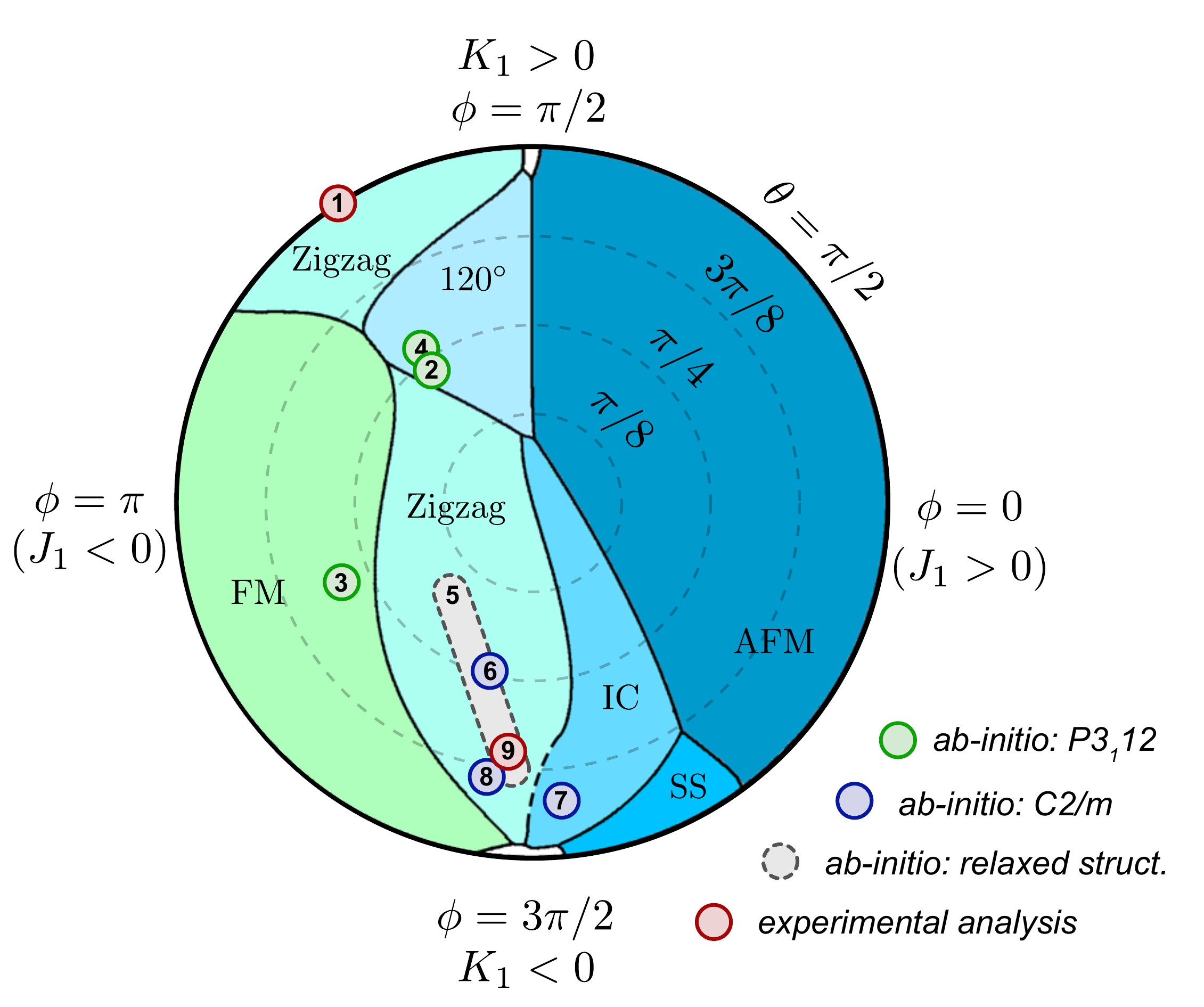}
\caption{\label{fig-rucl3-phasediag} Phase diagram of the $(J_1,K_1,\Gamma_1)$ model (with $J_3=0$) from Ref.~\onlinecite{winter2017breakdown}, using $J_1 = \cos \phi \sin \theta, K_1 = \sin \phi \sin \theta$, and $\Gamma_1 = \cos \theta$. Here, ``FM'' = ferromagnet, ``AFM'' = Neel antiferromagnet, ``IC'' = incommensurate spiral, ``SS'' = stripy order, and the white regions near $\theta=\pi/2, \phi=\pm\pi/2$ are the Kitaev spin-liquids. Reported interactions for $\alpha$-RuCl$_3$ in Table \ref{table_RuCl} are marked by numbered points, corresponding to references: (1)\cite{banerjee2016proximate}, (2)\cite{PhysRevB.93.155143}, (3)\cite{yadav2016kitaev}, (4)\cite{PhysRevB.93.214431}, (5)\cite{PhysRevB.93.155143}, (6)\cite{PhysRevB.93.214431}, (7)\cite{yadav2016kitaev}, (8)\cite{hou2016unveiling}, and (9)\cite{winter2017breakdown}. For (5), the range of values for various relaxed structures is indicated. Although the interactions in the real material are still under debate, the most recent works (5-9) agree $K_1<0$, with $\Gamma_1>0$.}
\end{figure}

In the last several years, one of the major barriers to understanding $\alpha$-RuCl$_3$ has been the wide variety of claims regarding the magnetic interactions, as summarized in Table \ref{table_RuCl} and Fig.~\ref{fig-rucl3-phasediag}. 
From the standpoint of theoretical approaches, discrepancies between various studies have arisen mainly from two factors: i) experimental uncertainty regarding the crystal structure of $\alpha$-RuCl$_3$, and ii) inherent complications that arise in the absence of a small parameter, i.e. when $\lambda \sim \Delta \sim J_H$. This latter condition increases the sensitivity of {\it ab-initio} estimates of the interactions to methodological details.

As with Na$_2$IrO$_3$, the first inelastic neutron experiments\cite{banerjee2016proximate} on $\alpha$-RuCl$_3$ were analyzed in terms of a Heisenberg-Kitaev model with $K_1 > 0$ and $J_1 < 0$, as required to stabilize zigzag order in the absence of other terms. However, such a combination of interactions is unlikely to appear in $\alpha$-RuCl$_3$ from a microscopic perspective; as discussed in Sec.~\ref{sec:theory}, an antiferromagnetic $K_1$ is likely to be realized (in edge-sharing $d^5$ systems) only in conjunction with a large off-diagonal $\Gamma_1$ interaction, as both rely on large direct metal-metal hopping. Interestingly, the first {\it ab-initio} studies of $\alpha$-RuCl$_3$, carried out on the outdated $P3_112$ structure, predicted precisely this situation.\cite{PhysRevB.93.155143,yadav2016kitaev,PhysRevB.93.214431} The anomalously small Ru-Cl-Ru bond angle of 89$^\circ$ in this structure likely overestimates direct hopping effects, leading to $K_1 > 0$, and $|\Gamma_1|\sim |J_1| \sim |K_1|$. However, since the availability of the updated $C2/m$ or $R\bar{3}$ structures, all {\it ab-initio} estimates have been in line with the original Jackeli-Khaliullin mechanism.\cite{PhysRevB.93.155143,PhysRevB.93.214431,yadav2016kitaev,hou2016unveiling} That is, $K_1$ is expected to be ferromagnetic, and to represent the largest term in the Hamiltonian. This is likely supplemented primarily by a large $\Gamma_1>0$ with $|\Gamma_1/K_1| \sim 0.5$, which leads to the observed anisotropy in the Weiss constant $\Theta$. These conclusions are strongly supported by the analysis of Ref.~\onlinecite{winter2017breakdown}, which demonstrated close theoretical agreement with the observed neutron response, when such terms are included. 

In Ref.~\onlinecite{winter2017breakdown}, the authors also offered an alternative interpretation of the observed neutron spectra. They noted that the presence of off-diagonal $\Gamma_1$ interactions lifts underlying symmetries that would otherwise protect conventional magnon excitations. In the absence of such symmetries, the magnons may decay into a broad continuum of multi-magnon states, with characteristics matching the continuum observed in $\alpha$-RuCl$_3$. Since this effect occurs independent of proximity to the Kitaev spin-liquid, the authors concluded that proximity to the Kitaev state does not appear necessary to explain the unconventional continuum in $\alpha$-RuCl$_3$ -- in contrast with previous assertions.\cite{banerjee2016proximate,banerjee2016neutron} In fact, strong damping of the magnons should be considered a general feature of anisotropic magnetic interactions, suggesting similar excitation continua may appear in all materials discussed in this review. An interesting question is to what extent such overdamped magnons resemble the Majorana excitations of the pure Kitaev model?\cite{hermanns2017physics}

Finally, we note that more recent interest has turned to the response of $\alpha$-RuCl$_3$ in an external magnetic field, which suppresses the zigzag order at roughly $B_c \sim 7$ T for in-plane fields.\cite{PhysRevB.92.235119} Interest in the high-field phase is partially motivated by predictions of a field-induced spin-liquid state.\cite{yadav2016kitaev} A picture of this high-field state is now emerging from neutron,\cite{sears2017phase,banerjee2017b} NMR,\cite{baek2017observation,zheng2017gapless,jansa2017} specific heat,\cite{baek2017observation,sears2017phase,wolter2017field} magnetization,\cite{PhysRevB.92.235119,PhysRevB.91.094422} dielectric,\cite{aoyama2017} and thermal transport\cite{hentrich2017large,leahy2016anomalous} measurements, as well as from THz and electron spin resonance\cite{ponomaryov2017,wang2017} spectroscopies. 

In the vicinity of the critical field, phononic heat transport is strongly suppressed, indicating a multitude of low-lying magnetic excitations consistent with the closure of an excitation gap.\cite{hentrich2017large,leahy2016anomalous} This result is supported both by specific heat data\cite{baek2017observation,sears2017phase,wolter2017field} and by a strong increase of the NMR relaxation rate near $B_c$ at low temperatures.\cite{baek2017observation} The closure of the gap likely demonstrates the existence of a field-induced quantum critical point, which has been suggested to be of Ising type\cite{wolter2017field} based on the magnetic interactions of Ref.~\onlinecite{winter2017breakdown}. For $B>B_c$, NMR,\cite{baek2017observation} thermal transport,\cite{hentrich2017large} and specific heat\cite{baek2017observation,sears2017phase,wolter2017field} measurements all demonstrate the opening of an excitation gap that increases linearly with field. In this field range, the specific heat shows no peak on decreasing the temperature. This has been suggested as evidence that this gapped state is a quantum spin-liquid connected to the Kitaev state, thus implying the emergence of fractionalized excitations at high field.\cite{banerjee2017b}
However, recent consideration of the relevant microscopic interactions have indicated that the high-field state may instead represent a quantum paramagnetic state supporting non-fractionalized excitations and lacking direct connection to the Kitaev spin-liquid.\cite{winter2017b} The nature of the excitations close to the critical field $B\approx B_c$ remains an interesting subject of future study, particularly given the possibility of quantum critical behaviour.\cite{baek2017observation,wolter2017field}

\subsection{Beyond 2D: $\beta$- and $\gamma$-Li$_2$IrO$_3$}

\begin{figure}
\includegraphics[width=\linewidth]{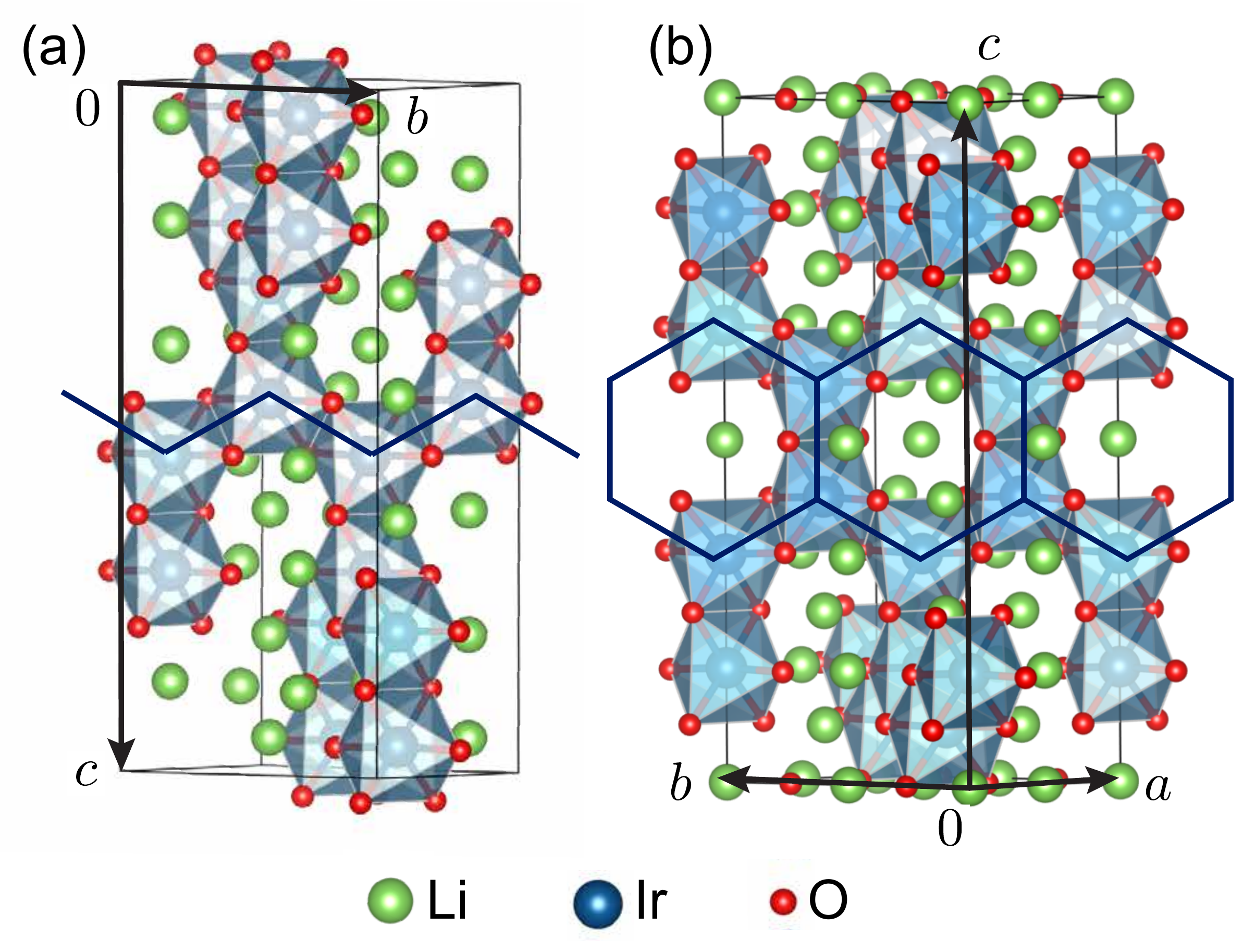}
\caption{Crystal structures of (a) $\beta$- and (b) $\gamma$-phases of Li$_2$IrO$_3$. The structures feature crossed zigzag and honeycomb chains, respectively, running in the $ab$-plane. These are emphasized in each case. }
\label{fig-3dir-struc}
\end{figure}
\label{sec:beta-gamma}
The planar honeycomb iridate $\alpha$-Li$_2$IrO$_3$ can be seen as a toolbox for designing further Kitaev materials. Its $\beta$- and $\gamma$-polymorphs represent three-dimensional (3D) varieties of the honeycomb lattice. Similar to the original (planar) honeycomb version, each site of the lattice is three-coordinated, but the bonds are no longer coplanar - forming, instead, 3D networks that are coined ``hyper''-honeycomb ($\beta$-Li$_2$IrO$_3$, $\mathcal H^0$) and ``stripy''- or ``harmonic''-honeycomb ($\gamma$-Li$_2$IrO$_3$, $\mathcal H^1$) lattices. Here, $\mathcal H$ stands for a single stripe of hexagons, and $\mathcal H^{\infty}$ denotes planar honeycomb lattice. By changing the superscript at $\mathcal H$, an infinitely large number of such lattices can be constructed.\cite{modic2014} 

\subsubsection{Crystal structures and synthesis}

On the structural level, the polymorphism of Li$_2$IrO$_3$ stems from the fact that the A$_2$MO$_3$ oxides are ordered versions of the rocksalt structure, where oxygen ions form close packing, with A and B cations occupying octahedral voids.\cite{hauck1980} By changing the sequence of the A and B ions, crystal structures hosting any given $\mathcal H^n$ spin lattice can be generated, although under real thermodynamic conditions only a few of them are stable. The discovery of three different well-ordered polymorphs in Li$_2$IrO$_3$ seems to be a result of extensive crystal growth attempts inspired by prospects of studying Kitaev physics. Other A$_2$MO$_3$ compounds are also known in multiple polymorphs, although many of them are fully or partially disordered versions of the $\alpha$- and $\beta$-type structures.\cite{hauck1980}

The hyperhoneycomb $\beta$-phase of Li$_2$IrO$_3$ is a high-temperature polymorph that forms upon heating the $\alpha$-phase above 1000\,$^{\circ}$C.\cite{Mannithesis} Tiny single crystals with the size of few hundred $\mu$m are obtained by annealing in air, similar to Na$_2$IrO$_3$,\cite{biffin2014b,takayama2015,ruiz2017} whereas larger crystals can be grown by vapor transport from separated educts.\cite{freund2016} $\beta$-Li$_2$IrO$_3$ crystallizes in the orthorhombic space group $Fddd$, with zigzag chains running in alternating directions in the $ac$-plane (Fig.~\ref{fig-3dir-struc}).\cite{takayama2015,biffin2014b} In the language of the Kitaev interactions, these chains form the X- and Y-bonds, while the Z-bonds (parallel to the $b$-axis) link together adjacent layers of chains. For the initially reported structure of Ref.~\onlinecite{takayama2015}, the Ir-O-Ir bond angles are all $\sim 94^\circ$, indicating a similar degree of trigonal compression of the local IrO$_6$ octahedra as in the $\alpha$-phase. 

The stripy-honeycomb $\gamma$-phase is instead grown at lower temperatures from the LiOH flux,\cite{modic2014} yielding crystals with largest dimension $\sim$ 100 $\mu$m. Its thermodynamic stability with respect to the other two polymorphs has not been investigated.\footnote{Note that in chemistry literature $\gamma$-phase typically refers to the disordered rocksalt polymorph of A$_2$MO$_3$ compounds.} $\gamma$-Li$_2$IrO$_3$ crystallizes in the orthorhombic $Cccm$ space group, with crossed stripes of honeycomb plaquettes running in the $ac$-plane (Fig.~\ref{fig-3dir-struc}). Each stripe is composed of pairs of zigzag chains, containing the X- and Y-bonds, in the Kitaev terminology. There are two crystallographically unique Z-bonds: those within each honeycomb stripe, and those linking adjacent stripes. Unlike the $\alpha$- and $\beta$-phases, the distortion of the IrO$_6$ octahedra is quite asymmetric, leading to a range of Ir-O-Ir bond angles between $\sim 90^\circ$ and $\sim97^\circ$. On this basis, the magnetic properties can be expected to be complex, as discussed below.

\subsubsection{Electronic Properties}

Given their more recent discovery, significantly less is known regarding the electronic structure of the 3D Li$_2$IrO$_3$ phases, although many aspects are expected to resemble their 2D counterparts. Both are known to be electrical insulators on the basis of DC resistivity.\cite{takayama2015,modic2014} {\it Ab-initio} estimates of the crystal field splitting in the hyperhoneycomb $\beta$-phase have suggested it to be on the same order as in the 2D honeycomb materials,\cite{katukuri2016,kim2015} based on the crystal structure of Ref.~\onlinecite{takayama2015}. This seems to be consistent with the results of x-ray magnetocircular dichroism (XMCD) experiments that observe a pronounced difference in the intensities at the L$_2$ and L$_3$ edges, in agreement with the $j_\text{eff}$ predictions.\cite{takayama2015} In contrast, the trigonal crystal field terms in the $\gamma$-phase are estimated to be much larger, $\Delta \sim 0.2$ eV, based on the reported crystal structure.\cite{Yingli2015} The optical conductivity of $\gamma$-Li$_2$IrO$_3$ has been reported, and shows a similar dominant peak near 1.5 eV as for the 2D iridates due to intersite $j_\text{eff} = \frac{3}{2} \rightarrow \frac{1}{2}$ excitations.\cite{PhysRevB.92.115154} However, enhanced intensity at lower frequency is suggestive of some departures from ideality, which might be consistent with the larger distortion of the IrO$_6$ octahedra.\cite{Yingli2015} This places some importance on establishing the validity of the $j_\text{eff}$ picture in these materials.

\subsubsection{Magnetic Properties}
Both $\beta$- and $\gamma$-Li$_2$IrO$_3$ are readily distinguishable from planar honeycomb iridates by the sharply increasing magnetic susceptibility that becomes constant below $T_N=37$\,K ($\beta$)~\cite{biffin2014b,takayama2015} and 39.5\,K ($\gamma$).\cite{modic2014,biffin2014g} This increase appears to be highly anisotropic and occurs only for the magnetic field applied along the $b$ direction in both compounds.\cite{modic2014,ruiz2017} Indeed, the Curie-Weiss temperatures of both materials are highly anisotropic. For $\beta$-Li$_2$IrO$_3$, fitting of the susceptibility above 150 K yielded $\Theta_a \sim -94$\,K, $\Theta_b \sim +18$\,K, and $\Theta_c\sim 0$, with somewhat anisotropic effective moments in the range $\mu_\text{eff} \sim 1.7-2.0$\,$\mu_B$.\cite{ruiz2017} 
In contrast, strong deviations from Curie-Weiss behaviour were reported for the $\gamma$-phase,\cite{modic2014} albeit with a similar level of anisotropy of the $g$-values in the range $\sim 1.9-2.4$.\cite{kimchi2014} These values are suggestive of strongly anisotropic magnetic interactions, with some deviations from the ideal $j_\text{eff}$ picture. 

Comparing to the $\alpha$-phase, the $\Theta$ values are shifted toward positive (ferromagnetic) values. The highest (most ferromagnetic) value is observed for $\Theta_b$ identifying the $b$ direction as most polarizable. Isothermal magnetization measured for this field direction increases sharply in low fields for both the $\beta$- and $\gamma$-phases mirroring the susceptibility upturn. In both cases, a kink slightly below 3\,T indicates suppression of the zero-field ordered state, consistent with the vanishing of the  $\lambda$-type anomaly in the specific heat at $T_N$.\cite{takayama2015,modic2016,ruiz2017} 

While the thermodynamic properties set $\beta$- and $\gamma$-phases apart from $\alpha$-Li$_2$IrO$_3$, the ordered states of all three polymorphs share a lot of commonalities.\cite{williams2016,biffin2014b,biffin2014g} All three order as incommensurate spiral phases, featuring counter-rotating spirals, which are hallmarks of the Kitaev exchange.\cite{biffin2014b,biffin2014g} The $\beta$- and $\gamma$-phases additionally share the same propagation vector $\mathbf k=(0.57(1),0,0)$ but differ in their basis vector combinations: $(iA_x,iC_y,F_z)$ and $i(A,-A)_x, i(-1)^m(F,-F)_y,(F,F)_z$ ($m=1,2$), respectively.\footnote{Note that the $\gamma$-phase features two nonequivalent Ir sites in the $Cccm$ structure, as opposed to a single Ir site in the $Fddd$ structure of the $\beta$-phase.} As noted in section \ref{sssec_ir_rh_thermag}, the complexity of these magnetic structures leaves room for interpretation regarding the underlying magnetic interactions. Phenomenologically, it is known that the ordered states of both $\beta$- and $\gamma$-phases can be reproduced for a nearest-neighbor Heisenberg-Kitaev model supplemented by an additional Ising anisotropy $I_c$ along the Z-bonds only.\cite{kimchi2014,kimchi2015} However, it has also been shown that such phases appear in the absence of $I_c$, within the $(J_1,K_1,\Gamma_1)$-model studied in Ref.~\onlinecite{kim2015} and~\onlinecite{lee2015}. In both cases, a dominant ferromagnetic Kitaev $K_1$ term is required to stabilize the observed order. For a complete discussion of these two approaches, the reader is referred to Ref.~\onlinecite{lee2016}.

Several {\it ab-initio} studies of $\beta$-Li$_2$IrO$_3$ concur on the ferromagnetic nature of the Kitaev term $K_1$ and on the relevance of the off-diagonal anisotropy $\Gamma_1$, which may be on par with $K_1$.\cite{kim2015,kim2016,katukuri2016} The weak distortions of the hyperhoneycomb lattice appear to play a minor role, leading to roughly similar interactions on the X-, Y-, and Z-bonds.\cite{kim2015,katukuri2016} In this sense, the $(J_1,K_1,\Gamma_1)$-model appears to provide an adequate starting point for understanding $\beta$-Li$_2$IrO$_3$. However, further work will be required to fully establish the minimal interaction model. For example, the authors of Ref.~\onlinecite{katukuri2016} emphasized the role of longer-range interactions, with the inclusion of a $J_2$ term. Considering the symmetry of the crystal structure, such long-range terms might also include Dzyaloshinskii-Moriya interactions, which typically stabilize incommensurate states, as noted for the $\alpha$-phase.\cite{PhysRevB.93.214431} To date, no significant {\it ab-initio} studies of the magnetic interactions have been reported on the structurally more complex $\gamma$-phase, which still evades detailed microscopic analysis.

A fruitful approach in the study of the 3D Kitaev systems has been the use of external pressure~\cite{breznay2017,veiga2017,takayama2015} and magnetic fields\cite{modic2016,ruiz2017} to tune the magnetic response. Like any three-coordinated lattice, the hyperhoneycomb and stripy-honeycomb geometries give rise to spin-liquid states when purely Kitaev interactions are considered.\cite{kimchi2014,mandal2009,nasu2014} On the other hand, realistic models including $J$, $K$, and $\Gamma$ terms for nearest-neighbor interactions turn out to be quite complex hosting multiple ordered states of different nature along with a few regions where spin-liquid states might occur.\cite{lee2014a,lee2015,lee2016} The prospects of tuning $\beta$- and $\gamma$-Li$_2$IrO$_3$ toward a disordered, possibly spin-liquid state are actively explored both experimentally~\cite{takayama2015,modic2016,breznay2017} and theoretically.\cite{kim2016} The zero-field incommensurate states are indeed quite fragile and can be suppressed by either pressure~\cite{breznay2017} or magnetic field applied along a suitably chosen direction.\cite{modic2016,ruiz2017} Understanding the nature of emerging new phases, and their relationship to the underlying microscopic description, represents an interesting venture that requires further investigation.

The 2D and 3D honeycomb-like systems are easily distinguishable by their Raman response.\cite{perreault2015} As with $\alpha$-RuCl$_3$, a continuum is observed extending over a broad frequency range. Polarization dependence of the experimental Raman spectra for both $\beta$- and $\gamma$-Li$_2$IrO$_3$ is indeed consistent with predictions for the Kitaev model,\cite{glamazda2016,perreault2015} whereas the temperature-dependence of the spectral weight has been conjectured as a signature of fractionalized excitations.\cite{glamazda2016} As with $\alpha$-RuCl$_3$, this interpretation is considered controversial, but the similarities of the observations clearly place the 3D iridates on the same grounds as 2D systems. 

\section{Extending to Other Lattices} \label{sec:othermat}
Half a decade of intense research has shown that realising purely Kitaev interactions may not be feasible in any real material, but extended models including more realistic interactions host a plethora of interesting states and phenomena of their own. This has stimulated investigations of a broader class of $4d$ and $5d$ transition-metal compounds, where frustrated anisotropic interactions have been suggested to play a significant role. While the full relevance of Kitaev interactions and the Jackeli-Khaliullin mechanism in these materials remains under debate, we briefly review here a selection of these systems with a focus on the future prospects of their research.

\subsection{Hyperkagome Na$_4$Ir$_3$O$_8$: \\ A Possible 3D spin-liquid}

\begin{figure}
\includegraphics[width=0.95\linewidth]{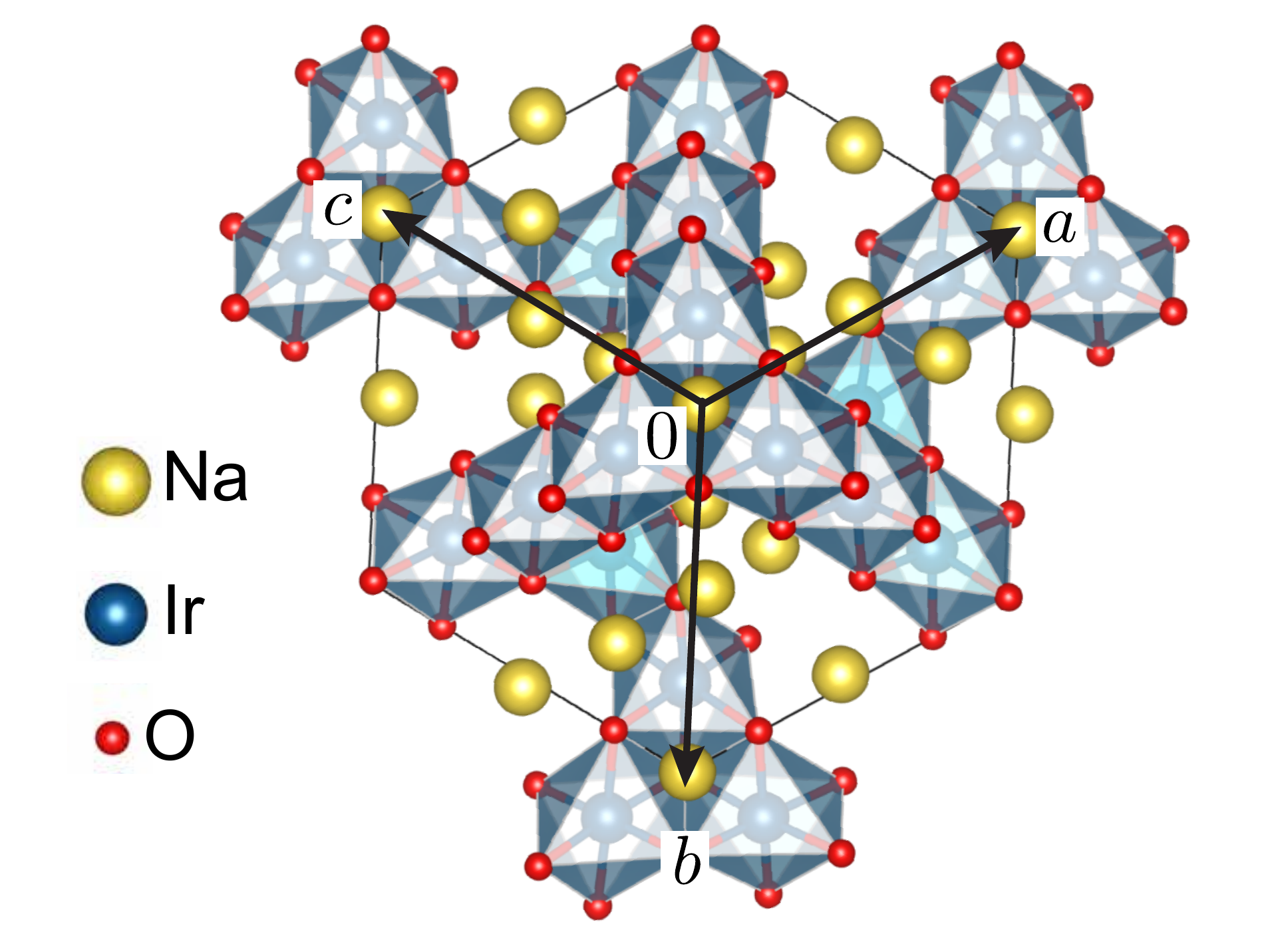}
\caption{Idealized crystal structure of hyperkagome compound Na$_4$Ir$_3$O$_8$ viewed along the chiral 3-fold symmetry axis.}
\label{fig-hyperkag-struc}
\end{figure}
The hyperkagome material Na$_4$Ir$_3$O$_8$ holds a special place in the study of Kitaev interactions, as it represents one of the first $5d$ materials for which bond-dependent Kitaev-like terms were discussed.\cite{chen2008} Its study also triggered experimental work on honeycomb iridates, as Na$_2$IrO$_3$ has been obtained~\cite{singh2010} as a side product of (unsuccessful) crystal growth for Na$_4$Ir$_3$O$_8$. The non-trivial chiral $P4_132 / P4_332$ crystal structure of Na$_4$Ir$_3$O$_8$ hosts a hyperkagome lattice of Ir$^{4+}$ ions, a 3D analog of planar kagome lattice,\cite{okamoto2007} as shown in Fig.~\ref{fig-hyperkag-struc}. Following early theoretical interest in this system~\cite{hopkinson2007,lawler2008a,lawler2008b,zhou2008,bergholtz2010,podolsky2011,singh2012,chen2013}, magnetic exchange parameters were assessed microscopically arriving at somewhat conflicting results on the nature of anisotropy and its role in this material.\cite{chen2008,norman2010,micklitz2010} Recent RIXS measurements\cite{takayama2014} can be interpreted in the $j_\text{eff}$ picture, but quantum chemistry calculations have also suggested significant crystal-field splitting.\cite{katukuri2014quantum}

Experimental data do not resolve the controversy over the magnetic interactions. Na$_4$Ir$_3$O$_8$ exhibits strong antiferromagnetic coupling, as reflected by the Curie-Weiss temperature $\Theta=-650$\,K, and exhibits a peak in the magnetic specific heat around 30\,K. The linear term in the low-temperature specific heat~\cite{singh2013} and the broad excitation continuum observed by Raman scattering~\cite{gupta2016} are reminiscent of a gapless spin liquid.\cite{forte2013} On the other hand, spin freezing is observed at 6\,K,\cite{dally2014,shockley2015} about the same temperature as in Li$_2$RhO$_3$.\cite{khuntia2015} Recent theoretical works have reconsidered the phase diagram of the honeycomb-inspired nearest neighbour $(J_1,K_1,\Gamma_1$) model on the hyperkagome lattice,\cite{PhysRevB.94.064416,PhysRevB.93.094419} with the inclusion of a symmetry-allowed DM-interaction. These works found a variety of incommensurate states suggesting a complex energy landscape with only discrete symmetries. Such a situation has been argued to promote glassy spin-freezing. 

Given these observations, the spin freezing may also be promoted by weak structural disorder in Na$_4$Ir$_3$O$_8$. In the stoichiometric compound, the Na sites are likely disordered.\cite{okamoto2007} Moreover, single crystal growth for Na$_4$Ir$_3$O$_8$ was not successful so far, most likely because sodium is easily lost to produce mixed-valence Na$_{4-x}$Ir$_3$O$_8$.\cite{takayama2014} The Na deficiency may extend to $x=1.0$, manifesting a rare example of doping an Ir$^{4+}$-based insulator into a semi-metallic state.\cite{proepper2014,fauque2015,yoon2015,balodhi2015} Were Na$_4$Ir$_3$O$_8$ available in very clean form, it would be a natural candidate for spin-liquid behavior on the 3D hyperkagome lattice, but chemistry has so far been a major obstacle in achieving clean single crystals.

\subsection{Quasi-1D CaIrO$_3$: Failure of the $j_\text{eff}$ Picture}
The post-perovskite phase of CaIrO$_3$ was first discussed in the Jackeli-Khaliullin context in Ref.~\onlinecite{PhysRevLett.110.217212}. Earlier work had established the material as a magnetic insulator with a charge gap of $\sim 0.17$ eV, which displays antiferromagnetic order below $T_N = 115$\,K.\cite{PhysRevB.74.241104} While the crystal structure features edge-sharing Ir$^{4+}$ octahedra, it is now established that the crystal-field splitting associated with tetragonal distortions is sufficiently large to quench the $j_\text{eff}$ state. In this sense, CaIrO$_3$ stands as a primary counterexample to the other materials presented in this review.

\begin{figure}
\includegraphics[width=\linewidth]{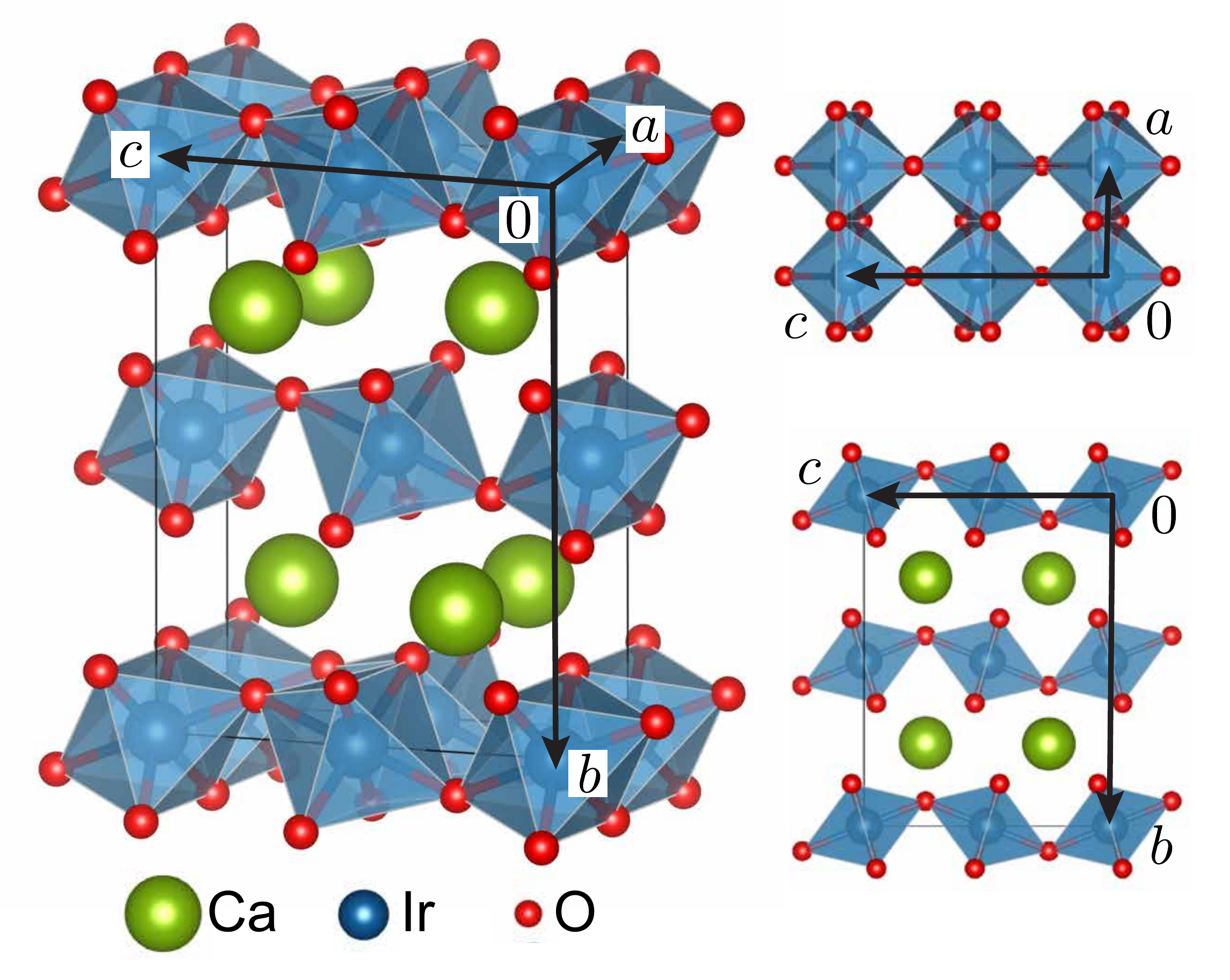}
\caption{\label{fig-cairo-struc} 
Different views of the unit cell of CaIrO$_3$ showing a combination of edge- and corner-sharing octahedra.}
\end{figure}
Within the orthorhombic $Cmcm$ structure of CaIrO$_3$, the Ir$^{4+}$ ions form
decoupled layers of IrO$_6$ octahedra lying within the
$ac$-plane, as shown in Fig.~\ref{fig-cairo-struc}. Along the $c$-axis, the octahedra are linked by
a tilted corner sharing geometry, and are therefore expected to display large
antiferromagnetic Heisenberg-type magnetic interactions. In contrast, the bonds
along the $a$-axis are edge-sharing type, having the potential to realize
weaker ferromagnetic Kitaev interactions.\cite{PhysRevLett.110.217212} This
view is indeed consistent with the observed magnetic order, in which spins
adopt a canted antiferromagnetic state with antiferromagnetic alignment along
the $c$-axis bonds, and ferromagnetic alignment for $a$-axis bonds. Provided
the $a$-axis bonds featured dominant Kitaev couplings, the tilting of the
octahedra would lead to a spontaneous canted moment along the $b$-axis; such a
moment is indeed clearly observed in magnetization measurements. Moreover,
initial evidence for the $j_\text{eff}$ picture was taken from the absence of
resonant x-ray scattering (RXS) intensity at the L$_2$ edge, which
would be suppressed for large $j_\text{eff} = \frac{1}{2}$ character in the
$t_{2g}$ hole. 

Despite such positive evidence for $j_\text{eff}$ physics in CaIrO$_3$, there remained several discrepancies. {\it Ab-initio} calculations suggested large crystal field splittings on the order of $0.6-0.8$ eV (on par with $\lambda$), associated with the tetragonal distortions.\cite{Bogdanov12,PhysRevLett.115.096401} Such splittings were predicted to largely quench the orbital moment in the ground state, leading to predominantly Heisenberg-type interactions, with small additional anisotropies. Interestingly, the interactions along the corner sharing $c$-axis bonds were estimated to be larger than the $a$-axis interactions by nearly $|J_c/J_a| \sim 20$, emphasizing the suppression of interactions for edge-sharing bonds. Subsequent RIXS experiments strongly confirmed the results of the {\it ab-initio} calculations, through the observation of a large splitting of the $t_{2g}$ states consistent with $|\Delta/\lambda| > 1$.\cite{PhysRevLett.112.176402} These observations highlight the sensitivity of the low-energy spin-orbital coupled states to crystal field splitting.

\subsection{Double perovskites: \\ Complex magnetism on an fcc lattice}
\begin{figure}
\includegraphics[width=\linewidth]{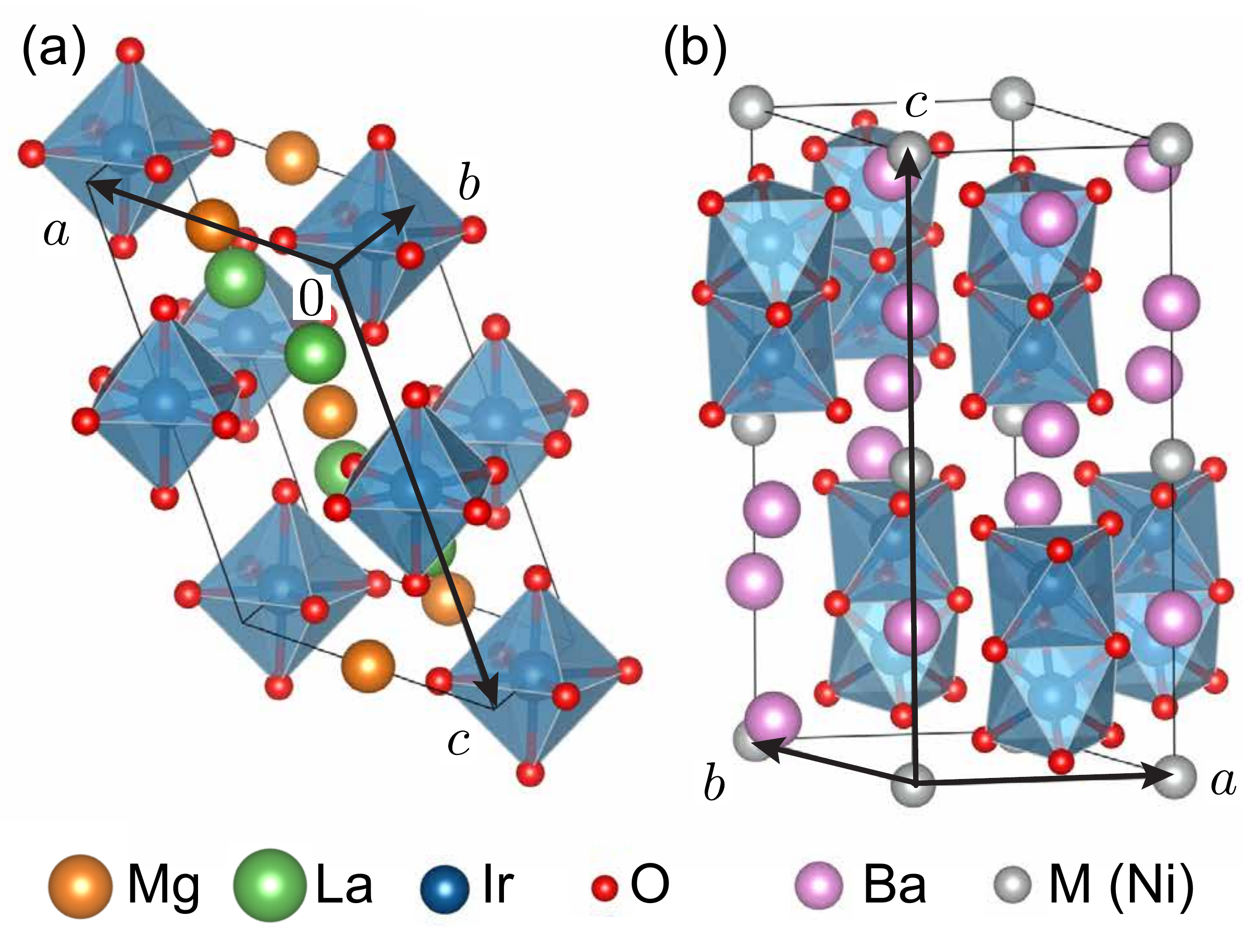}
\caption{Crystal structures of (a) the double perovskite La$_2$MgIrO$_6$ and (b) the hexagonal perovskite Ba$_3$IrTi$_2$O$_9$. In the former, the IrO$_6$ octahedra are isolated but interact via long-range coupling. In the latter, the IrO$_6$ octahedra form face-sharing dimers with properties varying with oxidation state.}
\label{fig-double-struc}
\end{figure}

La$_2$MgIrO$_6$ and La$_2$ZnIrO$_6$ are double perovskites with the checkerboard ordering of the Ir and Mg/Zn atoms (Fig.~\ref{fig-double-struc}).\cite{ramos1994,currie1995,battle1996} The Ir$^{4+}$ ions are well separated by non-magnetic ``spacers'' (Mg$^{2+}$, Zn$^{2+}$) that bring the energy scale of magnetic couplings down to 10\,K or less,\cite{currie1995,powell1993} and presumably restrict interactions to nearest neighbors. Spatial arrangement of the magnetic ions is described by an fcc lattice~\cite{cook2015} with a minor distortion arising from monoclinic symmetry of the underlying crystal structure.

Interactions between the Ir$^{4+}$ ions are predominantly antiferromagnetic.\cite{cao2013} Long-range order sets in below $T_N=12$\,K in La$_2$MgIrO$_6$ and 7.5\,K in La$_2$ZnIrO$_6$. Interestingly, the magnetic structure of La$_2$MgIrO$_6$ is purely collinear, A-type antiferromagnetic, whereas La$_2$ZnIrO$_6$ features a similar, but canted ordered state with the sizable net moment of 0.22\,$\mu_B$/Ir.\cite{cao2013} While the microscopic origin of this difference remains unsettled,\cite{battle1996,cook2015,zhu2015,aczel2016} the similarity between La$_2$MgIrO$_6$ and La$_2$ZnIrO$_6$ is reinforced by a gapped and dispersionless excitation observed in both systems taken as possible evidence for dominant Kitaev interactions in Ir$^{4+}$-based doubled perovskites.\cite{aczel2016} Sr$_2$CeIrO$_6$ with the non-magnetic Ce$^{4+}$ is a further member of the same family.\cite{harada1999,harada2000,kanungo2016}

Whereas high connectivity of the fcc lattice is probably detrimental for the spin-liquid physics, the $J-K-\Gamma$ model on the fcc lattice hosts a variety of interesting ordered states even in the classical limit.\cite{cook2015} On the experimental side, double perovskites are very convenient for chemical modifications, such as electron/hope doping~\cite{zhu2015} or tailoring magnetic behavior by replacing Mg or Zn with $3d$ ions.\cite{powell1993,currie1995} Multiple examples of Ir-containing double perovskties have been reported. However, many of them involve charge transfer~\cite{kolchinskaya2012} resulting in the non-magnetic Ir$^{5+}$, or feature $3d$ ions with high magnetic moments that obscure the $4d/5d$ magnetism.\cite{narayanan2010,manna2016}

Cleaner examples of anisotropic magnetism on the fcc lattice may be found in hexahalides~\cite{roessler1977} like K$_2$IrCl$_6$, where cubic symmetry keeps the lattice undistorted and ensures the pure $j_{\rm eff}=\frac12$ state of Ir$^{4+}$. Magnetic behavior of hexahalides shows salient signatures of magnetic frustration,\cite{cooke1959,griffiths1959,hutchings1967,lindop1970,moses1979} and the high symmetry of the lattice prevents the appearance of Dzyaloshinkii-Moriya interactions between select Ir centers. These materials were studied long before the Kitaev era and warrant re-evaluation in the context of current knowledge on the magnetism of Ir$^{4+}$ compounds.

\subsection{Hexagonal perovskites}
Hexagonal perovskites are derivatives of the cubic perovskite structure, in which half of the octahedra are partly replaced by dimers, trimers, and, in more exotic cases, larger ``stacks'' of face-sharing octahedra (Fig.~\ref{fig-double-struc}). According to their name, these structures (at least in their simplest and largely idealized version) feature hexagonal symmetry that facilitates formation of triangular and hexagonal lattice geometries.

A naive attempt of incorporating Ir$^{4+}$ into hexagonal perovskite structure results in Ba$_3$IrTi$_2$O$_9$,\cite{byrne1970} which unfortunately exhibits structural disorder,\cite{dey2012} in addition to the promising feature of absent magnetic order. An idealized, structurally ordered version of this structure would entail sizable Kitaev interactions,\cite{catuneanu2015} but in reality Ti$^{4+}$ and Ir$^{4+}$ are heavily mixed within the dimers.\cite{dey2012,kumar2016,lee2017} Since Ir$^{4+}$ is unlikely to occupy the single octahedra, accommodating two Ir atoms within the dimer and leaving non-magnetic ions to the single octahedra turns out to be a more viable approach. 

Such Ba$_3$MIr$_2$O$_9$ oxides are more likely to form ordered crystal structures indeed.\cite{doi2004,sakamoto2006} Interesting low-temperature magnetism will generally appear only in the mixed-valence case of Ir$^{4.5+}$ that corresponds to trivalent M ions. The purely Ir$^{4+}$ systems should be mundane spin dimers entering singlet state already at high temperatures.\cite{doi2004} The formally non-magnetic Ir$^{5+}$ may, however, exhibit vague signatures of weak magnetism in the same type of structure.\cite{nag2016} At least one of these compounds, Ba$_3$InIr$_2$O$_9$, lacks long-range magnetic order and reveals persistent spin dynamics down to 20\,mK potentially showing quantum spin liquid behavior,\cite{dey2017} whereas Ba$_3$YIr$_2$O$_9$~\cite{panda2015} may be magnetically ordered below 4\,K.\cite{dey2013,dey2014} 

The mixed-valence Ir$^{4.5+}$ state entails magnetic electrons occupying molecular orbitals of the Ir--Ir dimer. Correlations, covalency, and spin-orbit coupling select among several electronic states~\cite{streltsov2016} and define interactions between such dimers. The exact nature of these electronic states, the relevance of Kitaev terms in ensuing magnetic interactions, and even the geometry of magnetic couplings (hexagonal, triangular, or both~\cite{dey2017}) remain to be established. 

The diverse structural chemistry with a choice of more than 10 different elements on the M site~\cite{doi2004,sakamoto2006} and feasibility of Ir$_3$O$_{12}$ trimers replacing the dimers in Ba$_3$MIr$_2$O$_9$~\cite{shimoda2009,shimoda2010} result in a much higher flexibility of hexagonal perovskites compared to the honeycomb iridates, which are essentially restricted to only two compounds with Li and Na. Hexagonal perovskites with $4d$ and $5d$ metals other than Ir show low ordered moments~\cite{senn2013} or even formation of disordered magnetic states,\cite{ziat2017} which may be of interest too. On the downside, hexagonal perovskites are prone to structural distortions~\cite{zurloye2009} sometimes accompanied by tangible disorder.\cite{ling2010} In mixed-valence systems, charge-transfer or charge-ordering processes may additionally occur.\cite{miiller2012,kimber2012,terzic2015}

\subsection{Other materials}
Interesting physics of the Kitaev-Heisenberg model on the triangular lattice\cite{li2015,rousochatzakis2016,PhysRevB.91.155135,PhysRevB.92.184416} and the dearth of compounds being representative of this model call for a further materials search, extending to new classes of compounds and employing advanced synthesis techniques. Exotic and fairly expensive rhodium compounds might come for help here, because experimental procedures for synthesizing K$_x$RhO$_2$ oxides are well established.\cite{zhang2013} The ultimate limit of Rh$^{4+}$-based layered RhO$_2$ is probably unfeasible, given the fact that a layered structure collapses upon the complete deintercalation of the alkaline-metal cation.\cite{mikhailova2016} On the other hand, such materials could be good candidates for Kitaev-like models on the triangular lattice in the electron-doped regime. For the undoped regime, other structure types should be searched for.

Elaborate chemistry tools may be used for deliberate preparation of new $4d$ and $5d$ transition-metal compounds. The first step in this direction is incorporating Ru$^{3+}$ into metal-organic frameworks,\cite{yamada2017} which are known for their high flexibility and tunability and may potentially realize spin lattices beyond honeycombs in 2D or 3D.\cite{obrien2016,yamada2017b} However, further work will be needed to assess the magnitude of Kitaev terms in such compounds, where the linkage between the Ru$^{3+}$ ions is significantly more complex than in $\alpha$-RuCl$_3$.

\section{Outlook}

The experimental explorations on $4d$ and $5d$ transition-metal-based Mott-insulating materials with frustrated
anisotropic interactions reviewed in this paper validate the realization of
  the Jackeli-Khaliullin mechanism, i.e. 
 there are now many candidate materials 
 with strong evidence for dominant ferromagnetic Kitaev-like interactions in all such cases.
 However, the current studies also emphasize the difficulty of realizing the
idealized pure Kitaev model in real materials. Nonetheless, the complex
properties of such systems have proven to host a variety of surprises and
associated physical and synthetic questions that need to be resolved: \begin{itemize}
\item How can the magnetic interactions be more strictly controlled via 
 external parameters such as chemical and/or physical pressure, strain or magnetic field?
\item Given the strong sensitivity of the magnetic interactions to structural details, what is the role of structural disorder and magnetoelastic coupling?
\item How can such anisotropic (Kitaev) interactions be synthetically extended to other lattices? 
\item What role can the further development of anisotropic experimental probes (such as polarization-sensitive RIXS or Raman scattering, other spectroscopic probes)  play in the study of such magnetism?
\item How can one describe the dynamical response of strongly anisotropic magnets, where there is emerging experimental evidence for a clear breakdown of the conventional magnon picture? 
\item To what extent are the interactions beyond the Kitaev terms responsible for the observed properties of the known materials?
\item What insights into the real materials can be gained from exact results (e.g. for the pure Kitaev model)? Are there additional exactly solvable points in the extended phase diagram?
\item Given the potential to realize a variety of anisotropic magnetic Hamiltonians in real materials, are exotic states other than the Kitaev spin liquid accessible? Where should one look?
\item What new avenues can we expect when driving anisotropic magnetic materials out of equilibrium? Mapping magnetic dynamics onto charge excitations may be a suitable way to proceed.\cite{alpichshev2015,nembrini2016}
\end{itemize}

Given the plethora of essential questions, both theoretical and experimental, there is no doubt that the study of Kitaev-Jackeli-Khaliullin materials will continue to inspire for years to come.

\section{Acknowledgements}
The field of Kitaev materials attracted hundreds of scientists over the last
decade, and it will not be possible to mention everyone who provided us with
new insights and inspiring ideas during conference talks and informal meetings.
Nevertheless, we would like to deeply acknowledge the teams in Augsburg
(Friedrich Freund, Anton Jesche, Rudra Manna, Soham Manni, and Ina-Marie
Pietsch), Dresden (Nikolay Bogdanov, Liviu Hozoi, Vamshi Katukuri, Satoshi
Nishimoto, and Ravi Yadav), Frankfurt (Harald Jeschke, Ying Li, and Kira
Riedl), and Mohali (Ashiwini Balodhi and Kavita Mehlawat), as well as Radu
Coldea, Giniyat Khaliullin, Daniel Khomskii, Igor Mazin, Ioannis Rousochatzakis, and Steph
Williams.
Last but not least, we are grateful to our funding agencies, Alexander von
Humboldt Foundation through the Sofja Kovalevskaya Award (AAT), Deutsche
Forschungsgemeinschaft through grants TRR49 (Frankfurt), TRR80 and SPP1666
(Augsburg), and SFB1143 (Dresden), as well as DST, India through Ramanujan
Grant No. SR/S2/RJN- 76/2010 and through DST Grant No. SB/S2/CMP-001/2013 (YS).

%

\end{document}